\documentclass[%
reprint,
superscriptaddress,
longbibliography,
amsmath,amssymb,
aps,
prb,
floatfix,
]{revtex4-2}

\usepackage{graphicx}%
\usepackage{xcolor}
\usepackage{dcolumn}%
\usepackage{bm}%
\usepackage[
colorlinks,
linkcolor=blue,
citecolor=blue,
urlcolor=blue
]{hyperref}%
\usepackage[
subrefformat=parens,
labelformat=parens,
caption=false
]{subfig}
\usepackage{booktabs}

\DeclareMathOperator{\csch}{csch}
\renewcommand{\Re}{\operatorname{Re}}

\begin{document}

\title{Thermal interferometry of anyons}
\author{Zezhu Wei}
\affiliation{Department of Physics, Brown University, Providence, Rhode Island 02912, USA}
\affiliation{Brown Theoretical Physics Center, Brown University, Providence, Rhode Island 02912, USA}
\author{Navketan Batra}\affiliation{Department of Physics, Brown University, Providence, Rhode Island 02912, USA}
\affiliation{Brown Theoretical Physics Center, Brown University, Providence, Rhode Island 02912, USA}
\author {V. F. Mitrovi{\'c}}\affiliation{Department of Physics, Brown University, Providence, Rhode Island 02912, USA}
\author{D. E. Feldman}
\affiliation{Department of Physics, Brown University, Providence, Rhode Island 02912, USA}
\affiliation{Brown Theoretical Physics Center, Brown University, Providence, Rhode Island 02912, USA}

\date{\today}

\begin{abstract}
Anyonic interferometry probes the braiding phases of excitations in topologically ordered matter. This technique is well established for charged quasiparticles in the fractional quantum Hall effect. We propose to extend it to neutral anyons, such as Ising anyons in Kitaev magnets and quasiparticles in other neutral spin liquids. We find that the thermal current through an interferometer is sensitive to the statistics of tunneling quasiparticles. We present a systematic investigation of signatures of various Abelian and non-Abelian topological orders in Fabry-P\'erot and Mach-Zehnder interferometers.
The heat current through a Fabry-P\'erot device is different for different topological orders and depends on the topological charge inside the interferometer. A Mach-Zehnder device shows interference in topologically trivial systems only. For a non-trivial statistics, the heat current reduces to the sum of the contributions from two constrictions in the interferometer.
Furthermore, we identify another probe of topological order that involves the scaling of the thermal current through a single tunneling contact at low temperatures. The current shows a universal temperature dependence, sensitive to the topological order in the system. 

\end{abstract}

\maketitle

\section{Introduction}

A key feature of topological order is the existence of anyons obeying fractional statistics~\cite{review-FH}. 
The statistics of Abelian anyons manifests itself in braiding phases, accumulated by particles traveling around each other. To define the statistics of non-Abelian anyons one also needs to know how different particles fuse into composite anyons. Fractional statistics has been discussed for decades in the context of the fractional quantum Hall effect (QHE), and multiple probes of anyons in the quantum Hall effect have been proposed~\cite{review-FH}. Several of them have recently been implemented~\cite{review-FH}.

The bulk-edge correspondence hypothesis connects the statistics of anyons in the bulk of a 2D system with the structure of a 1D gapless edge theory \cite{WenBook}. The latter determines the quantized thermal conductance at the temperatures much below the bulk energy gap \cite{t1,t2,t3}. Thus, the experimentally measured thermal conductance \cite{texp1,texp2,texp3} gives an evidence of fractional statistics. In particular, fractional quantization of thermal conductance has brought experimental evidence of non-Abelian statistics in the QHE at $\nu=5/2$ in GaAs \cite{texp2}. Thermal conductance yields however a rather indirect evidence of statistics. A more direct approach involves anyon collision experiments \cite{collider-exp}. Arguably, the most direct approach is anyonic interferometry~\cite{chamon1997:PhysRevB.55.2331,stern2006:PhysRevLett.96.016802,eo2,
halperin2011:PhysRevB.83.155440,law2006:PhysRevB.74.045319,MZ2,feldman2007:shot_noise,willett10,manfra19,manfra20}.

\begin{figure}[htbp]
    \centering
    \includegraphics[scale=0.8]{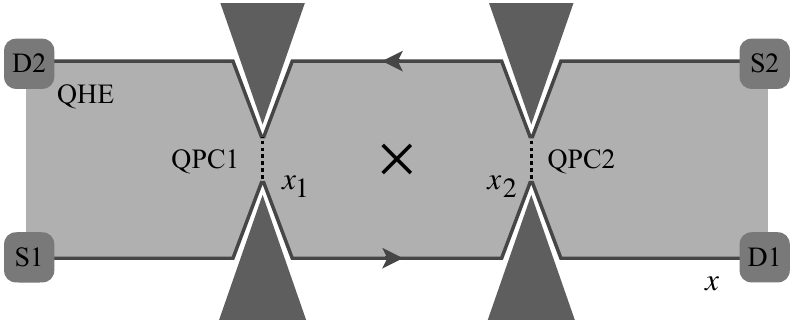}
    \caption{Schematics of an electronic interferometer. Constrictions bring the QHE edges closer. This facilitates the tunneling of charge from one edge to the other. Interference between the two paths that connect source S1 to drain D2 gives information about a localized anyon indicated with a $\times$ symbol.}
    \label{fig:electronic_FP_interferometer}
\end{figure}

The idea of anyonic interferometry naturally follows from the definition of the braiding phase. The schematics of the setup is illustrated in Fig.~\ref{fig:electronic_FP_interferometer}. Two QHE edges are brought close to each other at two constrictions. Charge tunnels between the edges at the constrictions and hence two paths emerge, which connect source S1 and drain D2
via one of the two constrictions. The electric current in D2 depends on the phase difference of the two trajectories. The phase difference includes a sample-dependent contribution from the constrictions, the Aharonov-Bohm phase, proportional to the device area, and the statistical phase, which depends on the number of anyons localized in the device. Only the sum of the three phases is observed, but they can be disentangled by changing the magnetic field. Indeed, the non-universal phase shows a weak field dependence, the Aharonov-Bohm phase changes continuously, and the statistical phase jumps every time a new anyon enters the device. This has been observed~\cite{manfra20} at $\nu=1/3$, and interesting interferometry data exist at other filling factors~\cite{review-FH}.

Quantum Hall anyons carry charge. What about neutral systems such as Kitaev magnets~\cite{kitaev2006:anyons}? The Aharonov-Bohm technique is no longer applicable in the absence of an electric current. We propose to employ thermal currents instead. In contrast to an electric, spin, or any other current, an energy current can flow in any system. This idea was introduced for one particular anyon type,
Ising anyons in Kitaev magnets, in our earlier Letter~\cite{wei2021:PhysRevLett.127.167204} and in a rather different form  in Ref.~\cite{prx-t-int}.
In this article we systematically address signatures of various possible Abelian and non-Abelian fractional statistics in thermal interferometry experiments. The idea applies to both charged and neutral anyons, but has
to be implemented in somewhat different ways in those two cases. Indeed, the magnetic field may not be a useful knob in chargeless systems, and hence a different knob is needed. As we show, useful information comes instead from the comparison of different interferometer topologies.

A possibility~\cite{RuCl3} that $\alpha$-RuCl$_3$ hosts a neutral Kitaev liquid with non-Abelian Ising statistics has attracted much interest recently. The existing experimental results are controversial~\cite{jclub-lee}, and one of our motivations consists in the development of a probe of anyonic statistics, suitable to a Kitaev spin liquid.

Thermal interferometry of fermions has been previously investigated for heterostructures based on topological superconductors~\cite{ti-te}. We demonstrate that heat transport changes dramatically for anyons in comparison with fermions and bosons.

We introduce two interferometer topologies: Fabry-P\'erot and Mach-Zehnder in Sec.~\ref{sec:model}. 
We also briefly review Ising topological liquids~\cite{kitaev2006:anyons} in that section, since the Ising topological order is particularly important for us.  
An interferometer is made of two constrictions. As a starting point it is hence essential to analyze a single constriction. We do that in Sec.~\ref{subsec:boson_single} in the simplest problem of tunneling between two separate topological liquids. In that case, only bosons can tunnel through a point contact.  We consider a single tunneling contact in an Ising liquid in Sec.~\ref{sec:Ising-singleQPC} and address a general statistics in Sec.\ref{subsec:arbitary_single}. In the low-energy limit, the heat current through a constriction exhibits scaling as a function of the temperature at a fixed ratio of the temperatures on the two sides of the device. It is easy to identify the scaling exponent for an arbitrary anyon statistics. The exponent is different for different anyon types and hence can be used as a probe of topological order. In the most interesting case of a Kitaev spin liquid, we go beyond scaling analysis and derive a general expression for the heat current  as a function of the two temperatures. 

A single-constriction probe of statistics is indirect. The bulk of the paper focuses on double-constriction geometries, which allow probes of anyon braiding. 
Sec.~\ref{subsec:boson_double} considers the Fabry-P\'erot and Mach-Zehnder interferometers made of two separate topological liquids so that only bosons can tunnel through the vacuum between the two liquids. 
After that warming-up exercise, Sec.~\ref{sec:ising} contains a detailed study of the two interferometer geometries for Ising anyons in Kitaev liquids.   
We discover that the heat current depends on the topological charge localized inside a Fabry-P\'erot interferometer. This can be used to probe statistics provided that we can control the trapped topological charge. If such control is unavailable, a dramatically different behavior of the heat current in a Mach-Zehnder interferometer can be combined with Fabry-P\'erot data to identify fractional statistics.

Sec.~\ref{sec:arbitary} investigates interferometry for an arbitrary topological order. The case of Abelian statistics is straightforward in both interferometer geometries. 
For non-Abelian statistics we have to separately consider the tunneling of anyons which are identical and different from their antiparticles. 
Sec.~\ref{sec:noise} addresses another signature of topological order: telegraph noise of heat current in Fabry-P\'erot devices with a hole. 
We discuss experimental realizations and summarize in Sec.~\ref{sec:conclusion}.

Several Appendixes contain technical details. Appendix~\ref{appendix:psi} addresses a toy model of fermion tunneling in interferometers for Kitaev spin liquids. Appendix~\ref{appendix:psi-dx-psi} contains detailed calculations for an interferometer made of two separate Kitaev magnets. Appendix~\ref{appendix:ising-correlation} deals with correlation functions of Ising anyon operators in Kitaev systems. Appendix~\ref{appendix:sigma} goes through the tedious calculations of the heat current in a Fabry-P\'erot interferometer made of a single piece of a Kitaev material. Appendix~\ref{appendix:Mach-zehnder_phase} discusses subtleties of Mach-Zehnder interferometry. Appendix~\ref{appendix:anyon_example} considers an exotic topological order that defies naive expectations of how interference of anyons works in a Fabry-P\'erot device. Appendix~\ref{appendix:noise} contains detailed calculations of telegraph noise. Appendix~\ref{appendix:inter_size} addresses the dependence of the heat current on the interferometer size for systems with a single edge mode.

\section{Models}\label{sec:model}

In this section we introduce the three basic models we consider below: a single constriction between two edges of a topological liquid, a Fabry-P\'erot interferometer, and a Mach-Zehnder interferometer.

We focus on systems with a bulk energy gap and gapless edge states~\cite{review-FH,nayak2008:RevModPhys.80.1083}. 
The fractional quantum Hall effect gives rise to many such systems. Another relevant situation involves some spin liquids, a Kalmeyer-Laughlin liquid \cite{semion2} being the simplest example.
Non-Abelian statistics in Kitaev magnets has recently attracted much interest, and we will pay particular attention to Kitaev magnets \cite{kitaev2006:anyons}. Their edge theory contains a single chiral Majorana mode.
The low-energy Hamiltonian of one right(left)-moving edge is~\cite{kitaev2006:anyons,nayak2008:RevModPhys.80.1083}
\begin{equation}\label{eq:free_field_ham}
H = \mp \frac{i v}{4\pi} \int dx\, \psi \partial_x \psi,
\end{equation}
where $v$ is the edge velocity.
The Majorana fermion $\psi$ satisfies the anticommutation relation,
\begin{equation}
\{ \psi(x), \psi(y) \} = 2 \pi  \delta(x-y).
\end{equation}

Kitaev magnets contain three types of anyons: trivial bosons $\bm{1}$, Majorana fermions $\psi$, and Ising anyons $\sigma$. These  quasiparticles obey the following fusion rules:
\begin{align}
    \psi \times \psi = \bm{1}; ~~~~ \psi \times \sigma = \sigma; ~~~~ \sigma \times \sigma = \bm{1}+\psi,
\end{align}
where the final equality represents two possible fusion outcomes. The topological spin $\theta_x$ of these quasiparticles is given by $\theta_{\bm{1}}=1$, $\theta_{\psi}=-1$, and $\theta_{\sigma}=e^{i\pi/8}$. The topological spin determines the phase $\phi_{ab}^c$ accumulated when a quasiparticle of type $a$ encircles a quasiparticle of type $b$ in counterclockwise sense
under the assumption that they fuse to a quasiparticle of type $c$: 
$\exp(i\phi_{ab}^{c})=\theta_c/(\theta_a\theta_b)$.  Another important piece of information is the quantum dimension of anyons. It is 1 for $ \bm{1} $ and $\psi$, and $\sqrt{2}$ for $\sigma$.  
More details and a discussion of other topological orders can be found in Refs.~\cite{review-FH,bonderson2007:non-abelian}.

\subsection{Single point contact}
\begin{figure}[!htb]
	\bigskip
	\centering\scalebox{1.3}[1.3]
	{\includegraphics{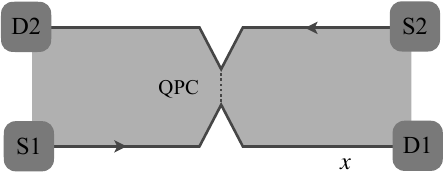}}
	\caption{{A single quantum point contact (QPC) is shown, where the two counter-propagating edges of the spin liquid come close and the tunneling between the edge modes takes place as indicated with a dashed line.}}
	\label{fig2:thermal}
\end{figure}
To discuss the tunneling behavior in the presence of point contacts between two different edges, we consider the model depicted in Fig.~\ref{fig2:thermal}. 
The two edges host counter-propagating edge modes, and could either be edges of the same spin liquid, or different spin liquids.
The lower and upper edges have two different temperatures, $ T_1 $ and $ T_2 $. 
This is achieved by bringing either of them in thermal equilibrium with its source S1 or S2, maintained at the temperature $T_1$ or $T_2$.

The Hamiltonian of this single-point-contact tunneling model is of the form
\begin{equation}
H = H_1 + H_2 +  H_T, 
\end{equation}
where $ H_{1,2} $ are the free Hamiltonians for the two chiral edges, and $ H_T $ is the tunneling Hamiltonian describing the tunneling of quasiparticles through the quantum point contact.

In general, the tunneling Hamiltonian can be written in the following form,
\begin{equation}\label{eq:hamiltonian_not_own_anti}
H_T = \Gamma \hat{T} + \Gamma^{*} \hat{T}^{\dagger},
\end{equation}
where $ \hat{T} $ is a tunneling operator responsible for transporting one quasiparticle from
the lower edge to the upper edge.
When the tunneling quasiparticle is its own antiparticle, $ \hat{T} $ and $ \hat{T}^{\dagger} $ are equivalent, hence,
\begin{equation}\label{eq:hamiltonian_own_anti}
H_T = \Gamma \hat{T},
\end{equation}
where $ \Gamma $ is a real tunneling amplitude that ensures Hermiticity of the tunneling Hamiltonian $ H_T $.
The Hamiltonians (\ref{eq:hamiltonian_not_own_anti},\ref{eq:hamiltonian_own_anti}) include only one most relevant tunneling process. This is justified at low energies for many topological orders.

In the presence of a tunneling point contact, the heat current flowing along the edges can tunnel across the contact, thus introducing a tunneling heat current $ I_T $. 
Here we treat the contact Hamiltonian $H_T$ as a perturbation and assume a small $ \Gamma $.
We can find the operator $ \hat{I}_T $ for the heat current using the Heisenberg equation of motion,
\begin{equation}\label{eq:thermal-current-operator}
\hat{I}_T = \frac{\partial H_1}{\partial t} = -i[H_1, H_1+H_2+H_T] = -i[H_1,H_T].
\end{equation}
Strictly speaking, the above expression gives the energy current. It is the same as the heat current provided that the electric current is zero. Otherwise, a correction, proportional to the square of the electric current, is conventionally subtracted~\footnote{The heat current is usually defined as the difference between the energy current and the product of the chemical potential and the particle current. The heat and energy currents become the same if the product of the particle number and the chemical potential is subtracted from the Hamiltonian. In the case of the Ising topological order, the number of the Majorana fermions is not defined and the chemical potential is always zero}. 
We will ignore this complication, that is, we will assume a zero electric current.
In the most interesting case of spin liquids, the electric current is constrained to be zero.
In quantum Hall systems with a non-zero electric current, the energy and heat currents can be related to each other via known results for the electric current.

To the lowest order in perturbation theory, the expectation value of the heat current is
\begin{equation}\label{eq:thermalcurrent-perturbationtheory}
\langle I_T(t) \rangle =  -i \int_{-\infty}^{t} d t'  \langle{[\hat{I}_T(t),  H_T(t')]}\rangle .
\end{equation}
The tunneling heat current should be proportional to $ |\Gamma|^2 $, $I_T=r(T_1,T_2)|\Gamma|^2$, where the factor 
 $r(T_1,T_2)$ depends on the details of the edge theory and the nature of tunneling quasiparticles.
 Typically, the tunneling of one quasiparticle type dominates in the low-energy regime of interest for this paper. 
 The dominant tunneling operator is the most relevant tunneling operator in the renormalization group sense. 
 We will assume below that only one quasiparticle type together with its anti-particle can tunnel. 
\begin{figure}[htbp]
\bigskip
\centering\scalebox{1}[1]
{\includegraphics[scale=1.3]{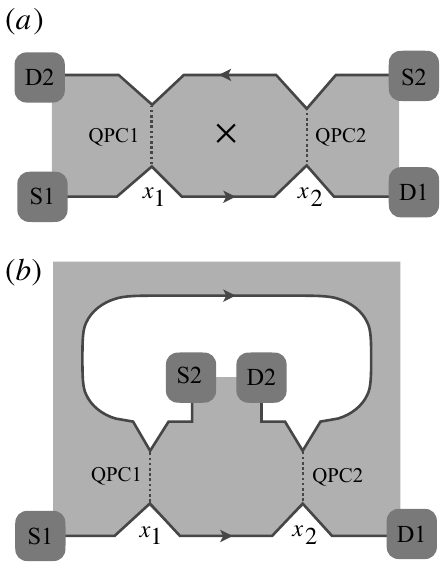}}
\caption{{ Schematics of (a) Fabry-P\'erot and (b) Mach-Zehnder interferometers. Heat travels from sources S1 and S2 to drains D1 and D2 along chiral edges and tunnels between the edges at the two point contacts shown with dashed lines. A localized anyon is marked with a $\times$ symbol.}}
\label{fig1:thermal}
\end{figure}

\subsection{Fabry-P\'erot interferometer}\label{sec:model_FP}
Following the discussion of the single-point-contact tunneling, we are in a position to consider thermal interferometers.
Electronic Fabry-P\'erot interferometers have been introduced to study the Aharonov-Bohm effect and fractional statistics in the quantum Hall regime~\cite{review-FH}. This paper considers a thermal Fabry-P\'erot interferometer shown in Fig.~\ref{fig1:thermal}(a).
The device is similar to its electronic counterpart.
It contains two point contacts, QPC1 and QPC2, located at coordinates $ x_{1} $ and $ x_{2} $.
When a heat-carrying quasiparticle tunnels through the interferometer from the lower edge to the upper edge, it could take either of the two paths S1-QPC1-D2 or S1-QPC2-D2.
This results in quantum interference, whose phase depends on the quasiparticle excitations in the central region between QPC1 and QPC2 as well as the device's size and the tunneling contacts' details.
For electronic interferometers in QHE systems, the interference phase contains a sample-dependent contribution from QPCs, an Aharonov-Bohm phase, proportional to the magnetic flux through the interferometer, and a statistical phase due to quasiparticle excitations inside the interference loop. 
On the other hand, for a thermal interferometer probing charge-neutral excitations, the Aharonov-Bohm phase is absent.

The tunneling Hamiltonian in a Fabry-P\'erot interferometer is 
\begin{equation}\label{eq:FP_tunneling_not_own_anti}
H_T = \Gamma_1 \hat{T}_1 + \Gamma_2 \hat{T}_2 + \mathrm{H.c.},
\end{equation}
where $ \Gamma_{1,2} $ are two generally complex tunneling amplitudes, $ \hat{T}_1 $ is the tunneling operator that corresponds to the transfer of a quasiparticle from the lower to upper edge through QPC1, and $ \hat{T}_2 $ corresponds to the transfer through QPC2.
In the absence of localized quasiparticles in the interference loop, the tunneling thermal current takes the general form,
\begin{align}
I_T ={} & (|\Gamma_1|^2 + |\Gamma_2|^2)  r(T_1,T_2)\nonumber\\ 
&+ 2\Re{[\Gamma_1 \Gamma_2^{*} \tilde{r}(T_1, T_2, L_1 + L_2) ]},
\end{align}
where $\Re$ stands for the real part, $ L_{i} $ denotes the distance between the two point contacts along edge $i=1,~2$, and
$ \tilde{r}(T_1, T_2, L_1 + L_2) $ describes the interference term in the thermal current.
In the presence of bulk quasiparticles, this term depends additionally on the statistical phase induced by these localized quasiparticles.
The sum of the two distances enters the coefficient $\tilde{r}$, if only one edge mode exists and the edge velocities are identical on the two edges of the device. Otherwise,
the length dependence of $\tilde{r}$ is more complicated. See Appendix~\ref{appendix:inter_size} for
the derivation of the length dependence and a detailed discussion of the relevant assumptions.

In an Abelian quantum Hall liquid with the filling factor $ \nu=1/(2m+1) $, the statistical phase accumulated by an anyon of charge $\nu e$ around $n$ localized anyons of the same charge is equal to $\phi= 2\pi \nu n$.
However, for the tunneling of non-Abelian quasiparticles, we need to take the anyon fusion rule into consideration since it is possible for two non-Abelian anyons to fuse in more than one fusion channel. A detailed discussion can be found in Sec.~\ref{sec:arbitary}.

\subsection{Mach-Zehnder interferometer}\label{sec:model_mach-zehnder}

A Mach-Zehnder interferometer \cite{review-FH,MZ-heiblum,MZ-review} shown in Fig.~\ref{fig1:thermal}(b)  is superficially similar to a Fabry-P\'erot interferometer. It also has two tunneling contacts and two interfering paths from S1 to D2. Yet, it has an important topological difference from the Fabry-P\'erot setup: Drain D2 is inside the interference loop.
Hence, in contrast to Fabry-P\'erot interferometry, each tunneling event changes the localized topological charge in the central region.
As a result, the statistical phase $ \phi$, accumulated by an anyon on the interference loop,  changes after each quasiparticle tunneling event.
Also, as seen from Fig.~\ref{fig1:thermal}(b), the two edges in a Mach-Zehnder interferometer have the same propagation direction, thus, the thermal current depends on the difference of the distances between the point contacts on the two edges and not the sum, see Appendix~\ref{appendix:inter_size} and Ref.~\cite{law2006:PhysRevB.74.045319}.

The state of a Mach-Zehnder interferometer is characterized by the localized topological charge in the central region.
Transitions between different states can be analyzed through a continuous-time Markov chain model.

The situation is particularly interesting if the tunneling anyon, whose topological charge is denoted as $ x $, is its own antiparticle. The tunneling Hamiltonian for a single point contact is given in Eq.~(\ref{eq:hamiltonian_own_anti}).
The tunneling Hamiltonian in a Mach-Zehnder interferometer is
\begin{equation}\label{eq:MZ_Hamiltonian_own_anti}
H_T = \Gamma_{1} \hat{T}_{1} + \Gamma_{2} \hat{T}_{2} e^{i \alpha},
\end{equation}
where $ \Gamma_{1,2} $ are two real tunneling amplitudes at the two quantum point contacts, and the phase $ \alpha $ ensures Hermiticity~\cite{wei2021:PhysRevLett.127.167204}.
We compute the phase $\alpha$ in Appendix~\ref{appendix:Mach-zehnder_phase}.
When the tunneling of $ x $ changes the localized topological charge from $ a $ to $ b $, the tunneling rate can be expressed as~\cite{feldman2007:shot_noise}
\begin{equation}\label{eq:general_MZ_rate}
p_{xa}^{b} = P_{xa}^{b} \tilde{p} (T_1, T_2, \Gamma_1, \Gamma_2, \phi_{xa}^{b}, |L_1 - L_2|),
\end{equation}
where 
$ P_{xa}^{b} = N_{xa}^b {d_b}/({d_x d_a})$ is the fusion probability for $ x\times a \to b $,
$N_{xa}^b$
	is the fusion multiplicity, $d_\alpha$ is the quantum dimension of anyon $\alpha$, 
and $ \tilde{p} $ incorporates the dependence on the temperatures $ T_{1,2} $, the tunneling amplitudes $ \Gamma_{1,2} $, the statistical phase $\phi_{xa}^{b}$, and the interferometer size.
When the path of topological charge $ x $ encloses localized charge $ a $, according to the algebraic theory of anyons, the statistical phase is $ \exp(i \phi_{x a}^{b}) = {\theta_b}/(\theta_a \theta_x)$, where $\theta_{\alpha}$ are topological spins~\cite{review-FH,kitaev2006:anyons}. 

In what follows we will assume that $|L_{1}-L_{2}|$ is much shorter than the thermal length $ v/T$, where $v$ is the edge velocity. This will allow neglecting the interferometer size.

If we use $ f_{a} $ to denote the probability of the localized topological charge being $ a $,
we can write down the following kinetic equation~\cite{feldman2007:shot_noise},
\begin{align}\label{eq:kinetic}
\dot{f}_a 
=  \sum_b f_b p_{x b}^{a} - \sum_{c} f_{a} p_{x a}^c.
\end{align}
This equation can also be written in a matrix form,
$ \dot{f}_a  = M_{a b} f_b $,
where the matrix entries $ M_{ab} $ are the tunneling rates,
\begin{align}
M_{ab} = p_{x b}^a - \delta_{ab} \sum_{c} p_{xa}^c.
\end{align}
When the system is in dynamical equilibrium, $ \dot{f}_a=0 $, one can solve for $ f_a $.
The average thermal current is given by
\begin{align}\label{eq:MZ_thermalcurrent}
I^{\rm MZ} = \overline{\Delta E} \sum_{ab} f_a p_{xa}^b,
\end{align}
where $ \overline{\Delta E} $ is the average heat transferred across the point contact by a tunneling event. Since we neglected the interferometer size, the statistical distribution of the transferred energy in one tunneling event is the same for all combinations of the topological charges $a$ and $b$, and the same as for a single tunneling contact. Indeed, in our limit, the two constrictions can be fused into one.

As an example, a detailed calculation for interferometers built for probing the Ising topological order is shown in Sec.~\ref{sec:ising}.

\section{Tunneling Between Topological Liquids}\label{sec:tunneling_Between_Topological}

As a warming-up exercise, we consider tunneling between two Ising topological liquids through vacuum.

\subsection{Single constriction}\label{subsec:boson_single}

In the case of two adjacent spin liquids, some edge excitations may tunnel from one spin liquid to the other. Before we compute the thermal current in the two-constriction geometry, consider first the case of a single point contact. In this case, we will not obtain any interference effects. Nonetheless, we do obtain some non-trivial thermal transport. 

The edge Hamiltonians have the standard form (\ref{eq:free_field_ham}). 
Every contribution to the Hamiltonian must be topologically trivial. In particular, any allowed interaction between the two topological liquids is described by operators, which are products of two Bose fields acting on each of the liquids.
Hence, the form of the tunneling Hamiltonian is such that only pairs of Majorana fermions tunnel,
\begin{equation}
H_T = - \Gamma \psi_1(x_0)\partial_x\psi_1(x_0)\psi_2(x_0)\partial_x\psi_2(x_0).
\end{equation}
Here $x_0$ is the location of the constriction. 

We can now define the thermal current as the time derivative of the free Hamiltonian of one of the edges and use the Heisenberg equations of motion as in Eq.~(\ref{eq:thermal-current-operator}). One obtains
\begin{align}\label{eq:psi_dx_psi_current_operator}
\hat{I}_T = -v\Gamma &\left( \partial_x\psi_1(x_0)\partial_x\psi_1(x_0) + \psi_1(x_0)\partial^2_x\psi_1(x_0) \right) \nonumber\\ 
&\times \psi_2(x_0)\partial_x\psi_2(x_0).
\end{align}
Using perturbation theory (\ref{eq:thermalcurrent-perturbationtheory}), in Appendix~\ref{appendix:psi-dx-psi}, we compute the thermal current between two spin liquids to the lowest order in the tunneling amplitude $\Gamma$ as,
\begin{align}\label{eq:non-int_psi-dx-psi}
\langle I_{T} \rangle^{\text{non-int}}_{\Gamma}  = \frac{4\pi^9\Gamma^2}{2835v^8} &\Big[ 41(T_2^8 - T_1^8) + 62 T_1^2T_2^2(T_2^4-T_1^4)\Big]. 
\end{align}
This is the contribution to the thermal current due to a single constriction between two spin liquids. 
Its non-trivial temperature dependence is a signature of the Ising topological order.
It is instructive to compare the result with the heat current in a Luttinger liquid of topologically trivial bosons. The action density is $L_n\sim (\partial_t \theta_{n})^2+(\partial_x\theta_n)^2$ on edge $n=1,2$. The tunneling operator $\hat T\sim\partial_x\theta_1\partial_x\theta_2$. An easy calculation shows that the thermal current scales as the fourth power of the temperatures $\sim T_1^4, T_2^4$. For a 1D system of free electrons, the heat current scales as the square of the temperature. Thus, the $\sim T_1^8, T_2^8$ scaling in the above result is a signature of a non-trivial topological order. The same is true for the non-trivial coefficients 41 and 62.

In the two-constriction geometry, as considered in the next part of the section, the result we just obtained will exactly be the non-interference contribution from each of the constrictions. A detailed derivation of the above expression can be found as the non-interference part of the full calculation in Appendix~\ref{appendix:psi-dx-psi}, where we consider a  Fabry-P\'erot interferometer. The calculations rely on some equations from Appendix~\ref{appendix:psi}, where we consider a simpler model, in which single Majorana fermions are allowed to tunnel between the edges. This is only allowed when the edges surround the same spin liquid. In such a situation, fermion tunneling is not the most relevant tunneling process, as discussed in the next section. It can become the leading contribution to transport near a resonance.

\subsection{Double constriction}\label{subsec:boson_double}

We now turn our attention to the double constriction geometry. In this case, interference effects will come into play. The two-constriction geometry we consider here is the Fabry-P\'erot interferometer made of two spin liquids. 

\begin{figure}[!htb]
\bigskip
\centering
 {\includegraphics[width=0.9\columnwidth]{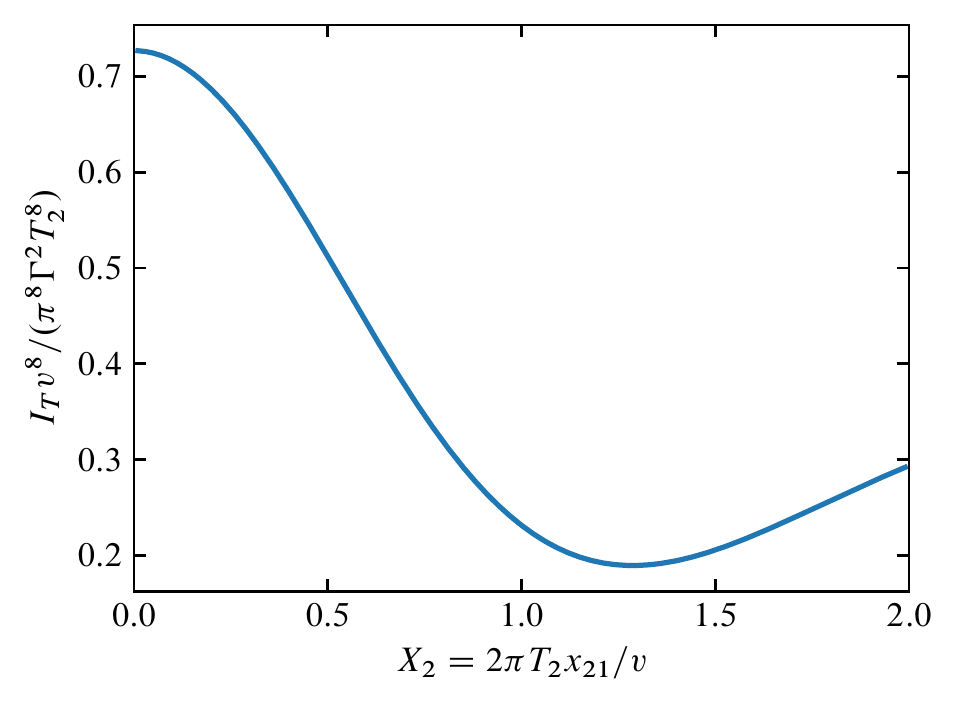}}
\caption{{This curve shows how the total thermal current $ I_T = 2 \langle I_T\rangle^{\text{non-int}}_{\Gamma} + \langle I_T \rangle^{\text{int}}$ varies with $ x_{21} $ when the limit $ T_1\to0 $ is taken.}}
\label{fig:current_psi_dx_psi}
\end{figure}

Similar to a single constriction geometry, the tunneling Hamiltonian is just the sum of individual tunneling Hamiltonians at the two constrictions, $x_1$ and $x_2$, 
\begin{align}
    H_T = - \sum_{i=1,2}\Gamma_i \psi_1(x_i)\partial_x\psi_1(x_i)\psi_2(x_i)\partial_x\psi_2(x_i).
\end{align}
Using this tunneling Hamiltonian, we find the tunneling thermal current operator as defined in Eq.~(\ref{eq:thermal-current-operator}),
\begin{align}
    \hat{I}_T = -\sum_{i=1,2}\Big[ \Gamma_i &\left( \partial_x\psi_1(x_i)\partial_x\psi_1(x_i) + \psi_1(x_i)\partial^2_x\psi_1(x_i) \right) \nonumber\\ 
&\times \psi_2(x_i)\partial_x\psi_2(x_i)\Big].
\end{align}
Using perturbation theory in Eq.~(\ref{eq:thermalcurrent-perturbationtheory}), one can now compute the thermal current between two spin liquids to the lowest order in the tunneling amplitude. In the $T_1=0$ limit and setting $\Gamma_1=\Gamma_2=\Gamma$, we find this expression to be,
\begin{align}
    \langle I_T\rangle = 2 \langle I_T\rangle^{\text{non-int}}_{\Gamma} + \langle I_T \rangle^{\text{int}}
\end{align}
where,
\begin{align}
    \langle I_T \rangle^{\text{int}} & =  \frac{8\pi^9 \Gamma^2 T^8_2 }{3v^8}  \Big[  (7-80\coth^2X_2 + 105 \coth^4X_2) \nonumber\\
    &\qquad\qquad\times \frac{1}{\sinh^4{X_2}}   - \frac{105}{X_{2}^8} + { \frac{10}{X_{2}^6}  } \Big].
\end{align}
Here we have defined $X_i = 2\pi T_i x_{21}/v$, where $x_{21}$ is the separation between the two constrictions. 
We assume the separation to be the same on both edges.
The non-interference term $\langle I_T \rangle^{\text{non-int}}_{\Gamma}$ is just the thermal current obtained in the single constriction geometry~(\ref{eq:non-int_psi-dx-psi}). The variation of the total thermal tunneling current with the separation between the two constrictions $X_2= 2\pi T_2 x_{21}/v$ is shown in Fig.~\ref{fig:current_psi_dx_psi}. The general expression for this tunneling thermal current in the case of $T_1 \ne 0$ and when the tunneling amplitudes at the two constrictions are not equal, i.e., $\Gamma_1\ne\Gamma_2$, is given in Appendix~\ref{appendix:psi-dx-psi}. 

A curious feature of the above result is a non-monotonous dependence of the heat current on the distance between the tunneling contacts, Fig.~\ref{fig:current_psi_dx_psi}.
Similar behavior would be natural for charged systems in a magnetic field due to Aharonov-Bohm oscillations. Significantly, we deal with a neutral spin liquid.
See Ref.~\cite{prx-t-int} for related curious behavior.

The analysis of the Mach-Zehnder geometry is very similar and does not add much to the Fabry-P\'erot case. 
This is a consequence of the trivial mutual statistics of tunneling bosons and confined anyons.

\section{Ising anyon tunneling}\label{sec:ising}

We now address the tunneling of Ising anyons between the two edges of a topological liquid.
As we will see, different tunneling currents correspond to different topological charges, trapped in an interferometer. It is not, however, obvious how to control the trapped charge. As an alternative approach to probing anyon statistics, one can compare heat currents through Fabry-P\'erot and Mach-Zehnder interferometers.

\subsection{Single constriction}\label{sec:Ising-singleQPC}

We start with a single constriction. We discover that at a fixed ratio of the edge temperatures $T_1/T_2$, the 
tunneling heat current scales as $\sim T_1^{1/4}$. This is a signature of fractional statistics.

The most relevant tunneling operator transfers a single Ising anyon from one edge of the spin liquid to the other. These two edges, being the edges of the same spin liquid, can be thought of as parts of a single edge connected by an infinitely remote section~\cite{bishara2008:PhysRevB.77.165302,nilsson2010:PhysRevB.81.205110}. 
Correlation functions for the entire system decompose as products of correlation functions on the upper and lower edges along with a phase factor that depends on topological order~\cite{fendley2007:PhysRevB.75.045317}. The decomposition of correlation functions is discussed in Appendix~\ref{appendix:ising-correlation}. 

\begin{figure}
	\centering
	\includegraphics[scale=1]{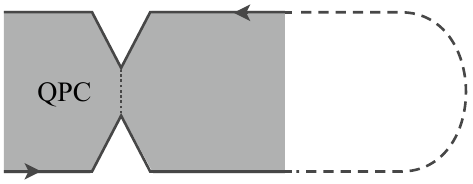}
	\caption{Geometry, in which the two edges of the spin liquid are treated as spatially separated parts of the same edge. The dashed line shows a long segment, of length $L$, connecting the two edges of the interferometer. 
	}
	\label{fig:infinite_segment_connecting_edges}
\end{figure}

Within the conformal-field-theoretic treatment, the correlation functions of the primary operators are holomorphic functions of the complex coordinate $w=v\tau\pm i x =i(vt\pm x)$, where $\tau$ is the imaginary time. We may compute the thermal correlation functions of the primary fields using the conformal transformation between coordinates $w$ on a cylinder~\cite{difrancesco1997:conformal,smits2014:PhysRevB.89.045308} and coordinates $z$ on a plane, given as $z=e^{2\pi i w / v\beta}$, where $\beta=1/T$ is the reciprocal of the edge temperature; then the 
two-point correlation functions satisfy the following relation, 
\begin{align}
\langle\sigma(w_1)\sigma(w_2) \rangle = \left( \frac{2\pi i z_1}{v\beta}\right)^{h}\left( \frac{2\pi i z_2}{v\beta} \right)^{h} \langle \sigma(z_1)\sigma(z_2) \rangle ,
\end{align}
where $h=1/16$ is the holomorphic conformal dimension of the Ising anyon field. With this we arrive at the thermal two-point correlation function of the Ising anyon field as,
\begin{align}
\langle \sigma(w_1) \sigma(w_2)\rangle = \left(\frac{\pi T}{v}\right)^{\frac{1}{8}}\frac{1}{\sin^{\frac{1}{8}} [\pi T (w_1-w_2)/v]}
\end{align}
The temperature $T$ appearing in the two-point function corresponds to one of the edges. Introducing a regulator $\epsilon$, the two-point function for the two edges can be written as,
\begin{eqnarray}\label{eq:ising-correlations}
    \langle \sigma_1(x,t) \sigma_1(0,0)\rangle = \frac{(\pi T_1/v)^{1/8}}{\sin^{1/8} [\pi T_1 (\epsilon+i(t-x/v))]} \\
    \langle \sigma_2(x,t) \sigma_2(0,0)\rangle = \frac{(\pi T_2/v)^{1/8}}{\sin^{1/8} [\pi T_2 (\epsilon+i(t+x/v))]}
\end{eqnarray}
We may now treat the two opposite edges of the same spin liquid as different parts of a single edge connected by a long edge segment of length $L$ (See Fig.~\ref{fig:infinite_segment_connecting_edges}). This allows us to use the above result and obtain the four-point correlation function in the limit of $L\rightarrow \infty$. This has been computed in Appendix~\ref{appendix:ising-correlation}. Once we obtain the thermal four-point correlation functions, we are in a position to use the perturbation theory, as considered previously, to compute the thermal tunneling current due to Ising anyon tunneling. To separate the interference effects from the rest of the physics of the problem, we first consider the case of a single point contact and later generalize the calculation to interferometers. 

The tunneling Hamiltonian creates anyons, which fuse to vacuum, on both sides of the constriction. Since the Ising anyon is its own antiparticle, the tunneling Hamiltonian creates anyons of the same type, and thus the form of the tunneling Hamiltonian can be written as,
\begin{align}
    H_T = e^{-i\pi/16} \Gamma \sigma_2(x_0,t)\sigma_1(x_0,t).
\end{align}
The subscripts on the Ising anyon field operators denote the side, on which the edge lies and are defined in the limit $L \rightarrow \infty$, as $\sigma_2(x)\equiv\sigma(L-x)$ and $\sigma_1(x)\equiv \sigma(x)$. The phase $e^{-i\pi/16}$ is fixed by the hermiticity condition for the tunneling Hamiltonian. Indeed, the average of $H_T$ must be real.
The form of the tunneling Hamiltonian suggests that the effect of a single point contact is to allow the tunneling of a single Ising anyon from one edge to the other. This is then treated perturbatively to the lowest order. The expectation value of the tunneling thermal current is given by Eq.~(\ref{eq:thermalcurrent-perturbationtheory}). The full calculation can be found as the non-interference part of the calculation in Appendix~\ref{appendix:sigma}. We find the thermal current to be,
\begin{align}\label{eq:non-int-current_anyon}
    \langle I_T \rangle^{\text{non-int}}_{\Gamma} = \Gamma^2(\pi T_1)^{\frac{1}{8}}(\pi T_2)^{\frac{1}{8}} \frac{\cos(3\pi/8)}{4\sqrt{2}v^{1/4}} F_{\text{non-int}}(T_2/T_1),
\end{align}
where the $F_{\text{non-int}}(n)$ function in the integral representation is given as,
\begin{align}\label{eq:func-f_non-int}
    F_{\text{non-int}}(n) = \int_{0}^{\infty} d\tau \left( \frac{1}{\tau^{\frac{5}{4}}n^{\frac{1}{8}}} - \frac{\cosh (\tau)}{\sinh^{\frac{9}{8}}(\tau) } \frac{1}{\sinh^{\frac{1}{8}}(n\tau)} \right)  .
\end{align}
See Fig.~\subref*{F_int:a} for the plot of $F_{\text{non-int}}$.
Note that the calculation given in Appendix~\ref{appendix:sigma} corresponds to the Fabry-P\'erot geometry, and the single-point-contact case corresponds to the non-interference part of the full calculation. A non-linear temperature dependence of the thermal current comes from the scaling dimension of the Ising anyon tunneling operator. Thus, it provides an experimental signature for fractional quasiparticle tunneling at a single point contact. In particular, at fixed $T_2/T_1$, the tunneling heat current scales as $T^{1/4}$.

\begin{figure}[!htb]
	\centering
	\subfloat[\label{F_int:a}]{%
		\includegraphics[width=0.9\columnwidth]{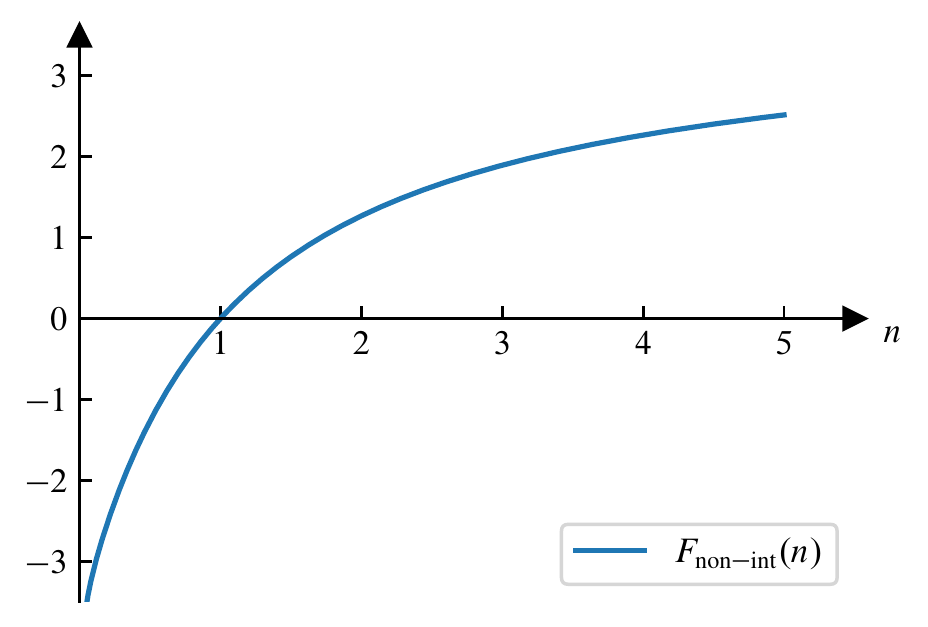}%
	}\\
	\subfloat[\label{F_int:b}]{%
		\includegraphics[width=0.9\columnwidth]{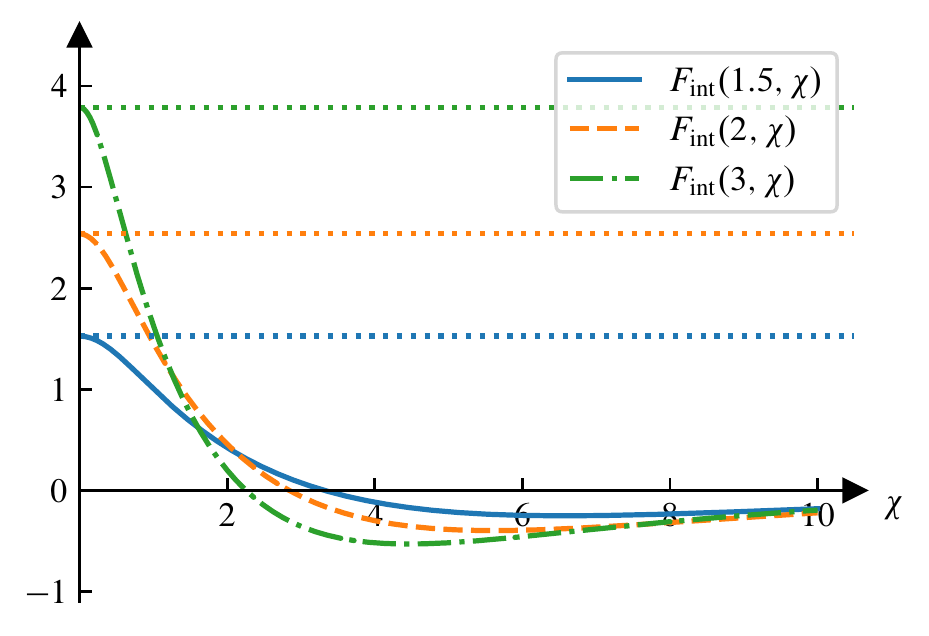}%
	}
	\caption{(Color online) (a) The behavior of the function $ F_{\text{non-int}}(n) $. (b) The behavior of the function $ F_{\text{int}}(n,\chi) $ for $n=1.5,2,3$. The dotted horizontal lines in (b) show the values of $2 F_{\text{non-int}}(n)$ for $n=1.5,2,3$, which become equal to $F_{\text{int}}(n,\chi)$ as $\chi \rightarrow 0$.}
\end{figure}

A universal scaling behavior has been predicted for tunneling electrical currents in fractional quantum Hall liquids, but theory rarely agrees with the data (for a review, see \cite{review-FH}). This has multiple reasons, at least one of which, specifically, long-range Coulomb forces \cite{tun-c,YF2013}, is absent in a spin liquid. Thus, there may be a better chance to observe scaling in the tunneling heat current  than for an electric current.

We now generalize our single-contact calculation to a Fabry-P\'erot interferometer geometry, where the fractional quasi-particle interference effects and anyon braiding effects become relevant. 

\subsection{Fabry-P\'erot geometry}

The behavior of the tunneling thermal current obtained in the previous subsection captures the non-trivial topological charge of an anyon. Single-point-contact measurements therefore provide a probe of the topological charge of the quasiparticles involved in the tunneling process. A more direct evidence of statistics involves the braiding of two quasiparticles in the two-point-contact geometry. In this section, we generalize the previous calculation to a Fabry-P\'erot interferometer geometry.

We use the two-point thermal correlation functions given in Eq.~(\ref{eq:ising-correlations}). These correlation functions are in turn used to compute the four-point correlation functions (Appendix~\ref{appendix:ising-correlation}). 
The calculation relies on the factorization into products of correlations functions on two separate edges~\cite{fendley2006:PhysRevLett.97.036801}.
To find the thermal current operator, we use the Heisenberg equation along with the tunneling Hamiltonian, which in this geometry takes the form,
\begin{align}
    H_T ={}& e^{-i\pi/16} \Gamma_1 \sigma_2(x_1,t)\sigma_1(x_1,t) \nonumber \\
    &+ e^{-i\pi/16} \Gamma_2 \sigma_2(x_2,t)\sigma_1(x_2,t),
\end{align}
where $x_i$ is the coordinate of the $i$th constriction, and correspondingly, $\Gamma_i$ is its tunneling amplitude. We now apply perturbation theory to compute the lowest order expectation value  using Eq.~(\ref{eq:thermalcurrent-perturbationtheory}). 

We start with the simplest case of a trivial topological charge trapped inside the device. We summarize the results here. Detailed calculations are contained in Appendix~\ref{appendix:sigma}.

In the simplest limit $\Gamma_1=\Gamma_2\equiv \Gamma$, we arrive at an expression that has the following form,
\begin{align}
    \langle I_T \rangle =  2\langle I_T \rangle^{\text{non-int}}_{\Gamma} + \langle I_T \rangle^{\text{int}},
\end{align}
where $\langle I_T \rangle^{\text{non-int}}_{\Gamma}$ is the contribution of a single point contact given in Eq.~(\ref{eq:non-int-current_anyon}), and $\langle I_T \rangle^{\text{int}}$ is the interference term that has the form (in the limit $\Gamma_1=\Gamma_2\equiv \Gamma$),
\begin{align}
    \langle I_T \rangle^{\text{int}} =  \Gamma^2(\pi T_1)^{\frac{1}{8}}(\pi T_2)^{\frac{1}{8}} \frac{\cos(3\pi/8)}{4\sqrt{2}v^{1/4}} F_{\text{int}}(T_2/T_1, \pi T_1x_{21}),
\end{align}
where the function $F_{\text{int}}(n,\chi)$ in the integral representation is given as,
\begin{align}\label{eq:func_F_int}
    F_{\text{int}}(n,\chi) =  \int_{0}^{\infty} d\tau &\Bigg( 
    \frac{1}{\tau^{ \frac{9}{8}}\sinh^{\frac{1}{8}}(2n\chi)} \nonumber\\
    &-\frac{\cosh (\tau)}{\sinh^{\frac{9}{8}}(\tau) } \frac{1}{\sinh^{\frac{1}{8}}(n\tau+2n\chi)} \nonumber \\
    &- \frac{\cosh (\tau+2\chi)}{\sinh^{\frac{9}{8}}(\tau+2\chi) } \frac{1}{\sinh^{\frac{1}{8}}(n\tau)}    \Bigg).
\end{align}
The function $F_{\rm int}$ is plotted in Fig.~\subref*{F_int:b}.
It can be checked that this function has the property that $\lim_{\chi\rightarrow 0}F_{\text{int}}(n,\chi) = 2F_{\text{non-int}}(n)$, where $F_{\text{non-int}}(n)$ is defined in Eq.~(\ref{eq:func-f_non-int}). Therefore in the $x_{21}\equiv x_2 - x_1 \rightarrow 0$ limit, the two-point-contact case reduces to that of a single point contact. Note that
the thermal current in the general case of $\Gamma_1\ne \Gamma_2$ is computed in Appendix~\ref{appendix:sigma}. The only difference from the above result is the substitution of $\Gamma^2$ in the non-interference contribution with $(\Gamma_1^2+\Gamma_2^2)/2$ and with $\Gamma_1\Gamma_2$
in the interference contribution.

So far we assumed a trivial trapped topological charge in the device.
The above result is easily extended to the other possible trapped topological charges $\psi$ and $\sigma$.
Since $\sigma$ accumulates the phase $\pi$ on a circle around $\psi$, the interference contribution changes its sign in comparison with the above equation. For the trapped charge $\sigma$, the even-odd effect is present~\cite{stern2006:PhysRevLett.96.016802,eo2}. The tunneling anyon 
has two equally likely fusion channels with the trapped anyons: $\bm{1}$ and $\psi$. The corresponding interference phases $\phi^{\bm{1}}_{\sigma\sigma}=-i\log(\theta_{\bm{1}}/\theta^2_\sigma)$ and 
$\phi^{\psi}_{\sigma\sigma}=-i\log(\theta_{\psi}/\theta^2_\sigma)$
differ by $\pi$.
Hence, only the non-interference contribution to the heat current survives.

If one can control the trapped topological charge, the existence of three different heat currents at the three possible trapped topological charges gives a signature of the Ising statistics. Presently, it is not clear how to control the trapped charge. As long as a device with the trapped charge $\sigma$ or $\psi$ can be fabricated, it will be possible to also obtain another trapped topological charge by doubling the size of the device:
for example, connecting two identical devices, each of which confines $\sigma$, yields the total confined charge $\sigma\times \sigma$, which can be either $\bm{1}$ or $\psi$. It might happen, however, that the trapped charge is always $\bm{1}$. In that case, the Fabry-P\'erot device would not be able to tell Ising anyons from bosons or fermions. To overcome this issue we consider a Mach-Zehnder interferometer.

\subsection{Mach-Zehnder geometry}\label{subsec:Ising_MZ}

As we already observed, Fabry-P\'erot interferometry, in the case of spin liquids, although sensitive to the enclosed topological charge, may not be very informative since it requires control over the topological charge enclosed between the two constrictions. In the case of electronic Fabry-P\'erot interferometry, this issue did not arise since anyonic excitations in the sample carry charge. 
The number of the trapped excitations depends on the magnetic flux.
Hence, the magnetic field becomes an external control parameter. Anyonic excitations in a Kitaev spin liquid, however, do not carry charge. Therefore magnetic field fails to provide such a control. It, instead, becomes useful to compare thermal transport in the Fabry-P\'erot and Mach-Zehnder interferometers~\cite{wei2021:PhysRevLett.127.167204}.

We thus turn to the double constriction case in the Mach-Zehnder geometry. In this geometry, one of the drains is topologically inside the loop, over which the interference phase is accumulated. Hence, with each tunneling event, the confined topological charge changes. The effective two-point tunneling Hamiltonian is given as,
\begin{align}\label{eq:mach-zehnder_tunneling-ham}
    H_T = \Gamma_1 \hat T_1 + \Gamma_2 \hat T_2 e^{i\alpha}.
\end{align}
where $\Gamma_i$, $i=1,2$ are real tunneling amplitudes, $\hat T_i$ are two tunneling operators that transfer Ising anyons from the outer edge to the inner edge, Fig.~\ref{fig1:thermal}. The phase $\alpha$ ensures the Hermiticity of the tunneling Hamiltonian $H_T$. This phase is computed in Appendix~\ref{appendix:Mach-zehnder_phase}. For Ising anyons, $\alpha=\pi/8$ as a consequence of the general rule that $\exp(i\alpha)$ equals the topological spin of the tunneling anyon (Appendix~\ref{appendix:Mach-zehnder_phase}).

As highlighted in Sec.~\ref{sec:model_mach-zehnder}, since in the Mach-Zehnder interferometer the tunneling probability changes with each tunneling event, the thermal current is defined in terms of the tunneling rates $p_{\sigma b}^{c}$ given by Eq.~(\ref{eq:general_MZ_rate}). For the case of Ising anyons, we find the following tunneling rates,
\begin{align}
    p_{\sigma\psi}^{\sigma} &=  \left[\Gamma_1^2+\Gamma_2^2-2\Gamma_1\Gamma_2\cos(\pi/8) \right]p(T_1,T_2), \\
    p_{\sigma \bm{1}}^{\sigma} &=  \left[\Gamma_1^2+\Gamma_2^2+2\Gamma_1\Gamma_2\cos(\pi/8) \right]p(T_1,T_2),
\end{align}
where $p(T_1,T_2)$ is computed in the single-point-contact geometry as,
\begin{align}
    p(T_1, T_2) & = {2 \pi} \sum_{m,n} \left|\langle m|e^{-i \pi/8} \sigma_2(x_0)\sigma_1(x_0) |n\rangle \right|^2 \nonumber\\
&\qquad\qquad\times \delta(E_{m}-E_{n}) P_{n}(T_1, T_2).
\end{align}
According to Eq.~(\ref{eq:non-int-current_anyon}),
\begin{equation}
	p(T_1, T_2) = (\pi T_1)^{\frac{1}{8}}(\pi T_2)^{\frac{1}{8}} \frac{\cos(3\pi/8)}{4\sqrt{2}v^{1/4}} F_{\text{non-int}}(T_2/T_1).
\end{equation}
We remind the reader that, to simplify equations, we assume the temperature low enough for the thermal length
$ v/T$ to exceed the interferometer size.

Since the Ising anyon $\sigma$ is its own anti-particle, i.e., one of the fusion channels gives vacuum on fusing two $\sigma$ anyons, we obtain the following relation from the algebraic theory of anyons [see Eq.~(\ref{E16})], in terms of the quantum dimensions, $p_{\sigma a}^{b} = p_{\sigma b}^{a} (d_b^2/d_a^2)$. This allows us to compute the remaining non-zero tunneling rates. Plugging this relation into Eq.~(\ref{eq:kinetic}), we solve for the probability $f_a$ in dynamical equilibrium,
\begin{align}
    f_{\sigma} = \frac{1}{2};~~~~~ f_{\bm{1}}=f_{\psi}=\frac{1}{4}.
\end{align}
Finally, the average heat $\overline{\Delta E}$ transferred in each tunneling event can be computed from a setup with a single point contact as a ratio between $r(T_1,T_2)$, calculated in Sec.~\ref{sec:Ising-singleQPC}, and $p(T_1,T_2)$. Given the tunneling rates $p_{\sigma b}^{c}$, the probabilities $f_{a}$, and the average transferred heat  $\overline{\Delta E}$, we can find the thermal current from Eq.~(\ref{eq:MZ_thermalcurrent}),
\begin{align}
    I_{T} = (\Gamma_1^2+\Gamma^2_2) (\pi T_1)^{\frac{1}{8}}(\pi T_2)^{\frac{1}{8}} \frac{\cos(3\pi/8)}{4\sqrt{2}v^{1/4}} F_{\text{non-int}}(T_2/T_1),
\end{align}
where the function $F_{\text{non-int}}$ is defined in Eq.~(\ref{eq:func-f_non-int}). We notice that the thermal current in the Mach-Zehnder geometry for Ising anyon tunneling is just the sum of the thermal current contributions from each of the single constrictions as found in Sec.~\ref{sec:Ising-singleQPC}, and the interference term is absent. This is a general feature of non-trivial topological orders as we will see in the next section where we treat an arbitrary anyon statistics.

The above behavior is quite different from the Fabry-P\'erot geometry with a trivial trapped topological charge. It is also quite different from the behavior expected from bosons and fermions in the Mach-Zehnder geometry. Indeed, bosons and fermions show the same interference pattern in both geometries. Thus, a comparison of  the two interferometer geometries provides a signature of the Ising statistics even if the trapped topological charge is bound to be trivial in the Fabry-P\'erot setup. Note that a comparison of theory and experiment requires the knowledge of $\Gamma_{1,2}$. It can be obtained from a single-point-contact geometry.

\section{Arbitrary anyon statistics}\label{sec:arbitary}
In addition to the Ising topological order in a Kitaev spin liquid, we would like to extend our theory of thermal interferometers to the case where anyons tunnel between two edges of an arbitrary topological phase of matter.
We will use the algebraic theory of anyons in Ref.~\cite{kitaev2006:anyons}, and
adopt the assumption that the distances $ L_{1,2} $ along the two edges between the two point contacts in the interferometer are much less than the thermal length $ l= v/ T $. This assumption allows us to use an effective one-point tunneling amplitude for the interferometer and simplify the expressions for the tunneling thermal current.

As in previous sections, we start with a single point contact, then consider the interferometry of Abelian anyons and the interferometry of non-Abelian anyons. Before addressing technical details, we list the key results in the next subsection.
	
\subsection{Summary of interferometry}

For a topologically trivial system, the same behavior is expected in the Fabry-P\'erot and Mach-Zehnder geometries. The interference phase does not depend on what particles are trapped in the device. Thus, one easily checks that in an interferometer with two tunneling amplitudes $\Gamma_{1,2}$ at the constrictions, the heat current scales as $|\Gamma_1+\Gamma_2|^2$. Indeed, this is obvious from applying the results of the subsection on Abelian statistics to the case of the trivial topological order.

On the other hand, for non-trivial statistics, the thermal current through a Fabry-P\'erot device 
depends on the trapped topological charge. The number of the possible values of the current reflects the number of possible anyon types. The current through the interferometer depends on the mutual statistical phases of the tunneling and trapped anyons. As is clear from Secs.~\ref{subsec:non_abelian_own_anti} and \ref{subsec:non_abelian_not_anti},
in the case of non-Abelian statistics, the current also depends on the fusion rules for the topological order. 

Thus, a Fabry-P\'erot device allows the identification of the topological order provided that one can control the trapped topological charge. If the trapped topological charge is always trivial, the Fabry-P\'erot approach cannot distinguish non-trivial orders from the trivial order. We have to rely instead on
the Mach-Zehnder geometry. Its behavior is strikingly simple and general: For any nontrivial topological order, the heat current scales as $|\Gamma_1|^2+|\Gamma_2|^2$.

\subsection{Single contact}\label{subsec:arbitary_single}

We rely on Eq.~(\ref{eq:thermal-current-operator}) for finding the thermal current operator and Eq.~(\ref{eq:thermalcurrent-perturbationtheory}) for the average heat current. 
For the purposes of this section,  the exact forms of the operators and currents are unimportant. It will suffice to know their scaling dimensions.
The current operator scales as the 
time derivative of the tunneling Hamiltonian. The  average current scales as the time integral of the product of that time derivative and the tunneling Hamiltonian. Thus, the average current scales as a power of the temperature $T^{2g}$, where $g$ is the scaling dimension of the tunneling operator. The exponent depends on statistics.  For Ising anyons, $g=1/8$ in agreement with the previous results. In the $\nu=1/3$ Laughlin state \cite{WenBook}
with the edge action $\frac{3}{4\pi}\int dx dt [\pm \partial_t\phi\partial_x\phi-v(\partial_x\phi)^2]$ 
and the charge density $e\partial_x\phi/(2\pi)$,
the anyon tunneling operator is $\Gamma\exp(i\phi_1+i\phi_2)+\text{H.c.}$, where the indices $1$ and $2$ refer to the two edges. Then $g=1/3$.

\subsection{Interferometry of Abelian anyons }\label{subsec:abelian}

In an Abelian topological order, the fusion channel of any pairs of anyons is unique,
therefore there exists the smallest integer $ m $, such that the fusion result of $ m $ anyons of type $x$ is in the vacuum sector. 
In the following subsections, we will discuss the two cases $m=2$ and $m>2$ separately.

\subsubsection{Tunneling anyon is its own antiparticle.}

We first consider the case when $ m=2 $; i.e. $ x \times x = \bm{1} $, where $ \bm{1} $ denotes vacuum. Obviously, $ x $ is its own antiparticle. One example is the semion topological order \cite{semion2,semion1}: it contains only one type of non-trivial topological charge $ s $, and the only non-trivial fusion rule is $ s \times s = \bm{1} $; the topological spin of $ s $ is $ \theta_{s}=i $.

Given the one-point contact Hamiltonian $ H_T = \Gamma \hat{T} $, where $ \Gamma $ is a real amplitude, the thermal current can be written as $ I_T =  r(T_1, T_2) \Gamma^2 $ as demonstrated in Sec.~\ref{sec:model}. 
The tunneling rate is $  p(T_1, T_2) \Gamma^2 $, where $ p(T_1, T_2) $ can be found with Fermi's golden rule:
\begin{equation}\label{eq:rate_abelian_own_anti}
p(T_1, T_2)  = {2 \pi}  \sum_{mn} |\langle m| \hat{T}  |n \rangle|^2 
 \delta(E_{m}-E_{n}) P_{n}(T_1, T_2),
\end{equation}
 $ |n\rangle $, $ |m\rangle $ are eigenstates of the unperturbed edge Hamiltonian, $ E_{n,m} $ are the total energies of the two edges for the corresponding state, and $ P_{n} $ is the Gibbs distribution. The average heat transferred in each tunneling event is $ \overline{\Delta E} = r(T_1, T_2)/p(T_1, T_2) $.

For a Fabry-P\'erot interferometer, the tunneling Hamiltonian is given as
\begin{equation}
H_{T} = \Gamma_{1} \hat{T}_1 + \Gamma_{2} \hat{T}_2,
\end{equation}
where $ \Gamma_{1,2} $ are real amplitudes. The localized topological charge has at least two possible values: $ \bm{1} $ or $ x $. Below we focus on these two possibilities only.
Depending on the sector, the tunneling current has two different values,
\begin{align}
I_{\bm{1}}^{\rm FP} & = (\Gamma_{1} + \Gamma_{2} )^2 r(T_1, T_2), \\
I_{x}^{\rm FP} & = |\Gamma_{1} + \Gamma_{2} e^{i\phi_{xx}^{\bm{1}}} |^2 r(T_1, T_2) ,
\end{align}
where $ \exp{(i \phi_{xx}^{\bm{1}} )}= \theta_{\bm{1}}/(\theta_{x} \theta_{x}) $ is the braiding phase. 
For semions, this phase factor is $-1$. 
This result actually applies to any Abelian statistics, as long as $x$ is its own antiparticle and not a fermion or boson. 
Indeed, the topological spin of vacuum should satisfy $\theta_{\bm{1}}=\theta_{x}^4=1$, hence $\theta_{x}=\pm i$ or $\pm 1$. 
The second option describes bosons and fermions and so we are left with $\theta_{x}=\pm i$ and $\exp(i\phi^{\bm{1}}_{xx})=-1$. 

In a Mach-Zehnder interferometer, the tunneling Hamiltonian in Eq.~(\ref{eq:MZ_Hamiltonian_own_anti}) contains an additional phase $ e^{i\alpha}=\theta_{x}$ in front of $\Gamma_2$, as discussed in Appendix~\ref{appendix:Mach-zehnder_phase}.
In the two sectors $ \bm{1} $ and $ x $, the effective one-point amplitudes are  $ \Gamma_{1}+\Gamma_{2} \theta_{x} $ and $ \Gamma_{1} + \Gamma_{2} \exp{(i\phi_{xx}^{\bm{1}}) } \theta_{x}$ respectively.
We further see that the tunneling rates are
\begin{align}
p_{\bm{1}} &= \left|\Gamma_{1} + \Gamma_{2} \theta_{x} \right|^2 p(T_1, T_2), \\
p_{x} &= \left|\Gamma_{1} - \Gamma_{2} \theta_{x} \right|^2 p(T_1, T_2).
\end{align}
Since $ \theta_{x}=\pm i $, the two rates are equal, but we will stick to the same notations as for other statistics.
The kinetic equation is now
\begin{equation}
\dot{f}_{x} = f_{\bm{1}} p_{\bm{1}} - f_{x} p_{x} = 0.
\end{equation}
Applying the requirement that the total probability is unity, we obtain the following solution:
\begin{equation}
f_{k} = \frac{1}{p_{k}} \left( \frac{1}{p_{\bm{1}}} + \frac{1}{p_{x}}\right)^{-1},\quad k=\bm{1},x.
\end{equation}
The thermal current reads
\begin{align}
I^{\rm MZ} & = \overline{\Delta E} \sum_{k=\bm{1},x} f_{k} p_{k}
= 2 \frac{r(T_1,T_2)}{p(T_1,T_2)} \left(\frac{1}{p_{\bm{1}}}+\frac{1}{p_a}\right)^{-1} \nonumber \\
&= (\Gamma_{1}^{2} + \Gamma_{2}^{2} ) r(T_1, T_2),
\end{align}
where the final equality only applies for $\theta_{x}=\pm i$.
The comparison with the experiment requires the knowledge of $\Gamma_{1,2}$ and $r$. 
They can be found from a single-contact setup.

\subsubsection{Tunneling anyon is not its own antiparticle}
\label{sec-V-C-2}
Now we consider the case when $ m>2 $, which implies that $ x $ is not the same as its antiparticle $ \bar{x} $.
To analyze the tunneling at a single point contact, we write down the tunneling Hamiltonian,
\begin{equation}
H_T = \Gamma \hat{T} + \mathrm{H.c.},
\end{equation}
where $ \hat{T} $ is the tunneling operator which creates a particle-antiparticle pair at the constriction. 
Similar to Eq.~(\ref{eq:rate_abelian_own_anti}), the thermal current can be expressed as
\begin{align}\label{eq:current_golden_not_anti}
I_T & = {2\pi} |\Gamma|^2  \sum_{m n} \Delta E \Big[ |\langle m | \hat{T} | n \rangle |^2  +  |\langle m | \hat{T}^\dagger| n \rangle |^2 \Big]  \nonumber\\
& \qquad \times \delta(E_m -E_n)  P_n (T_1, T_2),
\end{align}
where $ \Delta E $ is the energy change on the upper edge in the process $ |n\rangle \to | m \rangle $.
This equation contains two summations for $\hat{T}$ and $\hat{T}^\dagger$ respectively. The two sums are equal
in the low-temperature limit. 
Indeed, we expect that a conformal-field-theoretic description exists for the edge theory in that limit. 
The tunneling operator $\hat{T}$ for an Abelian topological order can be represented as an exponent $\exp(i\phi)$ of a Bose field $\phi$. The conjugate operator $\hat{T}^\dagger=\exp(-i\phi)$. The edge theory is quadratic in Bose fields and hence invariant with respect to the transformation $\phi\rightarrow -\phi$. This transformation exchanges $\hat{T}$ and $\hat{T}^\dagger$. Hence, the two contributions to $I_T$ are equal. The above argument may fail at a non-zero voltage in a conducting system if $a$ carriers electric charge. The reason is that changing the sign of the Bose field will then also require changing the sign of the voltage.
We assume zero voltage and find that
\begin{equation}
\label{eq:51}
I_{T} =  |\Gamma|^2 r(T_1,T_2),
\end{equation}
where the factor $ r(T_1,T_2) $ can be expressed as
\begin{align}\label{eq:r_not_own_anti}
	r(T_1,T_2) & = {4\pi} 
	\sum_{m n} \Delta E |\langle m | \hat{T} | n \rangle |^2   \delta(E_m -E_n) P_n (T_1, T_2).
\end{align}

A similar argument shows that the tunneling rates of $ x $ and $ \bar{x} $ in the same direction are the same.
The tunneling rates are $ |\Gamma|^2 p^{\pm}(T_1,T_2) $ for the tunneling of $ x $ ($ + $) and $ \bar{x} $ ($ - $), where, by an alternative form of Fermi's golden rule~\cite{stedman1971},
\begin{align}
p^{+} (T_1, T_2) &= \int_{-\infty}^{\infty} dt \,  \langle \hat{T}^{\dagger} (t) \hat{T}(0) \rangle ,\\
p^{-} (T_1, T_2) &= \int_{-\infty}^{\infty} dt \,  \langle \hat{T}(t) \hat{T}^{\dagger}(0) \rangle .
\end{align}
The above expressions change into each other when the sign of $\phi$ is changed.
Hence $p^{+} (T_1, T_2) = p^{-} (T_1, T_2)$. We will also use the notation $ p(T_1, T_2) = p^{+} (T_1, T_2) + p^{-} (T_1, T_2)$.

The above discussion shows that the tunneling of $ \bar{x} $ transfers the same amount of average heat as the tunneling of $ x $. The tunneling thermal current can be rewritten as
\begin{equation}
I_T = \overline{\Delta E} \, |\Gamma|^{2} p(T_1, T_2) ,
\end{equation}
where $ \overline{\Delta E} = {r(T_1,T_2)}/{p(T_1,T_2)}  $ is the average heat transferred.

It is easy to generalize from the single-constriction case to the Fabry-P\'erot geometry
with two tunneling contacts with the tunneling amplitudes $\Gamma_{1,2}$. 
One just needs to substitute 
$\Gamma\rightarrow \Gamma_1+\Gamma_2\exp(i\theta)$ in Eq.~(\ref{eq:51}), where $\theta$ is the statistical phase accumulated by
$a$ on a circle around the topological charge, trapped in the interferometer.

Now we come back to the Mach-Zehnder interferometer. We assume that the initial localized topological charge is $ \bm{1} $ (vacuum) and denote the localized topological charge
inside the device
after $ k $ tunneling events of $ x $ as $ b_{k} $, hence $b_{0} = \bm{1} $ and $ b_{k} = kx $. 
Clearly, $ b_{k} $ is periodic in $ k $, $ b_{k+m} = b_{k} $. As we will see below, the same result for the average heat current obtains for any initial trapped charge of the form $n x$. If the initial trapped charge is not $n x$,
our calculations require only a minor modification, and the final result for the heat current does not change.

For a Mach-Zehnder interferometer, its tunneling Hamiltonian is
\begin{equation}\label{eq:hamiltonian_MZ_not_own_anti}
H_T  = \Gamma_1 \hat{T}_1 + \Gamma_2 \hat{T}_2 + \mathrm{H.c.}
\end{equation}
When the localized topological charge is $ b_{k} $, the effective one-point tunneling amplitude is $ \Gamma_{1} +  \Gamma_{2} \exp{(i \phi_{k}^{k+1})} $ for the tunneling of $ x $, and $ \Gamma_{1}^{*} + \Gamma_{2}^{*} \exp{(i \phi_{k}^{k-1})} \theta_x^2 $ for the tunneling of $ \bar{x} $, where $ \exp{(i \phi_{k}^{k+1})} = {\theta_{b_{k+1}}}/({\theta_{x} \theta_{b_{k}}}) $ and $ \exp{(i \phi_{k}^{k-1})} = {\theta_{b_{k-1}}}/({\theta_{\bar{x}} \theta_{b_{k}}}) $. The origin of the factor $\theta_x^2$ in the tunneling amplitude for $\bar x$ can be understood in the spirit of Eqs. (\ref{E12}-\ref{E14}) as well as from physical considerations: $T^\dagger_2$ describes tunneling of $x$ from the inner edge of the interferometer to the outer edge. As a result, $\bar x$ is created and left behind on the inner edge. 
It affects the statistical phase in the tunneling amplitude.

The two tunneling rates for $ x $ and $ \bar{x} $ can be written down separately,
\begin{align}
p_{k}^{+} & = \frac{1}{2} \left|\Gamma_1+ \Gamma_2 e^{i \phi_{k}^{k+1}}\right|^2 p(T_1, T_2), \\
p_{k}^{-} & = \frac{1}{2} \left|\Gamma_1^*+ \Gamma_2^* e^{i \phi_{k}^{k-1}}\theta_x^2\right|^2  p(T_1, T_2) .
\end{align}
It is worth noting that the following equations hold in the algebraic theory of anyons~\cite{kitaev2006:anyons,preskill2004},
\begin{equation}
|\theta_{x}|=1,\quad \theta_{x} = \theta_{\bar{x}},\quad \theta_{b_{k}} = (\theta_{x})^{k^2}.
\end{equation}
Using these relations, we can find the phases in the tunneling rates to be complex conjugate to each other,
\begin{align}
e^{i \phi_{k}^{k+1}} & = \frac{\theta_{b_{k+1}}}{\theta_{x}\theta_{b_{k}}}
= (\theta_{x})^{2k}, \\
e^{i \phi_{k+1}^{k}} & = \frac{\theta_{b_{k}}\theta_x^2}{\theta_{\bar{x}}\theta_{b_{k+1}}}
= (\theta_{x})^{-2k} = (\theta_{x}^{*})^{2k}.
\end{align}
Hence $ p_{k}^{+} = p_{k+1}^{-} $, which shows that the tunneling rates for $x$ and $\bar{x}$ are the same, regardless of the topological charge of the localized anyon.
The kinetic equation now reads
\begin{equation}
\dot{f}_{k} = f_{k-1} p_{k-1}^{+} + f_{k+1} p_{k+1}^{-} - f_{k} ( p_{k}^{+} + p_{k}^{-} ),
\end{equation}
with $ k=0,1,\dots,m-1 $. 
After setting $ \dot{f}_{k}=0 $, the equation becomes
\begin{equation}
f_{k+1} p_{k}^{+} - f_{k} p_{k}^{+} =
f_{k} p_{k-1}^{+} - f_{k-1} p_{k-1}^{+},
\end{equation}
which means that $ f_{k}=1/m$ is independent of $k$.
Hence, the heat current is proportional to the average tunneling amplitude
\begin{equation}
I_T=r(T_1,T_2)\frac{1}{m}\sum_{k=0}^{m-1} |\Gamma_1+\Gamma_2\theta_x^{2k}|^2=
(|\Gamma_1^2|+|\Gamma_2|^2)r(T_1,T_2).
\end{equation}

For comparison, the thermal current in a Fabry-P\'erot interferometer has $ m $ distinct values
\begin{align}
I_{k}^{\rm FP} =  \left|\Gamma_1+ \Gamma_2 e^{i \phi_{k}^{k+1}}\right|^2 
 r(T_1, T_2).
\end{align}

\subsection{Tunneling of non-Abelian anyons, which are their own antiparticles}\label{subsec:non_abelian_own_anti}
Now we study the non-Abelian situation. 
As in the Abelian situation, we start with the case where the tunneling anyon is its own antiparticle.  
Examples of such anyons include Ising and Fibonacci anyons.
The fusion rule can be written as
\begin{equation}
x \times x = \bm{1} + \dots,
\end{equation}
where $\dots$ represents other possible fusion channels.
The one-point tunneling Hamiltonian should be 
\begin{equation}
H_T = \Gamma \hat{T} = \Gamma e^{-i \pi h_{x}} \hat{x}_{2}(x_{1}) \hat{x}_{1}(x_{1}),
\end{equation}
where $ \Gamma $ is a real tunneling amplitude, $ \hat{x}_{2} \hat{x}_{1} $ is the operator that creates a pair of anyons on the opposite edges of the interferometer, and $ h_{x} $ is the scaling dimension of the anyon field with topological charge $ x $.
Note that the topological spin $ \theta_{x} $ and the scaling dimension $ h_x $ are related \cite{kitaev2006:anyons} as $ \theta_{x} = e^{i 2\pi h_{x} } $ for a holomorphic field $ x $.
Similar to the Abelian case, we can find the thermal current in a one-point contact geometry using Eq.~(\ref{eq:thermalcurrent-perturbationtheory}), $ I_{T}=r(T_1, T_2) \Gamma^2$. The average heat transferred in each tunneling event is the ratio between $ r(T_1, T_2) $ and $ p(T_1, T_2) $, where $ p(T_1, T_2) $ is given by Eq.~(\ref{eq:rate_abelian_own_anti}).

In a Mach-Zehnder interferometer, the effective two-point tunneling Hamiltonian is given in Eq.~(\ref{eq:MZ_Hamiltonian_own_anti}), where the additional phase $ e^{i\alpha}$ is equal to $\theta_x $.
In a process where the tunneling anyon changes the trapped topological charge  from $ a $ to $ b $, the corresponding tunneling rate is
\begin{align}\label{eq:rate_self_anti}
p_{xa}^{b} = P_{xa}^{b} p (T_1, T_2) |\Gamma_1 + e^{i \phi_{xa}^{b} + i \alpha} \Gamma_2 |^2,
\end{align}
where $ P_{xa}^{b} =N_{xa}^b {d_b}/(d_x d_a)$ and $ \exp(i \phi_{x a}^{b}) = {\theta_b}/(\theta_a \theta_x) $, as in Eq.~(\ref{eq:general_MZ_rate}).

We can further write down and compare the tunneling rates for two different, yet related fusion routes for the localized anyon, $ x \times a \to b $, and $ x \times b \to a $,
\begin{align}
p_{x a}^{b} & =  N_{x a}^{b} \frac{d_{b}}{d_{x} d_{a}} 
\left|\Gamma_{1}+\Gamma_{2} \cdot \frac{\theta_{b}}{\theta_{x} \theta_{a}} \cdot \theta_{x} \right|^{2} p(T_1, T_2), \\
p_{x b}^{a} & =  N_{x b}^{a} \frac{d_{a}}{d_{x} d_{b}} 
\left|\Gamma_{1}+\Gamma_{2} \cdot \frac{\theta_{a}}{\theta_{x} \theta_{b}} \cdot \theta_{x} \right|^{2} p(T_1, T_2).
\end{align}
In the algebraic theory of anyons~\cite{kitaev2006:anyons}, $ N_{x a}^{b} = N_{\bar{x} b}^{a} = N_{x b}^{a} $ and $ |\theta_{a}|=|\theta_{b}| =1$. Since $ \Gamma_{1,2} $ are real amplitudes, by comparing the two equations we find that $ p_{x a}^{b} = (d_{b}/d_{a})^2 p_{x b}^{a} $.
Thus the kinetic equation~(\ref{eq:kinetic}) can be reduced to
\begin{equation}
\dot{f}_{a} = \sum_{b} p_{x b}^{a}  \Big[  f_b - \Big(\frac{d_{b}}{d_{a}}\Big)^{2} f_{a}  \Big] = 0.
\end{equation}
It is easy to read out the solution, $ f_a = d_{a}^{2} / D^2 $, where $ D =\sqrt{\sum_{a} d_{a}^2}$ is known as the global dimension.
The tunneling heat current through a Mach-Zehnder interferometer is 
\begin{align}\label{eq:thermal_current_self_anti}
I^{\rm MZ} = I_{0} + I_{1} + I_{1}^{*},
\end{align}
where
\begin{align}\label{eq:thermal_current_self_anti-1}
I_{0} & = (\Gamma_{1}^2 + \Gamma_{2}^{2} )r(T_1, T_2 ) \sum_{ab} N_{xa}^{b} \frac{d_{a} d_{b}}{d_{x}} \frac{1}{D^2}  \nonumber\\ 
&= (\Gamma_{1}^2 + \Gamma_{2}^{2} ) r(T_1, T_2 ),
\end{align}
and %
\begin{align}\label{eq:thermal_current_self_anti-2}
I_{1} & = \Gamma_{1} \Gamma_{2} r(T_1, T_2 ) \sum_{ab} N_{xa}^{b} \frac{d_{a} d_{b}}{d_{x}} \frac{1}{D^2} 
\frac{\theta_{b}}{\theta_{a}} \nonumber\\
&= \Gamma_{1} \Gamma_{2} r(T_1, T_2 )  \frac{\theta_{x}}{d_{x}} \sum_{a} s_{a \bar{x}} s_{1 a}
= 0 .
\end{align}
Here, we have used the identity $ d_{a} d_{b} = \sum_{c} N_{ab}^{c} d_{c} $. 
We also use the notion of the topological $S$-matrix $ s_{ab} =  \frac{1}{D} \sum_{c} N_{a \bar{b}}^{c} d_{c} \frac{\theta_{c}}{\theta_{a} \theta_{b}}$ and the orthogonality between $ s_{xa}/s_{x1} $ and $ s_{1a}/s_{11} $~\cite{kitaev2006:anyons,feldman2007:shot_noise}. Therefore, $ I^{\rm MZ} = (\Gamma_{1}^2 + \Gamma_{2}^{2} ) r(T_1, T_2 ) $, 
and the heat current shows no interference in a Mach-Zehnder interferometer.

For comparison, the Fabry-P\'erot current for a localized anyon of type $ a $ is
\begin{equation}\label{eq:FP_non_abelian_own_anti}
I_{a}^{\rm FP} = \sum_{b}
N_{x a}^{b} \frac{d_{b}}{d_{x} d_{a}} 
\left|\Gamma_{1}+\Gamma_{2} \cdot \frac{\theta_{b}}{\theta_{x} \theta_{a}} \right|^{2} r(T_1, T_2).
\end{equation}

Now we test our results for the Fibonacci topological order.
The Fibonacci topological order has only one non-trivial fusion rule: $ \tau \times \tau = \bm{1} + \tau $.
The quantum dimension of a Fibonacci anyon $\tau$ is $  d_\tau = ({1+\sqrt{5}})/{2} \equiv \varphi$,
and its topological spin is $ \theta_\tau = e^{4\pi i/5} $~\cite{nayak2008:RevModPhys.80.1083,preskill2004}. %
According to Eq.~(\ref{eq:rate_self_anti}), the tunneling rates in a Mach-Zehnder device are
\begin{align}
p_{\tau\tau}^{\tau} &= \frac{1}{\varphi} \left|\Gamma_1+\Gamma_2 \right|^2 p(T_1,T_2), \\
p_{\tau\tau}^{\bm{1}} &= \frac{1}{\varphi^2} \left|\Gamma_1+\Gamma_2 e^{-4\pi i/5}\right|^2 p(T_1,T_2),\\
p_{\tau \bm{1}}^{\tau} & =  \left|\Gamma_1 + \Gamma_2 e^{4\pi i/5}\right|^2 p(T_1,T_2).
\end{align}
The kinetic equations  are 
\begin{align}
    \dot{f}_\tau &= f_{\bm{1}} p_{\tau \bm{1}}^{\tau} + f_{\tau} p_{\tau\tau}^{\tau}
    - f_{\tau} (p_{\tau\tau}^{\tau}+p_{\tau\tau}^{\bm{1}}), \\
    \dot{f}_{\bm{1}} &= f_{\tau} p_{\tau\tau}^{\bm{1}} - f_{\bm{1}} p_{\tau \bm{1}}^{\tau}.
\end{align}
The solution is 
\begin{align}
    f_{\tau} &= \frac{p_{\tau \bm{1}}^{\tau}}{p_{\tau\tau}^{\bm{1}}+p_{\tau \bm{1}}^{\tau}}
    = \frac{\varphi^2}{1+\varphi^2}, \\
    f_{\bm{1}} &= \frac{p_{\tau \tau}^{\bm{1}}}{p_{\tau\tau}^{\bm{1}}+p_{\tau \bm{1}}^{\tau}} = \frac{1}{1+\varphi^2}.
\end{align}
After a substitution  back into the equation for the thermal current, one gets
\begin{align}
    I^{\rm MZ}  =  
    (\Gamma_1^2 + \Gamma_2^2)\, \overline{\Delta E} \, p(T_1,T_2)
    = (\Gamma_1^2 + \Gamma_2^2) r(T_1, T_2).
\end{align}
This confirms that Eqs.~(\ref{eq:thermal_current_self_anti}-\ref{eq:thermal_current_self_anti-2}) is a general solution for all kinds of anyons in this category. 

In summary, anyons that are their own antiparticles show no interference in the Mach-Zehnder geometry.

\subsection{Tunneling of non-Abelian anyons, which are not their own antiparticles}\label{subsec:non_abelian_not_anti}

Now let us consider the situation where the tunneling anyon $x$ is different from its antiparticle $\bar{x}$. More specifically, we consider a topological order in which the fusion results of $ x \times x$ and $ a \times  \bar{a}=\bm{1}+\dots$ share no common topological charges for all possible $ a $.
This condition is satisfied by many interesting topological orders, including the $\mathbb{Z}_3$-parafermions and the $k =3,M =1$ Read-Rezayi state~\cite{bonderson2007:non-abelian}. This property is automatically satisfied for charged anyons $x$ since $x\times x$ carriers electric charge and $a\times\bar a$ is neutral.

We will rely on the following statement for such systems: if $ x $ could fuse with $ a $ into $ b $, then $ N_{xb}^{a} = 0 $; that is, the available topological charges from the fusion between $ x $ and $ b $ do not include $ a $.
The proof of the statement is the following. Assume that $x\times a=b+\dots$ and $x\times b=a+\dots$, then
$x\times x\times a=x\times b+\dots=a+\dots$. Hence, $x\times x\times a\times \bar{a}=a\times\bar{a}+\dots=\bm{1}+\dots$
Hence, $x\times x$ contains the antiparticle for at least one fusion channel in $a\times\bar{a}$. Since $a\times\bar{a}$
contains each possible fusion outcome together with its antiparticle, we obtain the desired result.

This result means that the Hermitian conjugate tunneling operators $ \hat{T} $ and $ \hat{T}^{\dagger} $ that transfer $x$ between the two edges contribute independently to the tunneling process. This significantly simplifies our discussion below.
For completeness, we give an example of an  topological order that does not satisfy our condition in Appendix~\ref{appendix:anyon_example}.

The contributions to the thermal current from $ \hat{T} $ and $ \hat{T}^{\dagger} $ are denoted as $ |\Gamma|^2 r^{\pm} (T_1,T_2)$ respectively. Using the perturbation theory and conformal field theory (CFT), one can show that $ r^{+}(T_1, T_2) = r^{-}(T_1, T_2) $. 
First, we write the two quantities as integrals in the perturbation theory,
\begin{align}
	r^{+}(T_1, T_2) &= - i \int_{-\infty}^{t} \, d t' \langle [\hat{T'}^{\dagger}(t),  \hat{T}(t')] \rangle, \\
	r^{-}(T_1, T_2) &= - i \int_{-\infty}^{t} \, d t' \langle [\hat{T'}(t),  \hat{T}^{\dagger}(t')] \rangle,
\end{align}
where $ \hat{T} = \hat{x}_{2} (x,t) \hat{\bar{x}}_{1} (x,t) $ is the tunneling operator, and $ \hat{T'} = -\hat{x}_{2} (x_1,t) \partial_{t} \hat{\bar{x}}_{1} (x_1,t) $.
In 2D CFT, four-point correlation functions
with two fields on each edge
can be decomposed into products of two-point correlation functions
(Appendix~\ref{appendix:ising-correlation}); 
furthermore, the correlation function of two fields is fully determined by their conformal weights, or equivalently their scaling dimensions~\cite{difrancesco1997:conformal}. Using these two properties, one can find that $ \langle [\hat{T'}^{\dagger}(t),  \hat{T}(t')] \rangle = \langle [\hat{T'}(t),  \hat{T}^{\dagger}(t')] \rangle $, hence $ r^{+}(T_1, T_2) $ and $ r^{-}(T_1, T_2) $ are equal.
By the same argument, the tunneling rates given by $ p^{\pm}(T_1,T_2) |\Gamma|^2$, where $p^{+}(T_1,T_2)=\int dt\langle \hat{T}^\dagger(0) \hat{T}(t)\rangle$ and
$p^{-}(T_1,T_2)=\int dt \langle \hat{T}(0) \hat{T}^\dagger(t)\rangle$, are also equal to each other.
The average heat transferred in a tunneling event is 
\begin{equation}
	\overline{\Delta E} = \frac{r^{+}(T_1,T_2)}{p^{+}(T_1,T_2)} = \frac{r^{-}(T_1,T_2)}{p^{-}(T_1,T_2)}.
\end{equation}

In Mach-Zehnder interferometers, we consider again the two-point tunneling Hamiltonian given in Eq.~(\ref{eq:hamiltonian_MZ_not_own_anti}).
We assume that the temperature is so low that the distance between the two contacts along the edges is much shorter than the thermal length.
Then it is legitimate to ``fuse'' the two contacts into one. The effective tunneling amplitude depends on the statistical phase accumulated by an anyon around the interferometer.
When considering the process, in which the localized topological charge changes from $ a $ to $ b $, we need to find out whether it's $ x $ or $ \bar{x} $ tunneling that results in this fusion.
Moreover, if $ x $ and $ a $ can fuse to $ b $, then $ \bar{x} $ and $ b $ can fuse to $ a $ since $ N_{x a}^{b} = N_{\bar{x} b}^{a} $.
Now we could write the tunneling rates as 
\begin{align}
p_{x a}^{b} & =  N_{x a}^{b} \frac{d_b}{d_x d_a} \left|\Gamma_1+ \Gamma_2 e^{i \phi^{+}}\right|^2 p^{+}(T_1, T_2), \\
p_{\bar{x} b}^{a} & =   N_{\bar{x} b }^{a} \frac{d_a}{d_{\bar{x}} d_{b} }
\left|\Gamma_1^*+ \Gamma_2^* e^{i \phi^{-}}\theta_x^2\right|^2 p^{-}(T_1,T_2),
\end{align}
where $ e^{i\phi^+} =  {\theta_b}/({\theta_{x} \theta_a}) $, $ e^{i\phi^-} =  {\theta_a}/({\theta_{\bar{x}} \theta_b}) $, and the origin of the $\theta_x^2$ factor is the same as in the previous section. We observe that
\begin{equation}
\label{ratios}
p^b_{xa}=\frac{d_b^2}{d_a^2}p^a_{\bar x b}.
\end{equation}
The matrix $ M $ in the kinetic equation (\ref{eq:kinetic}) is of the following form,
\begin{align}
M_{ab} = p_{x b}^{a} + p_{\bar{x} b}^{a} - \delta_{a b} \sum_{c} (p_{x b}^{c} + p_{\bar{x} b}^{c} ),
\end{align}
but note that $ p_{x b}^{a} $ and $ p_{\bar{x} b}^{a} $ cannot be nonzero at the same time. The same applies to $ p_{x b}^{c} $ and $ p_{\bar{x} b}^{c} $.
Using Eq.~(\ref{ratios}) we solve the kinetic equation: $ f_a= d_a^2/D$. We now repeat the calculations from Eqs.~(\ref{eq:thermal_current_self_anti-1}) and (\ref{eq:thermal_current_self_anti-2})
with almost no changes. The final result is
\begin{equation}
I^{MZ}\sim|\Gamma_1|^2+|\Gamma_2|^2.
\end{equation}

The Fabry-P\'erot current for a localized anyon $ a $ in the fusion channel $ x \times a \to b $ is
\begin{equation}
I_{xa}^{b} = \left|\Gamma_{1}+\Gamma_{2} \cdot \frac{\theta_{b}}{\theta_{x} \theta_{a}} \right|^{2} r^{+}(T_1, T_2).
\end{equation}
The total thermal current in the Fabry-P\'erot setup is obtained by averaging over all possible fusion outcomes of $x$ ($ \bar{x} $) and the trapped anyon $a$:
\begin{equation}
    I_{a}^{\rm FP}=\sum_{b} \left(N^b_{xa}\frac{d_b}{d_x d_a}I^b_{xa} + N^b_{\bar{x} a}\frac{d_b}{d_{\bar{x}} d_a}I^b_{\bar{x}a}  \right).
\end{equation}
We again emphasize that for a fixed $ b $, it is impossible that both $ N_{x a}^{b}  $ and $ N_{\bar{x} a}^{b} $ are nonzero.

\section{Noise in Fabry-P\'erot interferometers}\label{sec:noise}

We now focus on the Ising statistics again. As we discussed, a Fabry-P\'erot interferometer
provides a clear signature of the Ising statistics as long as the trapped topological charge is under control. If that is not the case, we should compare the Fabry-P\'erot and Mach-Zehnder geometries. Fabricating a Mach-Zehnder device is more challenging than making a Fabry-P\'erot interferometer. In this section we propose an alternative approach to probing statistics on the basis of telegraph noise in a Fabry-P\'erot device with a hole.

\begin{figure}[!htb]
\bigskip
\centering%
{\includegraphics[scale=0.85]{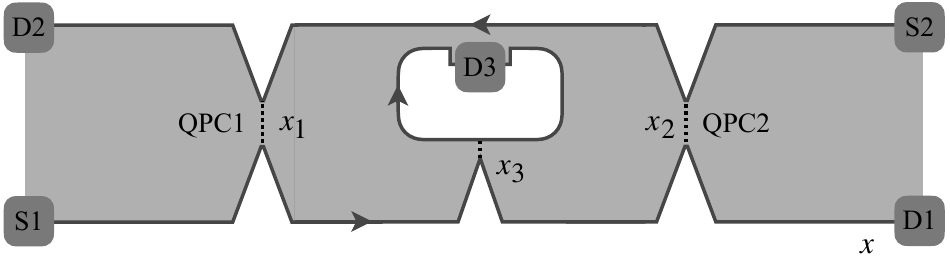}}
\caption{{Fabry-P\'erot interferometer with an inner closed edge inside the device. Telegraph noise due to weak tunneling between the bottom edge and the inner edge  at $x_3$ gives an alternative way to probe statistics. As in a Mach-Zehnder device, each tunneling event changes the enclosed topological charge.}}
\label{fig:hole_noise}
\end{figure}

We consider the case where an inner closed edge is present in a Fabry-P\'erot interferometer, and weak tunneling between the bottom edge and the inner edge is possible, as depicted in Fig.~\ref{fig:hole_noise}.
The average time between tunneling events at position $ x_3 $ is $ \tau $.
Each tunneling event at $x_3$ changes the topological charge trapped between QPC1 and QPC2.
Because of that, the thermal current from S1 to D2 is subject to change over time. We assume that $\tau$ is much longer than the average time between tunneling events at QPC1 and QPC2 (Fig.~\ref{fig:hole_noise}). As in the previous section, we assume that the interferometer is shorter than the thermal length.
As before, the tunneling amplitudes at QPC1 and QPC2 are $\Gamma_1$ and $\Gamma_2$. 
$\tau$ is determined by the tunneling amplitude at QPC3.

We consider the zero-frequency thermal shot noise given by the following equation~\cite{martin2005283:noise},
\begin{equation}\label{eq:noise_def}
S(\omega \to 0) = \lim_{\omega \to 0} \int_{-\infty}^{\infty} d t \, S(t)
 e^{i \omega t},
\end{equation}
where 
\begin{equation}
S(t) =  \langle \{ I_{T}(t), I_{T}(0)\}  \rangle
-2 \langle I_{T}(t) \rangle \langle I_{T}(0) \rangle ,
\end{equation}
the angular brackets denote averaging, the curly brackets stand for an anticommutator,
and $I_T$ is the total thermal current between the upper and lower edges through QPC1 and QPC2.

We will treat the tunneling process at QPC3 as a Poisson process.
We use the following notations:
$ t_0 <0$ is the start time, $ t_k $ is the time when the $ k $-th tunneling event happens, and the time interval between successive events follows an exponential distribution, whose rate parameter $ \lambda = 1/\tau $. 
A tunneling event results in a change of the topological charge of the inner edge.  
The charge changes from $ \sigma $ to $ \bm{1} $ or $ \psi $ with equal probability, but the next tunneling event must change it back to $ \sigma $.

In this picture, we assume that the initial topological charge of the inner edge is $ \sigma $, and the heat current at any time $ t $ is
\begin{equation}\label{eq:heat_current_rand}
I_{T}(t) = I_0 + \Delta I \sum_{k=1}^{\infty} s_k \theta(t-t_k),
\end{equation}
where $ I_0 =I_{\sigma}=r(T_1,T_2)(\Gamma_1^2+\Gamma_2^2)$ is the current at time $ t_0 $, $ \Delta I = 2 r(T_1, T_2) \Gamma_{1} \Gamma_{2}$, and the random variables $s_k=\pm 1$~\cite{kane2003:PhysRevLett.90.226802}.
We treat $s_{2k-1}$ as independent random variables with 
the probabilities
$P(s_{2k-1}=-1)=P(s_{2k-1}=1)=1/2$, while the value of $ s_{2k} $ completely depends on $ s_{2k-1} $ via the constraint $s_{2k-1}s_{2k}=-1$.
Note that the heat currents between the upper and lower edges for the inner-edge topological charges ${\bm{1}}$ and $\psi$ are $ I_{\bm{1}} = I_{\sigma}+\Delta I $ and $ I_{\psi} = I_{\sigma}-\Delta I $. Thus, our definition of the current satisfies all the requirements we set above.
The intervals $ \tau_k = t_k - t_{k-1} $ are independent and identically distributed exponential random variables,
\begin{equation}
P(\tau_k > t ) =  e^{-\lambda t}, \quad t\geq 0. %
\end{equation}
Let us define $N(t)$ as the total number of the tunneling events at QPC3 during the time interval $(t_0, t]$,
\begin{equation}\label{eq:number_events_def}
N(t) = \max \{k\geq0: t_k \leq t\}.
\end{equation}
This is known as a Poisson process, and $N(t)$ satisfies the Poisson distribution
\begin{equation}
P(N(t) = n) = e^{-\lambda(t-t_0)} \frac{(\lambda (t-t_0))^n}{n!}.
\end{equation}
After taking the average with respect to the random variables $ s_{2k-1} $ and $ \tau_{k} $, we find (see Appendix~\ref{appendix:noise})
\begin{equation}\label{eq:noise_result}
\lim_{\omega\to 0  }  \int_{-\infty}^{\infty} dt \, e^{i\omega t} 
S(t)
= 2 (\Delta I)^2 \tau (1- e^{\lambda t_0} ).
\end{equation}
In the limit $t_0\to -\infty$, the equation above reduces to $ 2 (\Delta I)^2 \tau $, and we can conclude that
\begin{equation}
S (\omega\to 0) 
= 8 \tau [r(T_1, T_2)]^2 \Gamma_1^2 \Gamma_2^2.
\end{equation}
If $ \tau $ is large, a high noise can be observed.

\begin{figure*}[t]
	\centering
	\includegraphics[scale=0.9]{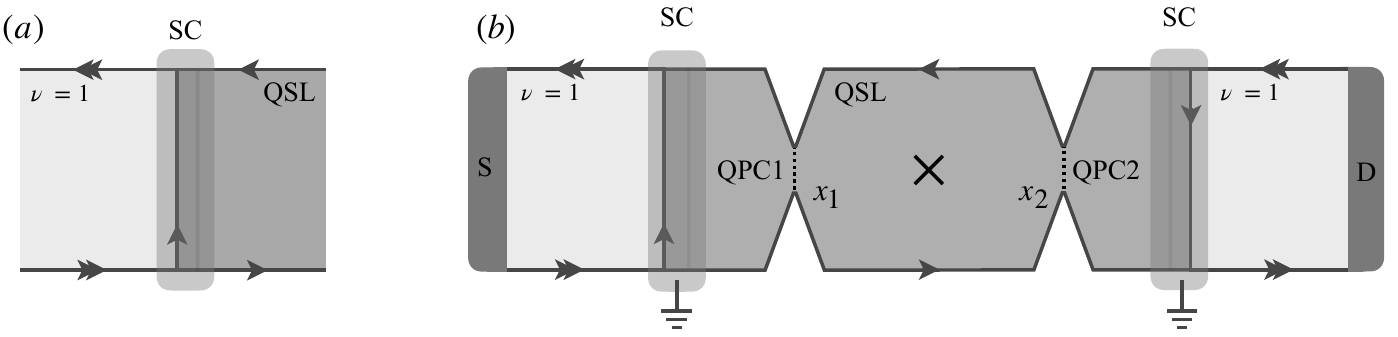}
	\caption{(a) The junction between a quantum spin liquid (QSL) and an integer quantum Hall liquid mediated by a superconductor (SC). The $\nu=1$ edge state (double arrow), under the influence of a superconductor, gets separated into two co-propagating Majorana fermions. Due to strong interactions with the QSL, one of the Majorana modes and the spin liquid's edge Majorana mode form a gapped fermion condensate (faded line under the superconductor without any arrow) which partially `sews' the two subsystems together. An emergent Majorana mode then continues on the edge of the QSL before recombining again with another Majorana mode on the other edge. Details of this process can be found in Ref.~\cite{aasen2020:PhysRevX.10.031014}. (b) Using the process described in (a), one can now consider a QSL interferometer between two superconducting junctions with two quantum Hall liquids. The whole system is then attached to an electrical source and drain and allows one to electrically probe the chiral Majorana edge states of the QSL.}
	\label{fig:IQH-SC-QSL}
\end{figure*}

\section{Conclusions}\label{sec:conclusion}
We have learned that interferometry can be done with neutral anyons as long as we focus on thermal current. The basic idea of a thermal interferometer probe is rather similar to electronic interferometry. However, details of the setup  differ. Controlling and measuring an electric current is relatively straightforward  even for low currents used in experiments with the fractional quantum Hall effect. Dealing with tiny thermal currents is harder. 

A breakthrough idea~\cite{Jezoin} allowed probing thermal currents in the quantum Hall effect. It turns out that thermal measurements can be reduced to purely electric measurements on the basis of Joule's law. This, of course, relies on a finite electrical conductance in QHE systems. Additional ideas are needed to deal with neutral systems such as Kitaev magnets. A crucial challenge is a very low temperature at which the experiment has to be conducted. Indeed,
the constriction size cannot be smaller than the unit cell size. This means the scale of the order of at least $a\sim 1$ nm in $ \alpha $-RuCl$_3$. The interferometer size should be at least an order of magnitude bigger, yet it should be shorter than the thermal length $\hbar v/T$, where the edge velocity $v\sim a \Delta /\hbar$, with $\Delta$ being the energy gap, is estimated at 
$\sim 10^5$ cm$/$s. Besides, the temperature should be much lower than the energy gap $\Delta$ estimated 
at a few Kelvin~\cite{RuCl3}. Thus, we have to deal with sub-Kelvin temperatures. Thermal currents $\sim T^2/h$ scale as the square of the temperature and are hard to probe in that temperature range.

One approach was advocated in Ref.~\cite{prx-t-int} . The idea builds on using edge-phonon interactions for a sufficiently long edge. Such interactions are highly irrelevant in the renormalization group sense, and we will not address that approach here. See Ref.~\cite{prx-t-int} for details. Two other approaches are possible.
Both reduce a thermal measurement to an electric noise measurement.

One approach requires a junction of a spin liquid and a quantum Hall systems. 
It was proposed that such a junction between a Kitaev magnet and an integer QHE liquid can be mediated by a 
superconductor~\cite{aasen2020:PhysRevX.10.031014}. The edge structure is illustrated in Fig.~\ref{fig:IQH-SC-QSL}(a). We can see that the QHE edge splits into a neutral channel running under a superconductor and a neutral edge of the Kitaev liquid. The modes recombine on the other side of the superconducting junction.
We now consider a setup with an interferometer between two superconducting junctions [Fig.~\ref{fig:IQH-SC-QSL}(b)]. If the QHE bars on the two sides are maintained at the same temperature, the noise in the drain equals the Nyquist noise at the common temperature. In a non-equilibrium situation, relevant for this paper, the noise is no longer 
given by the Nyquist formula and depends on the tunneling heat current.

\begin{figure}[!htb]
    \centering
    \includegraphics{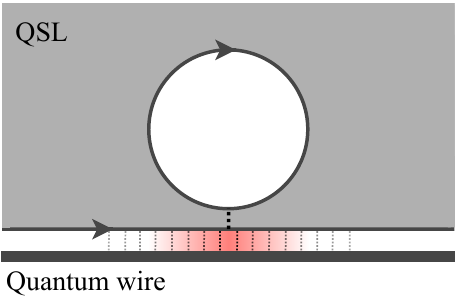}
    \caption{Interaction between the quantum wire and the spin liquid is more relevant in the renormalization group sense near the hole.}
    \label{fig:heat_transfer_QH_QSL}
\end{figure}

In another approach, heat is transferred between the interferometer edge and a quantum wire. The standard technique~\cite{Jezoin,texp1,texp2} can then be used to measure the heat current in the wire
by connecting it with a QHE system. 
The leading interaction of the wire and the Kitaev edge in contact, $\psi\partial_x\psi\partial_x\phi$, is proportional to the charge density $\sim \partial_x\phi$ in the wire and the energy density on the Kitaev edge. It is highly irrelevant in the renormalization group sense at low energies. Thus, a long region of contact may be needed. Two observations mitigate this challenge. First, the relevancy of the interaction can be increased by using the geometry from Fig.~\ref{fig:heat_transfer_QH_QSL}.
Here a hole is made near the Kitaev edge. 
The interaction of the wire with the Kitaev liquid near the point, where the hole approaches the edge, is $\sigma_1\sigma_2\partial_x \phi$ with $\sigma_{1,2}$ denoting anyon operators on the edge of the hole and the outer edge of the Kitaev liquid. 
This interaction is significantly more relevant than the interaction without the hole, provided that the system is tuned close to resonance so that the $\sigma_1\sigma_2$
interaction is suppressed.

It is also important that the interferometry experiment does not require  probing the precise energy current on the edge. It is enough to compare corrections to the edge thermal current due to the tunneling contacts at different trapped topological charges. This can be accomplished even if only a small portion of the heat current is transferred from the edge of the spin liquid to the wire. Indeed, the difference of the  transferred heat currents  with and without tunneling contacts in the Kitaev liquid is proportional to the tunneling heat current through the interferometer.

A Mach-Zehnder device cannot be confined in a single plane due to its complicated topology (Fig.~\ref{fig1:thermal}). This is a fabrication challenge. 
We would like to emphasize that this challenge has been overcome for quantum Hall liquids~\cite{MZ-heiblum,MZ-review}. On the other hand, quantum coherence persists in Mach-Zehnder interferometers at a longer size than in Fabry-P\'erot devices. Indeed, the current through a Fabry-P\'erot interferometer depends on the sum of the distances between the tunneling contacts along the two edges. In the Mach-Zehnder case, only the length difference matters. In particular, the heat current depends on the combination of the thermal length and that difference (Appendix~\ref{appendix:inter_size}). Thus, interference can be observed at shorter thermal lengths and higher temperatures than in a Fabry-P\'erot interferometer of a similar size.

Interferometry is the most direct probe of statistics. A less direct but simpler probe involves heat current through a single contact. It scales as a universal power of the temperature at a fixed ratio of the temperatures on the two edges across the contact.

In conclusion, Fabry-P\'erot interferometry shows distinct values of the heat current through the interferometer for each possible topological charge confined in the device. It may be challenging to control the trapped topological charge, in which case the heat current through the interferometer might be the same for anyons as for bosons and fermions. One can then distinguish anyons from fermions and bosons by using a Mach-Zehnder setup. 
A striking feature is the absence of interference signal in  a Mach-Zehnder device for
any non-trivial topological order.
Besides, anyonic statistic induces a high telegraph noise of the heat current in a Fabry-P\'erot device with  a hole.

\begin{acknowledgments}

This research was supported in part by the NSF under grant No. DMR-1902356.

\end{acknowledgments}

\newpage
\appendix
\onecolumngrid

\section{$\psi$ tunneling}\label{appendix:psi}

Here we compute, in the leading order, the tunneling thermal current in a two-constriction Fabry-P\'erot interferometer with single-Majorana-fermion tunneling from one edge of the spin-liquid to the other edge of the same spin-liquid, see also Ref.~\cite{prx-t-int}. Several equations from this appendix are used in the calculations presented in Appendix~\ref{appendix:psi-dx-psi}. The full Hamiltonian for the system is given as,
\begin{align}
H = \frac{v}{4\pi} \int_{-L/2}^{L/2} dx \Big[ -i : \psi_1\partial_x\psi_1: +i :\psi_2\partial_x\psi_2:\Big] + H_{T}.
\end{align}
where $H_{T}$ is the single-Majorana fermion tunneling Hamiltonian at the quantum point contacts. The geometry of the Fabry-P\'erot interferometer is shown in Fig.~\ref{fig1:thermal}(a). The subscripts 1 and 2 on the Majorana fermion field $\psi$ denote the two edges of the spin liquid. We work in the $L \rightarrow \infty$ limit and impose the periodic boundary conditions $\psi_a(x)\equiv \psi_a(x+L)$. The edges have different temperatures $T_{1,2}$.
The free field Hamiltonian can be diagonalized by working in the Fourier space, defining,
\begin{align}
\psi_1(x) &= \sqrt{\frac{2\pi}{L}} \sum_{k} e^{ikx} \psi_{1,k},   \\
\psi_2(x) &= \sqrt{\frac{2\pi}{L}} \sum_{k} e^{-ikx} \psi_{2,k}.
\end{align}
We first compute the Majorana-fermion two-point thermal correlation function,
\begin{align}
\langle \psi_1(x,t)\psi_1(0,0) \rangle &= \frac{2\pi}{L} \sum_{k,q} e^{ik(x-vt)}e^{-k\epsilon}\langle \psi_{k}\psi_{q}\rangle  = \frac{2\pi}{L} \sum_{k,q} e^{ik(x-vt)}e^{-k\epsilon} \frac{\delta_{k+q}}{e^{-\beta v k}+1} \\
& \xrightarrow[]{{L\to\infty}} \int dk ~\frac{e^{ik(x-vt+i\epsilon)}}{e^{-\beta v k}+1} = -\frac{\pi T_1}{v\sin[i\pi T_1(x/v-t+i\epsilon)]} .
\end{align}
The expression for $\psi_2$ is similar with a different temperature. The tunneling Hamiltonian is given as,
\begin{align}
H_{T} = -i \Gamma_1 \psi_2(x_1,t)\psi_1(x_1,t) -i \Gamma_2 \psi_2(x_2,t)\psi_1(x_2,t),
\end{align}
where $x_i$ is the location of the $i$th constriction (see Fig.~\ref{fig1:thermal}(a)), and $\Gamma_i$ is its real tunneling amplitude.
Using the equation of motion, we can identify the tunneling current operator as, 
\begin{align}
\hat{I}_{T}(t) = i \Gamma_1 \psi_2(x_1,t)\partial_t\psi_1(x_1,t) +i \Gamma_2 \psi_2(x_2,t)\partial_t\psi_1(x_2,t),
\end{align}
where the time derivatives are computed from the edge theory without tunneling.
An easy derivation consists in computing the commutator of the tunneling Hamiltonian with one of the edge Hamiltonians.
We now use perturbation theory to compute the average thermal current to the first non-zero order in the tunneling amplitudes,
\begin{align}
\langle I_{T} \rangle = - i \int_{-\infty}^{t} dt' \Big[& \Gamma_1^2 \langle [ \psi_2(x_1,t)\partial_t\psi_1(x_1,t),\psi_2(x_1,t')\psi_1(x_1,t')  ]\rangle + \Gamma_2^2 \langle [ \psi_2(x_2,t)\partial_t\psi_1(x_2,t),\psi_2(x_2,t')\psi_1(x_2,t')  ]\rangle  \nonumber\\
& +\Gamma_1\Gamma_2 \big( \langle [ \psi_2(x_1,t)\partial_t\psi_1(x_1,t),\psi_2(x_2,t')\psi_1(x_2,t')  ]\rangle +  \langle [ \psi_2(x_2,t)\partial_t\psi_1(x_2,t),\psi_2(x_1,t')\psi_1(x_1,t')  ]\rangle \big) \Big] ,
\end{align}
where the first two terms correspond to the independent single constriction contributions, and the last two terms are the interference terms between the two constrictions. Therefore, we can write the total thermal current as comprised of the non-interference and interference terms,
\begin{align}
\langle  I_{T} \rangle \equiv \sum_{j=1,2} \langle  I_{T} \rangle^{\text{non-int}}_{\Gamma_j}  + \langle  I_{T} \rangle^{\text{int}}.
\end{align} 
In the following two subsections, we individually focus on the non-interference and interference contributions. Note that the non-interference contribution $\langle I_{T} \rangle^{\text{non-int}}_{\Gamma_j}$ corresponds to the $j^{\text{th}}$ constriction, $j=1,2$ and is exactly what we obtain when we consider a single constriction geometry as shown in Fig.~\ref{fig2:thermal}. Therefore, the thermal current in the single constriction case can be found as the non-interference part, corresponding to one of the tunneling amplitudes $\Gamma_j$, of the full calculation that is presented in this appendix.

Our calculations assume that the two edges of the interferometer remain in equilibrium at the temperatures of the sources. This assumption is legitimate for weak tunneling despite the nonequilibrium nature of tunneling transport, if the average time between two consecutive tunneling events is much longer
then the travel time along the edges between the two constrictions \cite{law2006:PhysRevB.74.045319}.
See a detailed discussion of the validity conditions for the perturbation theory in Ref.~\cite{law2006:PhysRevB.74.045319}. Note that the physical situation is different from Ref.~\cite{Melcer}, where a strongly nonequilibrium edge state emerges due to the coexistence of counter-propagating channels of different temperatures on the same edge. In our case, the temperatures of the drain and source connected to the same edge are approximately equal.

Some topological orders allow counter-propagating edge modes on the same edge. Then equilibration processes on the edge change slightly the edge temperatures at small $\Gamma_i$. This effect is irrelevant in the lowest-order perturbation theory.

\subsection{Non-interference term}
\begin{figure}[!htb]
	\centering
	\subfloat[\label{subfig:contour_a}]{%
		\includegraphics[width=0.4\columnwidth]{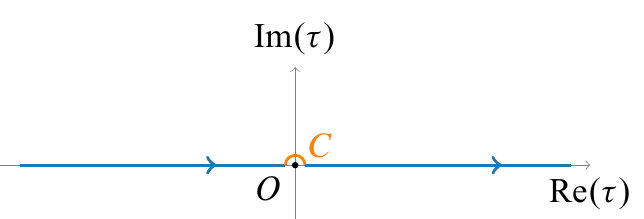}%
	}
	\subfloat[\label{subfig:contour_b}]{%
		\includegraphics[width=0.4\columnwidth]{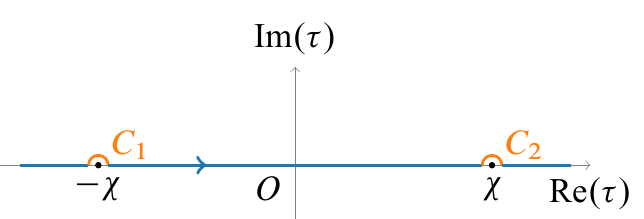}%
	}
	\caption{(Color online) Contours used to compute the integrals in Appendices~\ref{appendix:psi}, \ref{appendix:psi-dx-psi} and \ref{appendix:sigma}. Contour (a) is used for the non-interference terms, and contour (b) is used for the interference terms. $ \chi $ is proportional to the distance $ x_{21} $ between the constrictions. }
\end{figure}

We first compute the non-interference terms. 
We use the above derived thermal correlation functions, in which we set $v=1$,
\begin{align}
\langle \psi_1(x_1,t_1)\psi_1(x_2,t_2) \rangle =  \frac{\pi T_1}{\sin[\pi T_1 (\epsilon + i (t_1-t_2+x_2-x_1))]}, \\
\langle \psi_2(x_1,t_1)\psi_2(x_2,t_2) \rangle =  \frac{\pi T_2}{\sin[\pi T_2 (\epsilon + i (t_1-t_2+x_1-x_2))]}.
\end{align}
Plugging these into the non-interference contribution $\langle I_T\rangle^{\text{non-int}}_{\Gamma_j}$ to the thermal current due to the $j^{\text{th}}$ constriction, we obtain the expression,
\begin{align}
\langle I_{T}\rangle^{\text{non-int}}_{\Gamma_j} = \Gamma_j^2  \int_{-\infty}^{t} d t' \Bigg[ \frac{(\pi T_2)(\pi T_1)^2~\cos[\pi T_1 (\epsilon + i (t-t'))]}{\sin^2[\pi T_1 (\epsilon + i (t-t'))] \sin[\pi T_2 (\epsilon + i (t-t'))]} 
+(t\leftrightarrow t')
\Bigg], ~~~~~~~~(j=1,2) .
\end{align}
Defining $\tau\equiv \pi T_1(t-t')$ and $n = T_2/T_1$, we write the above integral as,
\begin{align}
\langle I_{T}\rangle^{\text{non-int}}_{\Gamma_j}
&=  \pi^2\Gamma_j^2 T_1 T_2  \int_{-\infty}^{\infty} d\tau ~\frac{\cos(-i\tau +\epsilon)}{\sin^2(-i\tau +\epsilon) \sin[n(-i\tau + \epsilon)]}.
\end{align}
Since the $\epsilon \rightarrow 0$ limit is to be taken after evaluating the integral, to evaluate the integral, we deform the integration contour as shown in Fig.~\subref*{subfig:contour_a},
\begin{align}
\langle I_{T}\rangle^{\text{non-int}}_{\Gamma_j} =  \pi^2 \Gamma_j^2 T_1 T_2 \left[ \int_{-\infty}^{-\epsilon} d\tau ~\frac{\cos(-i\tau)}{\sin^2(-i\tau) \sin(-in\tau)} + \int_{C} d\tau ~\frac{\cos(-i\tau)}{\sin^2(-i\tau) \sin(-in\tau)} + \int^{\infty}_{\epsilon} d\tau ~\frac{\cos(-i\tau)}{\sin^2(-i\tau) \sin(-in\tau)} \right].
\end{align}
Here $C$ is a half-circle near the origin. 
Pictorially, the integration contour is deformed as shown in Fig.~\subref*{subfig:contour_a}. Close to the origin, the integration variable can be defined as $\tau = \epsilon e^{i\theta}~$. 

On inspection, it's easy to see that the first and the last integrals cancel out,
hence only the integral along the contour $C$ gives a non-zero contribution,
\begin{align}
\int_{C}d\tau \frac{\cos(-i\tau)}{\sin^2(-i\tau) \sin(-in\tau)}
= \int_{\pi}^{0} d\theta \frac{\tau \cosh\tau}{\sinh^2{\tau} \sinh(n\tau)}.
\end{align}
Since $\epsilon$ is small, we may use the expansion,
\begin{align}
\frac{\cosh\tau}{\sinh^2{\tau} \sinh(n\tau)} = \frac{1}{n\tau^3} + \frac{1-n^2}{6n \tau} + \mathcal{O}(\tau).
\end{align}
Using this, we evaluate the integral in the $\epsilon\rightarrow 0$ limit,
\begin{align}
\int_{\pi}^{0} d\theta \frac{\tau \cosh\tau}{\sinh^2{\tau} \sinh(n\tau)} = \frac{\pi(n^2-1)}{6n} = \frac{\pi(T_2^2-T_1^2)}{6T_1T_2}.
\end{align}
Putting the result of the integral back into the expression for the non-interference terms in the thermal current we obtain,
\begin{align}
\sum_{j=1,2}\langle I_{T}\rangle^{\text{non-int}}_{\Gamma_j} = \frac{\pi^3(\Gamma_1^2+\Gamma_2^2)}{6} (T_2^2-T_1^2) .
\end{align}
This is the contribution to the thermal current from two independent constrictions. Calculation of the thermal current in the single constriction geometry would involve only the non-interference contribution $\langle I_T \rangle^{\text{non-int}}_{\Gamma_j} $ due to one of the constrictions.

\subsection{Interference term}
In the double constriction Fabry-P\'erot geometry, there is an interference contribution as well, which we compute now,
\begin{align}
\langle I_{T}\rangle^{\text{int}} = \Gamma_1\Gamma_2  \int_{-\infty}^{t} dt' \Bigg[& \frac{(\pi T_2)(\pi T_1)^2~~\cos[\pi T_1 (\epsilon + i (t-t'+x_{21}))]}{\sin^2[\pi T_1 (\epsilon + i (t-t'+x_{21}))] \sin[\pi T_2 (\epsilon + i (t-t'-x_{21}))]} \nonumber\\
&+\frac{(\pi T_2)(\pi T_1)^2 ~~\cos[\pi T_1 (\epsilon + i (t'-t-x_{21}))]}{\sin^2[\pi T_1 (\epsilon + i (t'-t-x_{21}))] \sin[\pi T_2 (\epsilon + i (t'-t+x_{21}))]} + (x_{21}\leftrightarrow -x_{21}) \Bigg] .
\end{align}
Like before, for convenience, we define $\tau\equiv \pi T_1(t-t')$, $\chi\equiv  \pi T_1 x_{21}$, and $n = T_2/T_1$,
\begin{align}
\langle I_{T}\rangle^{\text{int}} ={}& \pi^2\Gamma_1\Gamma_2T_1T_2\int_{0}^{\infty} d\tau \left[ \frac{\cos[i(\tau+\chi)+\epsilon]}{\sin^2[i(\tau+\chi)+\epsilon]\sin[n(i(\tau-\chi)+\epsilon)]} + \frac{\cos[-i(\tau+\chi)+\epsilon]}{\sin^2[-i(\tau+\chi)+\epsilon]\sin[n(-i(\tau-\chi)+\epsilon)]}\right] \nonumber \\
& + \pi^2\Gamma_1\Gamma_2T_1T_2\int_{0}^{\infty} d\tau \left[ \frac{\cos[i(\tau-\chi)+\epsilon]}{\sin^2[i(\tau-\chi)+\epsilon]\sin[n(i(\tau+\chi)+\epsilon)]} + \frac{\cos[-i(\tau-\chi)+\epsilon]}{\sin^2[-i(\tau-\chi)+\epsilon]\sin[n(-i(\tau+\chi)+\epsilon)]}\right] \\
={}& \pi^2\Gamma_1\Gamma_2T_1T_2\int_{-\infty}^{\infty} d\tau  \left[ \frac{\cos[-i(\tau+\chi)+\epsilon]}{\sin^2[-i(\tau+\chi)+\epsilon]\sin[n(-i(\tau-\chi)+\epsilon)]} + \frac{\cos[-i(\tau-\chi)+\epsilon]}{\sin^2[-i(\tau-\chi)+\epsilon]\sin[n(-i(\tau+\chi)+\epsilon)]}\right] .
\end{align}
The integral is evaluated using the contour defined in Fig.~\subref*{subfig:contour_b}. After deforming the contour around $\tau = \pm\chi$ as shown in Fig.~\subref*{subfig:contour_b}, we observe that the only surviving terms are
\begin{align}
\langle I_{T}\rangle^{\text{int}} =-i\pi^2\Gamma_1\Gamma_2T_1T_2 \int_{C_1+C_2}d\tau \left[ \frac{\cosh(\tau+\chi)}{\sinh^2(\tau+\chi)\sinh[n(\tau-\chi)]} + \frac{\cosh(\tau-\chi)}{\sinh^2(\tau-\chi)\sinh[n(\tau+\chi)]} \right],
\end{align}
The integration variable near $-\chi$ is given by $\eta=\epsilon e^{i\theta}-\chi$, and the integration variable near $\chi$ is given by $\eta=\epsilon e^{i\theta}+\chi$. Again, since $\epsilon$ is small, we use the series expansions to perform the integrals and then take the $\epsilon \rightarrow 0$ limit,
\begin{align}
\frac{\cosh\tau}{\sinh^2\tau} &= \frac{1}{\tau^2} + \frac{1}{6} + \mathcal{O}(\tau^2), \\
\frac{1}{\sinh\tau} &= \frac{1}{\tau} -\frac{\tau}{3} + \mathcal{O}(\tau^3).
\end{align}
Using these expansions, we arrive at the expression,
\begin{align}
\langle I_T \rangle^{\text{int}} = 2\pi^3 \Gamma_1\Gamma_2 \left [ T_2^2 \frac{\cosh(2\pi T_2 x_{21})}{\sinh^2(2\pi T_2x_{21})} - T_1^2 \frac{\cosh(2\pi T_1 x_{21})}{\sinh^2(2\pi T_1 x_{21})} \right]  .
\end{align} 
Combining the interference and non-interference terms together, we arrive at the expression for the thermal current,
\begin{align}
\langle I_T \rangle = \frac{\pi^3(\Gamma_1^2+\Gamma_2^2)}{6} (T_2^2-T_1^2)  +2\pi^3 \Gamma_1\Gamma_2 \left [ T_2^2 \frac{\cosh(2\pi T_2 x_{21})}{\sinh^2(2\pi T_2x_{21})} - T_1^2 \frac{\cosh(2\pi T_1 x_{21})}{\sinh^2(2\pi T_1 x_{21})} \right].   
\end{align}

\section{$\psi\partial_x\psi$ tunneling} \label{appendix:psi-dx-psi}
In this appendix, we compute the tunneling thermal current between adjacent spin liquids as discussed in the main text in Sec.~\ref{sec:tunneling_Between_Topological}. Note that the calculation for the single constriction geometry, as discussed in Sec.~\ref{subsec:boson_single}, corresponds to the non-interference part $\langle I_{T}\rangle^{\text{non-int}}_{\Gamma} $ (here $\Gamma_i = \Gamma$) of the full calculation in the double constriction Fabry-P\'erot geometry which we present in this appendix. We set the edge velocity $v=1$. Therefore, the final answers should be divided by $v^8$. We rely on Green's functions, computed in Appendix~\ref{appendix:psi}.

From perturbation theory and the expression of the  tunneling thermal current operator, Eq.~(\ref{eq:psi_dx_psi_current_operator}), the leading contribution to the tunneling thermal current is
\begin{align}
\langle I_T(t)\rangle &= - 4(\Gamma_1^2+\Gamma_2^2)\pi^9T_1^5T_2^4 \int_{-\infty}^{\infty}d\tau \frac{\cos[i\pi T_1(\tau+i\epsilon)]}{\sin^5[i\pi T_1(\tau+i\epsilon)]}\frac{1}{\sin^4[i\pi T_2(\tau+i\epsilon)]} \nonumber \\
&\qquad -4\Gamma_1\Gamma_2\pi^9T_1^5T_2^4 \Bigg[ \int_{-\infty}^{\infty}d\tau \frac{\cos[i\pi T_1(\tau+\chi+i\epsilon)]}{\sin^5[i\pi T_1(\tau+\chi+i\epsilon)]}\frac{1}{\sin^4[i\pi T_2(\tau-\chi+i\epsilon)]} + (\chi\rightarrow -\chi)\Bigg] \\
&= 4i(\Gamma_1^2+\Gamma_2^2)\pi^9T_1^5T_2^4 \int_{-\infty}^{\infty}d\tau \frac{\cosh[\pi T_1(\tau+i\epsilon)]}{\sinh^5[\pi T_1(\tau+i\epsilon)]}\frac{1}{\sinh^4[\pi T_2(\tau+i\epsilon)]} \nonumber \\
&\qquad + 4i\Gamma_1\Gamma_2\pi^9T_1^5T_2^4 \Bigg[ \int_{-\infty}^{\infty}d\tau \frac{\cosh[\pi T_1(\tau+\chi+i\epsilon)]}{\sinh^5[\pi T_1(\tau+\chi+i\epsilon)]}\frac{1}{\sinh^4[\pi T_2(\tau-\chi+i\epsilon)]} + (\chi\rightarrow -\chi)\Bigg],
\end{align}
where $\chi$ is the distance between the two tunneling contacts divided by $v$. We assume $\chi>0$.
We redefine the variables for convenience, $\eta = \pi T_1 \tau$, $\lambda = T_2/T_1$, and $\chi\rightarrow \pi T_1\chi$. This gives,
\begin{align}
\langle I_T(t)\rangle &= 4i(\Gamma_1^2+\Gamma_2^2)\pi^8T_1^4T_2^4 \int_{-\infty}^{\infty}d\eta \frac{\cosh\eta}{\sinh^5(\eta+i\epsilon)}\frac{1}{\sinh^4(\lambda\eta+i\epsilon)} \nonumber \\
&\qquad + 4i\Gamma_1\Gamma_2\pi^8T_1^4T_2^4 \Bigg[ \int_{-\infty}^{\infty}d\tau \frac{\cosh(\eta+\chi)}{\sinh^5(\eta+\chi+i\epsilon)}\frac{1}{\sinh^4(\lambda\eta-\lambda\chi+i\epsilon)} + (\chi\rightarrow -\chi)\Bigg].
\end{align}
This expression consists of the non-interference contributions due to each of the constrictions independently, and the interference contribution. Therefore, the total tunneling thermal current can be written as $\langle  I_T\rangle = \sum_{j=1,2}\langle I_T \rangle^{\text{non-int}}_{\Gamma_j} +\langle I_T \rangle^{\text{int}}  $, where the non-interference and interference contributions are given as,
\begin{align}\label{eq:psi_dx_psi_non_int_integral}
    \langle I_{T}\rangle^{\text{non-int}}_{\Gamma_j} &=  4i\Gamma_j^2\pi^8T_1^4T_2^4 \int_{-\infty}^{\infty}d\eta \frac{\cosh\eta}{\sinh^5(\eta+i\epsilon)}\frac{1}{\sinh^4(\lambda\eta+i\epsilon)}, ~~~~~~~~~~j=1,2, \\ 
    \langle I_{T}\rangle^{\text{int}} &= 4i\Gamma_1\Gamma_2\pi^8T_1^4T_2^4 \Bigg[ \int_{-\infty}^{\infty}d\tau \frac{\cosh(\eta+\chi)}{\sinh^5(\eta+\chi+i\epsilon)}\frac{1}{\sinh^4(\lambda\eta-\lambda\chi+i\epsilon)} + (\chi\rightarrow -\chi)\Bigg]. \label{eq:psi_dx_psi_int_integral}
\end{align}
In the following two subsections, we focus on the non-interference and interference contributions separately. Note that the non-interference contribution $\langle I_{T}\rangle^{\text{non-int}}_{\Gamma_j}$ corresponding to each $\Gamma_j$, $j=1,2$, is exactly what we obtain when we consider single constriction geometry between two different spin liquids. Therefore, the thermal current in the single constriction case, as discussed in Sec.~\ref{subsec:boson_single}, can be found as the non-interference part, corresponding to one of the tunneling amplitudes $\Gamma_i$, of the full calculation presented in this appendix.
\subsection{Non-interference term}
First we consider the non-interference contribution (Eq.~(\ref{eq:psi_dx_psi_non_int_integral})) to the tunneling thermal current. It can be rewritten as,
\begin{align}
\langle I_T \rangle^{\text{non-int}}_{\Gamma_j} &=  4i\Gamma_j^2 \pi^8 T_1^4T_2^4 \int_{-\infty}^{\infty} d\eta \frac{\cosh{\eta}}{\sinh^5{(\eta+i\epsilon)}\sinh^4(\lambda\eta+i\epsilon)}\\
&=4i\Gamma_j^2 \pi^8 T_1^4T_2^4 \Bigg[ \int_{-\infty}^{-\epsilon} d\eta \frac{\cosh{\eta}}{\sinh^5{(\eta)}\sinh^4(\lambda\eta)} + \int^{\infty}_{\epsilon} d\eta \frac{\cosh{\eta}}{\sinh^5{(\eta)}\sinh^4(\lambda\eta)} \nonumber\\
&\qquad\qquad\qquad\qquad\qquad\qquad + \int_{C} d\eta \frac{\cosh{\eta}}{\sinh^5{(\eta)}\sinh^4(\lambda\eta)}  \Bigg] \\
&=4i\Gamma_j^2 \pi^8 T_1^4T_2^4 \int_{C} d\eta \frac{\cosh{\eta}}{\sinh^5{(\eta)}\sinh^4(\lambda\eta)},  
\end{align}
where $j=1,2$.
Here, the integration contour is deformed as shown in Fig.~\subref*{subfig:contour_a}.
Close to the origin, the integration variable can be written as $\eta=\epsilon e^{i\theta}$. Since $\epsilon \ll 1$, we expand the integrand,
\begin{align}
\frac{\cosh{\eta}}{\sinh^5{(\eta)}\sinh^4(\lambda\eta)} = \dots + \frac{1}{\eta}\left[\frac{41 (\lambda^8 -1)}{2835~ \lambda^4} +\frac{62 (\lambda^4-1)}{2835~\lambda^2} \right] + \mathcal{O}(\eta^0).
\end{align}
All the terms in $\dots$ contain negative odd powers of $\eta$, and since $d\eta=i\epsilon e^{i\theta}d\theta$, on integration from $\pi$ to $0$, these terms give zero. The only non-zero contribution comes from the $1/\eta$ term. This allows us to perform the integral and take the $\epsilon \rightarrow 0$ limit, and we thus obtain,
\begin{align}
\sum_{j=1,2}\langle I_{T} \rangle^{\text{non-int}}_{\Gamma_j} = 4\pi^9(\Gamma_1^2+\Gamma_2^2)\frac{1}{2835} [ 41(T_2^8 - T_1^8) + 62 T_1^2T_2^2(T_2^4-T_1^4) ] .
\end{align}
This is the non-interference contribution to the thermal current due to two independent constrictions. The results discussed in the main text in Sec.~\ref{subsec:boson_single} correspond to $\langle I_T\rangle^{\text{non-int}}_{\Gamma_j}$ where $\Gamma_j$ is set to $\Gamma$.

\subsection{Interference term}
We now look at the interference contribution to the tunneling thermal current, Eq.~(\ref{eq:psi_dx_psi_int_integral}); this is given as,
\begin{align}
\langle I_{T} \rangle^{\text{int}} =4i\Gamma_1\Gamma_2\pi^8T_1^4T_2^4 \Bigg[ \int_{-\infty}^{\infty}d\eta \frac{\cosh(\eta+\chi)}{\sinh^5(\eta+\chi+i\epsilon)}\frac{1}{\sinh^4(\lambda\eta-\lambda\chi+i\epsilon)} + (\chi\rightarrow -\chi)\Bigg].
\end{align}
The contour can be deformed according to Fig.~\subref*{subfig:contour_b} and the integral rewritten as,
\begin{align}
\langle I_{T} \rangle^{\text{int}} = 4i\Gamma_1\Gamma_2\pi^8T_1^4T_2^4 \Bigg[& \int_{-\infty}^{-\chi-\epsilon}d\eta \frac{\cosh(\eta+\chi)}{\sinh^5(\eta+\chi)}\frac{1}{\sinh^4(\lambda\eta-\lambda\chi)} + \int_{C_1}d\eta \frac{\cosh(\eta+\chi)}{\sinh^5(\eta+\chi)}\frac{1}{\sinh^4(\lambda\eta-\lambda\chi)} \nonumber \\
& + \int_{-\chi+\epsilon}^{\chi-\epsilon}d\eta \frac{\cosh(\eta+\chi)}{\sinh^5(\eta+\chi)}\frac{1}{\sinh^4(\lambda\eta-\lambda\chi)} + \int_{C_2}d\eta \frac{\cosh(\eta+\chi)}{\sinh^5(\eta+\chi)}\frac{1}{\sinh^4(\lambda\eta-\lambda\chi)} \nonumber \\
& + \int_{\chi+\epsilon}^{\infty}d\eta \frac{\cosh(\eta+\chi)}{\sinh^5(\eta+\chi)}\frac{1}{\sinh^4(\lambda\eta-\lambda\chi)} + (\chi \rightarrow -\chi)\Bigg],
\end{align}
where the contours $C_1$ and $C_2$ represent respectively the integration variable going over the upper half-plane near $-\chi$ with $\eta=\epsilon e^{i\theta}-\chi$, and similarly the integration variable going over the upper half-plane near $\chi$ with $\eta=\epsilon e^{i\theta}+\chi$. The integrals are evaluated using the contour shown in Fig.~\subref*{subfig:contour_b}. Rearranging these integrals allows the cancellation of a few terms, and we are left with,
\begin{align}
\langle I_{T} \rangle^{\text{int}} = 4i\Gamma_1\Gamma_2\pi^8T_1^4T_2^4 \Bigg[&\int_{C_1}d\eta \frac{\cosh(\eta+\chi)}{\sinh^5(\eta+\chi)}\frac{1}{\sinh^4(\lambda\eta-\lambda\chi)} + \int_{C_2}d\eta \frac{\cosh(\eta+\chi)}{\sinh^5(\eta+\chi)}\frac{1}{\sinh^4(\lambda\eta-\lambda\chi)} + (\chi \rightarrow -\chi)\Bigg].
\end{align}
We now use the following expansions to compute the above integrals,
\begin{align}
\frac{1}{\sinh^4(\epsilon e^{i\theta}+2\chi)} &= \frac{1}{\sinh^4(2\chi)}-\frac{4\epsilon e^{i\theta}\cosh(2\chi)}{\sinh^5(2\chi)} + \frac{\epsilon^2 e^{2i\theta}}{\sinh^4(2\chi)}\left[ \frac{10\cosh^2(2\chi)}{\sinh^2(2\chi)} -2\right]  - \frac{4\epsilon^3 e^{3i\theta}}{3\sinh^4(2\chi)}\left[ \frac{15\cosh^3(2\chi)}{\sinh^3(2\chi)}- \frac{7\cosh(2\chi)}{\sinh(2\chi)}  \right] \nonumber\\
& \quad+ \frac{\epsilon^4 e^{4i\theta}}{3\sinh^4(2\chi)} \left[ \frac{105\cosh^4(2\chi)}{\sinh^4(2\chi)} - \frac{80\cosh^2(2\chi)}{\sinh^2(2\chi)} +7 \right] + \mathcal{O}(\epsilon^5).
\end{align}
Similarly, we have,
\begin{align}
\frac{\cosh(\epsilon e^{i\theta}+2\chi)}{\sinh^5(\epsilon e^{i\theta}+2\chi)} &= \frac{\cosh(2\chi)}{\sinh(2\chi)} + \epsilon e^{i\theta}\left[ \frac{1}{\sinh^4(2\chi)} - \frac{5\cosh^2(2\chi)}{\sinh^6(2\chi)} \right] - \epsilon^2 e^{2i\theta} \left[ \frac{7\cosh(2\chi)}{\sinh^5(2\chi)} - \frac{15\cosh^3(2\chi)}{\sinh^7(2\chi)}  \right] \nonumber \\
& \quad+ \frac{\epsilon^3 e^{3i\theta}}{3} \left[ \frac{80\cosh^2(2\chi)}{\sinh^6(2\chi)} - \frac{105\cosh^4(2\chi)}{\sinh^8(2\chi)} - \frac{7}{\sinh^4(2\chi)} \right] + \mathcal{O}(\epsilon^4).
\end{align}
Using these, we find the interference contribution to the thermal current,
\begin{align}\label{eq:appendix_a-interference-contribution}
\langle I_{T} \rangle^{\text{int}} = 4i\Gamma_1\Gamma_2\pi^8T_1^4T_2^4 &\Bigg[ \int_{\pi}^{0}d\theta i\epsilon e^{i\theta} \left[ \frac{1}{\epsilon^5 e^{5i\theta}} - \frac{1}{3\epsilon^3 e^{3i\theta}} +\mathcal{O}(\epsilon)\right]\frac{1}{\sinh^4(\lambda\epsilon e^{i\theta}-2\lambda\chi)} \nonumber\\ 
&+\int_{\pi}^{0}d\theta i\epsilon e^{i\theta}  \frac{\cosh(\epsilon e^{i\theta}+2\chi)}{\sinh^5(\epsilon e^{i\theta}+2\chi)} \left[ \frac{1}{\epsilon^4\lambda^4 e^{4i\theta}} - \frac{2}{3\epsilon^2\lambda^2 e^{2i\theta}} + \mathcal{O}(\epsilon^0) \right] \nonumber \\
&+\int_{\pi}^{0}d\theta i\epsilon e^{i\theta}  \frac{\cosh(\epsilon e^{i\theta}-2\chi)}{\sinh^5(\epsilon e^{i\theta}-2\chi)} \left[ \frac{1}{\epsilon^4\lambda^4 e^{4i\theta}} - \frac{2}{3\epsilon^2\lambda^2 e^{2i\theta}} + \mathcal{O}(\epsilon^0) \right] \nonumber\\
& + \int_{\pi}^{0}d\theta i\epsilon e^{i\theta} \left[ \frac{1}{\epsilon^5 e^{5i\theta}} - \frac{1}{3\epsilon^3 e^{3i\theta}} +\mathcal{O}(\epsilon)\right]\frac{1}{\sinh^4(\lambda\epsilon e^{i\theta}+2\lambda\chi)} \Bigg].
\end{align}
Now we observe that all the terms with negative odd powers of $\epsilon$ give non-zero contributions, since we integrate from $\pi$ to $0$ (even powers of $\epsilon$ come with $e^{2in\theta}$ that gives zero on integration unless $n=0$). The term of order $\epsilon^0$ will also give a non-zero contribution. All positive powers of $\epsilon$ vanish since we take the $\epsilon\to 0$ limit at the end of the calculation. We need to make sure that the remaining $\epsilon$-dependent terms with odd negative powers
cancel out so that in the $\epsilon \to 0$ limit, the integral remains well behaved. Indeed that happens since in Eq.~(\ref{eq:appendix_a-interference-contribution}), we note that in the first and fourth integrals (and similarly the second and third integrals), the negative odd powers of $\epsilon$ terms come with opposite signs, which is how they cancel. Out of all these, the only remaining terms that give a non-zero contribution are $\epsilon^0$-order terms,
\begin{align}
\langle I_{T} \rangle^{\text{int}} = \Lambda &\Bigg[\int_{\pi}^{0}d\theta i  \frac{2\lambda^4}{3\sinh^4(2\lambda\chi)} \left[ \frac{105\cosh^4(2\lambda\chi)}{\sinh^4(2\lambda\chi)} - \frac{80\cosh^2(2\lambda\chi)}{\sinh^2(2\lambda\chi)} +7 \right] -\int_{\pi}^{0}d\theta i \frac{2}{3} \frac{\lambda^2 }{\sinh^4(2\lambda\chi)}\left[ \frac{10\cosh^2(2\lambda\chi)}{\sinh^2(2\lambda\chi)} -2\right] \nonumber \\
& +\int_{\pi}^{0}d\theta i \frac{1}{3} \left[ \frac{80\cosh^2(2\chi)}{\sinh^6(2\chi)} - \frac{105\cosh^4(2\chi)}{\sinh^8(2\chi)} - \frac{7}{\sinh^4(2\chi)} \right] \frac{2}{\lambda^4} - \int_{\pi}^{0}d\theta i \left[ \frac{1}{\sinh^4(2\chi)} - \frac{5\cosh^2(2\chi)}{\sinh^6(2\chi)} \right] \frac{4}{3\lambda^2 } \Bigg],
\end{align}
where we defined $\Lambda = 4i\Gamma_1\Gamma_2\pi^8T_1^4T_2^4 $ for convenience. This integral can now be computed to give,
\begin{align}
\langle I_{T} \rangle^{\text{int}} = 4\Gamma_1\Gamma_2\pi^9T_1^4T_2^4 &\Bigg[  \frac{2\lambda^4}{3\sinh^4(2\lambda\chi)} \left[ \frac{105\cosh^4(2\lambda\chi)}{\sinh^4(2\lambda\chi)} - \frac{80\cosh^2(2\lambda\chi)}{\sinh^2(2\lambda\chi)} +7 \right] + \frac{2\lambda^2}{3\sinh^4(2\lambda\chi)} \left[ 2- \frac{10\cosh^2(2\lambda\chi)}{\sinh^2(2\lambda\chi)} \right]  \nonumber\\
& - \frac{2}{3\lambda^4\sinh^4(2\chi)} \left[ \frac{105\cosh^4(2\chi)}{\sinh^4(2\chi)} - \frac{80\cosh^2(2\chi)}{\sinh^2(2\chi)}  + 7 \right] - \frac{2}{3\lambda^2\sinh^4(2\chi)} \left[ 2- \frac{10\cosh^2(2\chi)}{\sinh^2(2\chi)} \right] \Bigg] .
\end{align}
Plugging back the values of $\chi$ and $\lambda$ gives,
\begin{align}
\langle I_{T} \rangle^{\text{int}} = \frac{8}{3}\Gamma_1\Gamma_2\pi^9 &\Bigg[ \frac{T_2^8}{\sinh^4(2\pi T_2x_{21})} \left[ \frac{105\cosh^4(2\pi T_2x_{21})}{\sinh^4(2\pi T_2x_{21})} - \frac{80\cosh^2(2\pi T_2x_{21})}{\sinh^2(2\pi T_2x_{21})} +7 \right] + \frac{T_1^2T_2^6}{\sinh^4(2\pi T_2x_{21})} \left[ 2- \frac{10\cosh^2(2\pi T_2x_{21})}{\sinh^2(2\pi T_2x_{21})} \right]  \nonumber\\
 -& \frac{T_1^8}{\sinh^4(2\pi T_1x_{21})} \left[ \frac{105\cosh^4(2\pi T_1x_{21})}{\sinh^4(2\pi T_1x_{21})} - \frac{80\cosh^2(2\pi T_1x_{21})}{\sinh^2(2\pi T_1x_{21})}  + 7 \right] - \frac{T_1^6T_2^2}{\sinh^4(2\pi T_1x_{21})} \left[ 2- \frac{10\cosh^2(2\pi T_1x_{21})}{\sinh^2(2\pi T_1x_{21})} \right] \Bigg] .
\end{align}

\begin{figure}[!htb]
\centering
 \subfloat[\label{subfig:a}]{%
   \includegraphics[width=0.4\columnwidth]{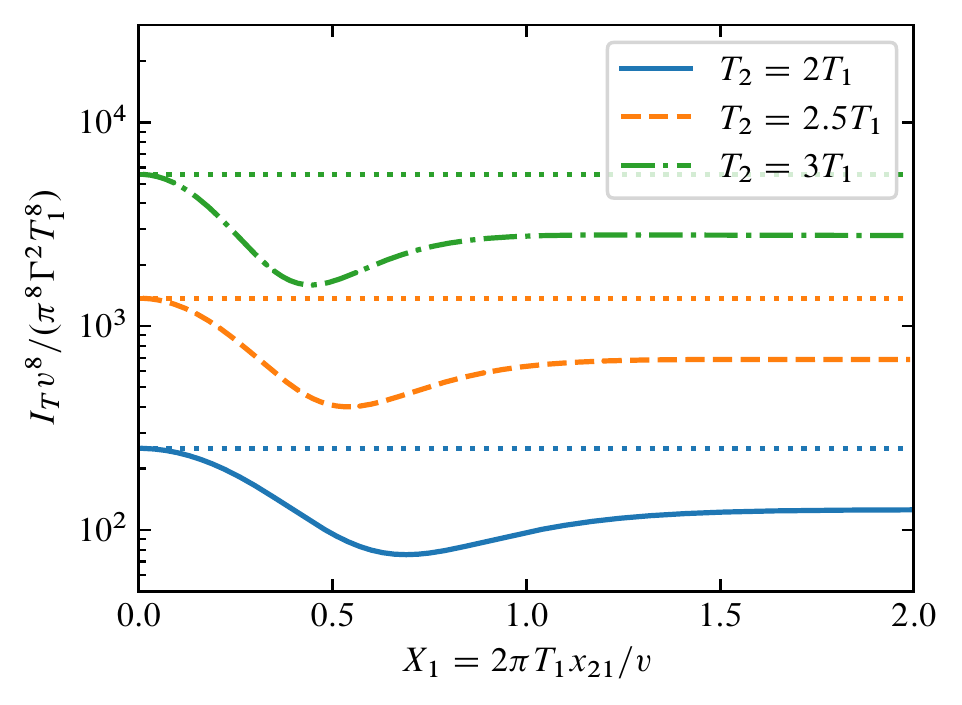}%
 }
 \subfloat[\label{subfig:b}]{%
   \includegraphics[width=0.4\columnwidth]{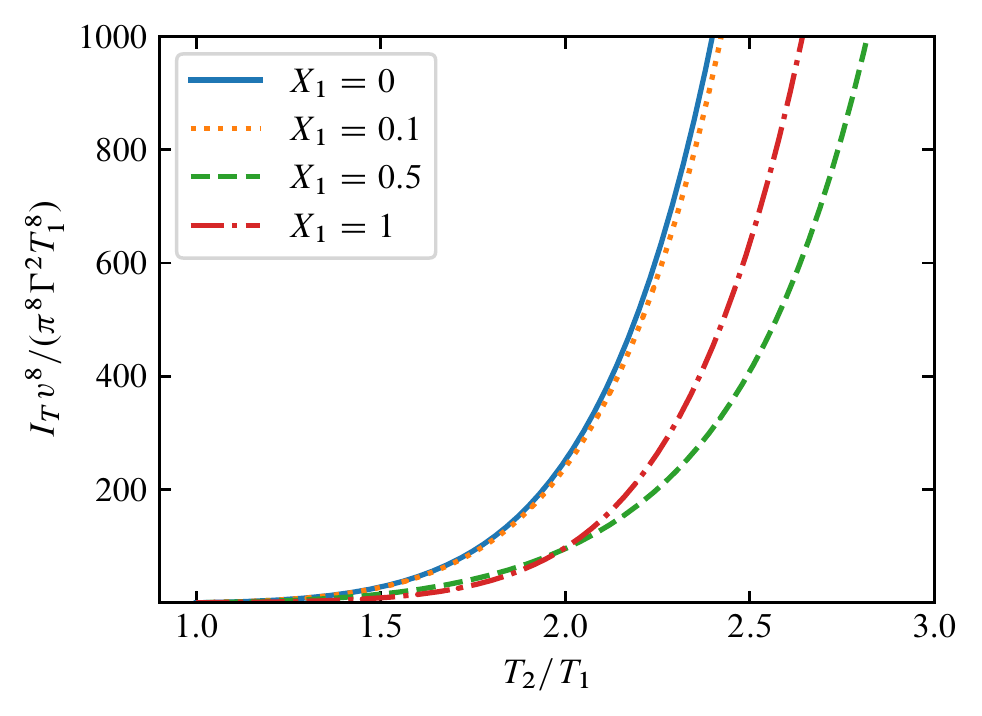}%
 }
\caption{(Color online) Thermal current between two spin liquids when $ \Gamma_1=\Gamma_2=\Gamma $. (a) Variation with $X_1$ at a fixed temperature ratio. The broken line shows the value of $2\langle I_T\rangle^{\text{non-int}}_{\Gamma}$. (b) Variation with $n=T_2/T_1$ shows the expected monotonic behavior.}
\end{figure}

Finally, combining the interference and non-interference contributions to the thermal current gives the total tunneling thermal current,
\begin{align}
\langle I_{T} \rangle & = \pi^9(\Gamma_1^2+\Gamma_2^2)\frac{4}{2835} [ 41(T_2^8 - T_1^8) + 62 T_1^2T_2^2(T_2^4-T_1^4) ]  + \frac{8}{3}\Gamma_1\Gamma_2\pi^9 \Bigg[ T_1^2T_2^6\csch^4(2\pi T_2x_{21}) \left[ 2- 10\coth^2(2\pi T_2x_{21}) \right] \nonumber \\
& - T_1^6T_2^2\csch^4(2\pi T_1x_{21}) \left[ 2- 10\coth^2(2\pi T_1x_{21}) \right] + T_2^8\csch^4(2\pi T_2x_{21}) \left[ 105\coth^4(2\pi T_2x_{21}) - 80\coth^2(2\pi T_2x_{21}) +7 \right]  \nonumber \\
& - T_1^8\csch^4(2\pi T_1x_{21}) \left[ 105\coth^4(2\pi T_1x_{21}) - 80\coth^2(2\pi T_1x_{21}) + 7 \right] \Bigg].
\end{align}
We now analyze this expression  by varying $X_1 \equiv 2\pi T_1 x_{21}/v$ at a fixed ratio $n\equiv T_2/T_1$ (Fig.~\subref*{subfig:a}), and then by varying the ratio of the temperatures at a fixed $X_1 \equiv 2\pi T_1 x_{21}/v$ (Fig.~\subref*{subfig:b}). If the two edges of the spin liquids are maintained at constant temperatures, the parameter $X_1$ is controlled by the distance between the two constrictions $x_{21}$. We also see from Fig.~\subref*{subfig:a} that as $X_1 \rightarrow 0$, at $\Gamma_1=\Gamma_2$, the tunneling thermal current equals twice the non-interference contribution $\langle I_T\rangle^{\text{non-int}}_{\Gamma}$ indicated by the broken line in Fig.~\subref*{subfig:a}.

\section{Decomposition of  correlation functions of Ising fields}\label{appendix:ising-correlation}
In this appendix, we demonstrate how the correlation functions of the fields defined on two edges can be decomposed into 
the product of two
correlation functions of the fields on the same edge. 
To achieve this, we assume that the two edges are connected by a long section of length $L$. We will take this length to be infinite to achieve the decomposition.

The tunneling operators create two excitations that fuse to vacuum on both sides of the constriction at $x_0$. Since Ising anyons are their own antiparticles, the form of the tunneling operator at the times $t$ and $t'$ is
\begin{subequations}
\begin{eqnarray}
    T_{12} &\equiv & \sigma_{2}(x_0,t)\sigma_1(x_0,t) \\
    T_{34} &\equiv & \sigma_{2}(x_0,t')\sigma_1(x_0,t'),
\end{eqnarray}
\end{subequations}
up to a constant factor ~$\sim\exp(-i\pi/16)$.
The subscript on the Ising field corresponds to the edge on which it is defined, and since the two edges are assumed to be connected, the two fields can be written in terms of each other as $\sigma_2(x_0) = \sigma_1(L-x_0)$. Notice however, this is well defined only in the $L \rightarrow \infty$ limit. We use the following coordinates,
\begin{subequations}
\begin{eqnarray}
    u_1 &=& t-x_0 \\
    u_2 &=& t-(-x_0+ L) \\
    u_3 &=& t' - x_0 \\
    u_4 &=& t' - (-x_0+ L)
\end{eqnarray}
\end{subequations}
and assume that the edge velocity $v=1$.
In this convention, the four-point functions can then be computed using the relation \cite{difrancesco1997:conformal}
\begin{align}
    \langle T_{12}T_{34}\rangle^2  = \frac{1}{2}\left[ \left(\frac{z_{13}z_{24}}{z_{12}z_{23}z_{34}z_{14}} \right)^{1/4} + \left( \frac{z_{14}z_{23}}{z_{13}z_{24}z_{12}z_{34}} \right)^{1/4} \right],
\end{align}
where $z_{ij} = {\sin[\pi T(\epsilon+iu_{ij})]}/{\pi T}$, and $u_{ij}\equiv u_i-u_j$. Notice that when we take the $ L\rightarrow \infty$ limit, the first term drops out and we arrive at the expression, 
\begin{align}
    \langle T_{12}T_{34}\rangle \sim  \frac{1}{\sqrt{2}}\frac{e^{i\pi/8}}{(z_{13}z_{24})^{1/8}} \left( \frac{\sinh(\pi T u_{14})\sinh(\pi T u_{32})}{\sinh (\pi T u_{12})\sinh(\pi T u_{34})} \right)^{1/8} \xrightarrow{ L \rightarrow \infty} \frac{e^{i\pi/8}}{\sqrt{2}(z_{13}z_{24})^{1/8}}
\end{align}
We now notice that the final expression, after taking the limit, can be written as a product of two-point functions defined on one of the edges, Eq.~(\ref{eq:ising-correlations}). Therefore,
\begin{align}
    \langle \sigma_{2}(x_0,t)\sigma_1(x_0,t) \sigma_{2}(x_0,t')\sigma_1(x_0,t') \rangle = \frac{e^{i\pi/8}}{\sqrt{2}} \langle \sigma_{2}(x_0,t)\sigma_{2}(x_0,t') \rangle \langle \sigma_1(x_0,t) \sigma_1(x_0,t')\rangle.
\end{align}
The choice of the branch in the above formula is dictated by the hermiticity of the tunneling operators $\sim\exp(-i\pi/16)\sigma_2\sigma_1$.
Indeed, the average of the square of a tunneling operator must be real and positive.

\section{$ \sigma $ tunneling}\label{appendix:sigma}

In this appendix, we compute the thermal tunneling current due to Ising anyon tunneling at two point contacts at $x_1$ and $x_2$ in the Fabry-P\'erot geometry (Fig.~\ref{fig1:thermal} (a)). The single constriction calculation is the non-interference part $\langle I_T\rangle^{\text{non-int}}_{\Gamma_j}$  of the full calculation presented in this appendix
with contributions from only one of the tunneling amplitudes $\Gamma_j$, $j=1,2$. 
We assume identical distances between the tunneling contacts along the two edges and identical edge velocities on the two edges. Since the result depends only on the sum of the edge lengths (Appendix~\ref{appendix:inter_size}), the first assumption can be easily removed.

In the calculation, we assume that the interferometer confines a trivial topological charge. We generalize to an arbitrary confined charge in the main text. 

The full tunneling Hamiltonian is given by,
\begin{align}
H_T =e^{-i\pi/16}\Gamma_1 \sigma_2(x_1,t)\sigma_1(x_1,t) + e^{-i\pi/16} \Gamma_2 \sigma_2(x_2,t)\sigma_1(x_2,t).
\end{align}
We define the current operator using the Heisenberg equation of motion,
\begin{align}
I_T^{(1)} = \frac{\partial H_1}{\partial t} = -i[H_1, H] = -i [H_1,H_T], 
\end{align}
where $H_1$ and $H_2$ correspond to the free field Hamiltonian (\ref{eq:free_field_ham}) defined on the left- and right-moving edges respectively. 
From here, we identify the current operator as,
\begin{align}
I^{(1)}_T = -e^{-i\pi/16}\Gamma_1 \sigma_2(x_1,t)\partial_t\sigma_1(x_1,t) - e^{-i\pi/16} \Gamma_2 \sigma_2(x_2,t)\partial_t\sigma_1(x_2,t),
\end{align}
where all Heisenberg operators are defined in terms of the edge Hamiltonian without tunneling.
Using perturbation theory, we obtain the expectation value of the thermal current due to the tunneling, to the lowest non-zero order,
\begin{align}
\langle I_T^{(1)}(t)\rangle = -i \int_{-\infty}^{t} dt' \langle [ I^{(1)}_T(t), H_T(t') ] \rangle.
\end{align}
Inserting the operators into the above expression gives,
\begin{align}
\langle I_T^{(1)}(t)\rangle = ie^{-i\pi/8} & \int_{-\infty}^{t} dt' \Big[\Gamma_1^2 \langle \big[ \sigma_2(x_1,t)\partial_t\sigma_1(x_1,t) , \sigma_2(x_1,t')\sigma_1(x_1,t') \big] \rangle + \Gamma^2_2 \langle \big[ \sigma_2(x_2,t)\partial_t\sigma_1(x_2,t) , \sigma_2(x_2,t')\sigma_1(x_2,t')\big] \rangle  \nonumber \\
+& \Gamma_1\Gamma_2  \langle \big[ \sigma_2(x_1,t)\partial_t\sigma_1(x_1,t), \sigma_2(x_2,t')\sigma_1(x_2,t') \big] \rangle  + \Gamma_1\Gamma_2 \langle \big[ \sigma_2(x_2,t)\partial_t\sigma_1(x_2,t), \sigma_2(x_1,t')\sigma_1(x_1,t') \big]\rangle\Big] .
\end{align}
From here we see that the thermal current is composed of non-interference and interference contributions, $\langle I_T \rangle = \sum_{j=1,2}\langle I_{T}\rangle^{\text{non-int}}_{\Gamma_j} + \langle I_{T} \rangle^{\text{int}}$. The thermal correlation functions are
\begin{align}
\langle \sigma_1(x_1,t_1)\sigma_1(x_2,t_2) \rangle =  \frac{(\pi T_1)^{1/8}}{\sin^{1/8}[\pi T_1 (\epsilon + i (t_1-t_2+x_2-x_1))]} \\
\langle \sigma_2(x_1,t_1)\sigma_2(x_2,t_2) \rangle =  \frac{(\pi T_2)^{1/8}}{\sin^{1/8}[\pi T_2 (\epsilon + i (t_1-t_2+x_1-x_2))]}
\end{align}
and, of course, the cross-correlations go to zero in the $L\rightarrow \infty$ limit (Appendix~\ref{appendix:ising-correlation}). 
To simplify notations, we set $v=1$.
In the following two subsections, we separately focus on the non-interference and interference contributions. 

\subsection{Non-interference term}
Using the above correlation functions and the results of Appendix~\ref{appendix:ising-correlation} on the four-point correlation functions, we first compute the expression for the non-interference contribution $(j=1,2)$.
\begin{align}
\langle I\rangle^{\text{non-int}}_{\Gamma_{j}} = \frac{\Gamma_j^2}{8\sqrt{2}}  \int_{-\infty}^{t} dt' \Bigg[ \frac{(\pi T_2)^{\frac{1}{8}}(\pi T_1)^{\frac{9}{8}}\cos[\pi T_1 (\epsilon + i (t-t'))]}{\sin^{\frac{9}{8}}[\pi T_1 (\epsilon + i (t-t'))] \sin^{\frac{1}{8}}[\pi T_2 (\epsilon + i (t-t'))]} +\frac{(\pi T_2)^{\frac{1}{8}}(\pi T_1)^{\frac{9}{8}} \cos[\pi T_1 (\epsilon + i (t'-t))]}{\sin^{\frac{9}{8}}[\pi T_1 (\epsilon + i (t'-t))] \sin^{\frac{1}{8}}[\pi T_2 (\epsilon + i (t'-t))]} \Bigg] .
\end{align}
Defining $\pi T_1(t-t')\equiv \tau$ and $n = T_2/T_1$, we can simplify the above integral,
\begin{align}
\langle I_T\rangle^{\text{non-int}}_{\Gamma_j} &= -\frac{\Gamma_j^2}{8\sqrt{2}} (\pi T_1)^{\frac{1}{8}} (\pi T_2)^{\frac{1}{8}}  \int_{\infty}^{0} d\tau \Bigg[ \frac{\cos(i\tau +\epsilon)}{\sin^{\frac{9}{8}}(i\tau +\epsilon) \sin^{\frac{1}{8}}[n(i\tau + \epsilon)]} +\frac{\cos(-i\tau +\epsilon)}{\sin^{\frac{9}{8}}(-i\tau +\epsilon) \sin^{\frac{1}{8}}[n(-i\tau + \epsilon)]} \Bigg]  
\nonumber\\
&=  \frac{\Gamma_j^2}{8\sqrt{2}} (\pi T_1)^{\frac{1}{8}} (\pi T_2)^{\frac{1}{8}}   \int_{-\infty}^{\infty} d\tau ~\frac{\cos(-i\tau +\epsilon)}{\sin^{\frac{9}{8}}(-i\tau +\epsilon) \sin^{\frac{1}{8}}[n(-i\tau + \epsilon)]}.
\end{align}
To evaluate the integral, we deform the contour as shown in Fig.~\subref*{subfig:contour_a},
\begin{align}
\langle I_T\rangle^{\text{non-int}}_{\Gamma_j} =  \frac{\Gamma_j^2}{8\sqrt{2}} (\pi T_1)^{\frac{1}{8}} (\pi T_2)^{\frac{1}{8}}  \Bigg[ \int_{-\infty}^{-\epsilon} \frac{d\tau~\cos(-i\tau)}{\sin^{\frac{9}{8}}(-i\tau) \sin^{\frac{1}{8}}(-in\tau)} + \int_{C} \frac{d\tau~\cos(-i\tau)}{\sin^{\frac{9}{8}}(-i\tau) \sin^{\frac{1}{8}}(-in\tau)}  + \int^{\infty}_{\epsilon}  \frac{d\tau~\cos(-i\tau)}{\sin^{\frac{9}{8}}(-i\tau) \sin^{\frac{1}{8}}(-in\tau)}  \Bigg].
\end{align}
Near the origin, the integration variable can be written as $\tau = \epsilon e^{i\theta}$. Carefully taking the $\epsilon \rightarrow 0$ limit and writing the integrands in terms of hyperbolic functions, we get a phase factor depending upon the location on the contour,
\begin{align}
\langle I_T\rangle^{\text{non-int}}_{\Gamma_j} &=  \frac{\Gamma_j^2}{8\sqrt{2}} (\pi T_1)^{\frac{1}{8}} (\pi T_2)^{\frac{1}{8}}  \Bigg[ \int^{\infty}_{\epsilon} \frac{d\tau~e^{-i\frac{5\pi}{8}}\cosh\tau}{\sinh^{\frac{9}{8}}\tau \sinh^{\frac{1}{8}}(n\tau)} + \int_{C} \frac{d\tau~\cos(-i\tau)}{\sin^{\frac{9}{8}}(-i\tau) \sin^{\frac{1}{8}}(-in\tau)}  + \int^{\infty}_{\epsilon}  \frac{d\tau~e^{i\frac{5\pi}{8}}\cosh\tau}{\sinh^{\frac{9}{8}}\tau \sinh^{\frac{1}{8}}(n\tau)}  \Bigg]\\
\label{D12}
\langle I_T\rangle^{\text{non-int}}_{\Gamma_j} &=  \frac{\Gamma_j^2}{8\sqrt{2}} (\pi T_1)^{\frac{1}{8}} (\pi T_2)^{\frac{1}{8}}  \Bigg[ \int^{\infty}_{\epsilon} \frac{d\tau~2\cos(5\pi/8)\cosh\tau}{\sinh^{\frac{9}{8}}\tau \sinh^{\frac{1}{8}}(n\tau)} + \int_{C} \frac{d\tau~\cos(-i\tau)}{\sin^{\frac{9}{8}}(-i\tau) \sin^{\frac{1}{8}}(-in\tau)}  \Bigg].
\end{align}
To evaluate the integral over the contour $C$, we note that $\tau \sim \epsilon$, so we expand the integrand as,
\begin{align}
\frac{\cos(-i\tau)}{\sin^{\frac{9}{8}}(-i\tau)\sin^{\frac{1}{8}}(-in\tau)} \Bigg|_{\tau = 0} = \frac{1}{n^{\frac{1}{8}}(-i\tau)^{\frac{5}{4}}}  + \mathcal{O}(\tau^{\frac{3}{4}}).
\end{align}
Since $d\tau = i\tau d\theta$, in the $\epsilon\rightarrow 0$ limit, higher order terms vanish. The only remaining term is
\begin{align}
\int_{C} \frac{d\tau~\cos(-i\tau)}{\sin^{\frac{9}{8}}(-i\tau) \sin^{\frac{1}{8}}(-in\tau)} = \int_{C} d\tau \frac{1}{n^{\frac{1}{8}}(-i\tau)^{\frac{5}{4}}} = e^{i\frac{5\pi}{8}}\int_{\pi}^{0} d\theta ~ i\epsilon e^{i\theta} \frac{1}{n^{\frac{1}{8}}\epsilon^{\frac{5}{4}}e^{i\frac{5\theta}{4}}} = -\frac{4}{n^{\frac{1}{8}}\epsilon^{\frac{1}{4}}}2\cos(5\pi/8).
\end{align}
We note that the divergence of this term cancels the divergence of the first integral
in Eq.~(\ref{D12}). We thus write it in a similar form,
\begin{align}
\frac{4}{\epsilon^{\frac{1}{4}}} = \int_{\epsilon}^{\infty}d\tau \frac{1}{\tau^{\frac{5}{4}}}.
\end{align}
Therefore the expression for the non-interference term is
\begin{align}
\sum_{j=1,2}\langle I_T\rangle^{\text{non-int}}_{\Gamma_j} &=  \frac{(\Gamma_1^2+\Gamma_2^2)}{4\sqrt{2}} (\pi T_1)^{\frac{1}{8}} (\pi T_2)^{\frac{1}{8}} \cos(5\pi/8) \int^{\infty}_{0} d\tau \Bigg[ \frac{\cosh(\tau)}{\sinh^{\frac{9}{8}}(\tau) \sinh^{\frac{1}{8}}(n\tau)} - \frac{1}{n^{\frac{1}{8}}\tau^{\frac{5}{4}}} \Bigg] \nonumber\\
&=  \frac{(\Gamma_1^2+\Gamma_2^2)}{4\sqrt{2}} (\pi T_1)^{\frac{1}{8}} (\pi T_2)^{\frac{1}{8}} \cos(3\pi/8) ~F_{\text{non-int}}(n),
\end{align}
where we defined $F_{\text{non-int}}(n)$ as in Eq.~(\ref{eq:func-f_non-int}). Note that the non-interference contribution, individually for each of  $\Gamma_j$, $j=1,2$, is exactly what we obtain when we consider single-constriction geometry. Results discussed in the
main text in Sec.~\ref{sec:Ising-singleQPC} corresponds to $\langle I_T \rangle^{\text{non-int}}_{\Gamma_j}$ where $\Gamma_j$ is set to $\Gamma$ of the single constriction.

\subsection{Interference term}
We now look at the interference contribution. 
\begin{align}
\langle I_T\rangle^{\text{int}} = \frac{\Gamma_1\Gamma_2}{8\sqrt{2}} (\pi T_1)^{\frac{9}{8}}(\pi T_2)^{\frac{1}{8}} &\int_{-\infty}^{t} d t'\Bigg[ \frac{\cos[\pi T_1(\epsilon +i(t-t'+x_{21}))]}{\sin^{\frac{9}{8}}[\pi T_1(\epsilon +i(t-t'+x_{21}))]\sin^{\frac{1}{8}}[\pi T_2(\epsilon +i(t-t'-x_{21}))]}  \nonumber \\
& + \frac{\cos[\pi T_1(\epsilon -i(t-t'+x_{21}))]}{\sin^{\frac{9}{8}}[\pi T_1(\epsilon -i(t-t'+x_{21}))]\sin^{\frac{1}{8}}[\pi T_2(\epsilon -i(t-t'-x_{21}))]} +(x_{21}\leftrightarrow - x_{21}) \Bigg]
\end{align}
Defining $\tau\equiv\pi T_1(t-t')$, $\chi\equiv\pi T_1 x_{21}$, and $n = T_2/T_1$, we may simplify the above integral,
\begin{align}\label{eq:FP_Ising_inter_integral}
\langle I_T\rangle^{\text{int}} = \frac{\Gamma_1\Gamma_2}{8\sqrt{2}}(\pi T_1)^{\frac{1}{8}}(\pi T_2)^{\frac{1}{8}} \int_{-\infty}^{\infty} d\tau \bigg[ \frac{\cos[\epsilon-i(\tau-\chi)]}{\sin^{\frac{9}{8}}[\epsilon-i(\tau-\chi)]\sin^{\frac{1}{8}}[n(\epsilon-i(\tau+\chi))]} + \frac{\cos[\epsilon-i(\tau+\chi)]}{\sin^{\frac{9}{8}}[\epsilon-i(\tau+\chi)]\sin^{\frac{1}{8}}[n(\epsilon-i(\tau-\chi))]} \Bigg]
\end{align}
Like before, we can now deform the contour as shown in Fig.~\subref*{subfig:contour_b},
\begin{align}
\label{D19}
\langle I_T\rangle^{\text{int}} = \frac{\Gamma_1\Gamma_2}{8\sqrt{2}}(\pi T_1)^{\frac{1}{8}}(\pi T_2)^{\frac{1}{8}} \Bigg[ \int_{-\infty}^{-\chi-\epsilon} d&\tau ~I + \int_{C_1} d\tau ~I + \int_{-\chi+\epsilon}^{\chi-\epsilon} d\tau ~I + \int_{C_2} d\tau ~I  + \int_{\chi+\epsilon}^{\infty} d\tau ~I  ~~+ ~~(\chi \leftrightarrow -\chi) \Bigg],
\end{align}
where the integrand $I$ is given as
\begin{equation}
    I = \frac{\cos[-i(\tau-\chi)]}{\sin^{\frac{9}{8}}[-i(\tau-\chi)]\sin^{\frac{1}{8}}[-in(\tau+\chi)]}.
\end{equation}
The integration variable near $-\chi$ is given by $\eta=\epsilon e^{i\theta}-\chi$, the integration variable near $\chi$ is given by $\eta=\epsilon e^{i\theta}+\chi$. We now look at each contribution to (\ref{D19}) individually. In particular,
\begin{align}
\label{D21}
\int_{-\infty}^{-\chi-\epsilon} d\tau  \frac{\cos[-i(\tau-\chi)]}{\sin^{\frac{9}{8}}[-i(\tau-\chi)]\sin^{\frac{1}{8}}[-in(\tau+\chi)]} =& \int_{\chi+\epsilon}^{\infty} d\tau \frac{e^{-i\frac{5\pi}{8}}\cosh(\tau+\chi)}{\sinh^{\frac{9}{8}}(\tau+\chi)\sinh^{\frac{1}{8}}[n(\tau-\chi)]} \\
\label{D22}
\int_{-\infty}^{-\chi-\epsilon} d\tau  \frac{\cos[-i(\tau+\chi)]}{\sin^{\frac{9}{8}}[-i(\tau+\chi)]\sin^{\frac{1}{8}}[-in(\tau-\chi)]} = &\int_{\chi+\epsilon}^{\infty} d\tau \frac{e^{-i\frac{5\pi}{8}}\cosh(\tau-\chi)}{\sinh^{\frac{9}{8}}(\tau-\chi)\sinh^{\frac{1}{8}}[n(\tau+\chi)]}
\end{align}
Similarly, 
\begin{align}
\label{D23}
\int^{\infty}_{\chi+\epsilon} d\tau  \frac{\cos[-i(\tau-\chi)]}{\sin^{\frac{9}{8}}[-i(\tau-\chi)]\sin^{\frac{1}{8}}[-in(\tau+\chi)]} =& \int_{\chi+\epsilon}^{\infty} d\tau \frac{e^{i\frac{5\pi}{8}}\cosh(\tau-\chi)}{\sinh^{\frac{9}{8}}(\tau-\chi)\sinh^{\frac{1}{8}}[n(\tau+\chi)]} \\
\label{D24}
\int^{\infty}_{\chi+\epsilon} d\tau  \frac{\cos[-i(\tau+\chi)]}{\sin^{\frac{9}{8}}[-i(\tau+\chi)]\sin^{\frac{1}{8}}[-in(\tau-\chi)]} = &\int_{\chi+\epsilon}^{\infty} d\tau \frac{e^{i\frac{5\pi}{8}}\cosh(\tau+\chi)}{\sinh^{\frac{9}{8}}(\tau+\chi)\sinh^{\frac{1}{8}}[n(\tau-\chi)]}
\end{align}
Combining Eqs.~(\ref{D21}-\ref{D24}) together gives,
\begin{align}
      \mathbb{I}_1 \equiv \int_{-\infty}^{-\chi-\epsilon} d\tau \left[I + (\chi \leftrightarrow-\chi)\right] + \int_{\chi+\epsilon}^{\infty} d\tau \left[I + (\chi \leftrightarrow-\chi)\right] =  2\cos(5\pi/8) \int_{\chi+\epsilon}^{\infty} d\tau \left[ \frac{\cosh(\tau-\chi)}{\sinh^{\frac{9}{8}}(\tau-\chi)\sinh^{\frac{1}{8}}[n(\tau+\chi)]} + (\chi\leftrightarrow -\chi) \right]
\end{align}
In the final expression, we make a variable shift $\tau \rightarrow \tau-\chi$, giving,
\begin{align}\label{eq:appendixD_int1}
    \mathbb{I}_1 =  2\cos(5\pi/8) \int_{\epsilon}^{\infty} d\tau \left[ \frac{\cosh(\tau)}{\sinh^{\frac{9}{8}}(\tau)\sinh^{\frac{1}{8}}[n(\tau+2\chi)]} + \frac{\cosh(\tau+2\chi)}{\sinh^{\frac{9}{8}}(\tau+2\chi)\sinh^{\frac{1}{8}}(n\tau)} \right]
\end{align}
We now look at the integrals corresponding to path $C_1$ in Eq.~(\ref{D19}), 
\begin{align}
\label{D27}
\int_{C_1} d\tau  \frac{\cos[-i(\tau-\chi)]}{\sin^{\frac{9}{8}}[-i(\tau-\chi)]\sin^{\frac{1}{8}}[-in(\tau+\chi)]} &\sim \int_{\pi}^{0} d\theta i\epsilon e^{i\theta} \frac{e^{-i\frac{5\pi}{8}}\cosh(2\chi)}{\sinh^{\frac{9}{8}}(2\chi)(n\epsilon e^{i\theta})^{\frac{1}{8}}} \xrightarrow{\epsilon \rightarrow 0} 0 \\
\label{D28}
\int_{C_1} d\tau  \frac{\cos[-i(\tau+\chi)]}{\sin^{\frac{9}{8}}[-i(\tau+\chi)]\sin^{\frac{1}{8}}[-in(\tau-\chi)]} & =\int_{\pi}^{0} d\theta~i\epsilon e^{i\theta} \frac{\cos(-i\epsilon e^{i\theta})}{\sin^{\frac{9}{8}}(-i\epsilon e^{i\theta})\sin^{\frac{1}{8}}[in(2\chi-\epsilon e^{i\theta})]} = \int_{\pi}^{0} d\theta~i\epsilon e^{i\theta} \frac{1}{(-i\epsilon e^{i\theta})^{\frac{9}{8}}\sin^{\frac{1}{8}}(in2\chi)} \nonumber\\
& = \frac{e^{i\frac{\pi}{2}}}{\sinh^{\frac{1}{8}}(2n\chi)}  \int_{\pi}^{0} d\theta ~i\epsilon e^{i\theta} \frac{1}{\epsilon^{\frac{9}{8}}e^{i\frac{9\theta}{8}}} = \frac{-e^{i\frac{\pi}{2}}(1-e^{-i\frac{\pi}{8}})}{\sinh^{\frac{1}{8}}(2n\chi)}\int_{\epsilon}^{\infty}d\tau \frac{1}{\tau^{9/8}}
\end{align}
Similarly, the integrals corresponding to path $C_2$,
\begin{align}
\label{D29}
\int_{C_2} d\tau  \frac{\cos[-i(\tau-\chi)]}{\sin^{\frac{9}{8}}[-i(\tau-\chi)]\sin^{\frac{1}{8}}[-in(\tau+\chi)]}  &= \int_{\pi}^{0} d\theta~i\epsilon e^{i\theta} \frac{\cos(-i\epsilon e^{i\theta})}{\sin^{\frac{9}{8}}(-i\epsilon e^{i\theta})\sin^{\frac{1}{8}}[-in(2\chi + \epsilon e^{i\theta})]} = \int_{\pi}^{0} d\theta~i\epsilon e^{i\theta} \frac{1}{(-i\epsilon e^{i\theta})^{\frac{9}{8}}\sin^{\frac{1}{8}}(-in2\chi)} \nonumber\\
& = \frac{e^{i\frac{5\pi}{8}}}{\sinh^{\frac{1}{8}}(2n\chi)}  \int_{\pi}^{0} d\theta ~i\epsilon e^{i\theta} \frac{1}{\epsilon^{\frac{9}{8}}e^{i\frac{9\theta}{8}}} = \frac{-e^{i\frac{5\pi}{8}}(1-e^{-i\frac{\pi}{8}})}{\sinh^{\frac{1}{8}}(2n\chi)}\int_{\epsilon}^{\infty}d\tau \frac{1}{\tau^{9/8}} \\
\label{D30}
\int_{C_2} d\tau  \frac{\cos[-i(\tau+\chi)]}{\sin^{\frac{9}{8}}[-i(\tau+\chi)]\sin^{\frac{1}{8}}[-in(\tau-\chi)]} &\sim  \int_{0}^{\pi} d\theta i\epsilon e^{i\theta} \frac{e^{i\frac{5\pi}{8}}\cosh(2\chi)}{\sinh^{\frac{9}{8}}(2\chi)(n\epsilon e^{i\theta})^{\frac{1}{8}}} \xrightarrow{\epsilon \rightarrow 0} 0 
\end{align}
Combining the non-zero contributions from Eqs.~(\ref{D27}-\ref{D30}),
\begin{align}\label{eq:appendixD_int2}
\mathbb{I}_2 \equiv 
\int_{C_1+C_2}  d\tau [I + (\chi \leftrightarrow -\chi)] =  \frac{-(e^{i\frac{\pi}{2}}-e^{i\frac{3\pi}{8}}+e^{i\frac{5\pi}{8}}-e^{i\frac{\pi}{2}})}{\sinh^{\frac{1}{8}}(2n\chi)}\int_{\epsilon}^{\infty}d\tau \frac{1}{\tau^{9/8}} = \frac{-2\cos(5\pi/8)}{\sinh^{\frac{1}{8}}(2n\chi)}\int_{\epsilon}^{\infty}d\tau \frac{1}{\tau^{9/8}}
\end{align}
Finally, the integrals over the interval $[-\chi+\epsilon,\chi-\epsilon]$ in Eq.~(\ref{D19}) are
\begin{align}
    \int_{-\chi+\epsilon}^{\chi-\epsilon} d\tau  \frac{\cos[-i(\tau-\chi)]}{\sin^{\frac{9}{8}}[-i(\tau-\chi)]\sin^{\frac{1}{8}}[-in(\tau+\chi)]} 
    &= e^{-i\pi/2} \int_{-\chi+\epsilon}^{\chi-\epsilon} d\tau
    \frac{\cosh(\chi-\tau)}{\sinh^{\frac{9}{8}}(\chi-\tau) \sinh^{\frac{1}{8}} [ n(\chi+\tau) ] } \\
    \int_{-\chi+\epsilon}^{\chi-\epsilon} d\tau  \frac{\cos[-i(\tau+\chi)]}{\sin^{\frac{9}{8}}[-i(\tau+\chi)]\sin^{\frac{1}{8}}[-in(\tau-\chi)]} &= e^{i\pi/2} \int_{-\chi+\epsilon}^{\chi-\epsilon} d\tau \frac{\cosh(\chi+\tau)}{\sinh^{\frac{9}{8}}(\chi+\tau)\sinh^{\frac{1}{8}}[n(\chi-\tau)]} 
\end{align}
Clearly the sum of these two integrals vanishes. Now we combine all non-zero contributions, specifically Eqs.~(\ref{eq:appendixD_int1}) and (\ref{eq:appendixD_int2}). %
Notice that the divergences are canceled out once these integrals are combined to give the interference contribution to the thermal current as,
\begin{align}
\langle I_T\rangle^{\text{int}} &= \frac{\Gamma_1\Gamma_2}{4\sqrt{2}}(\pi T_1)^{\frac{1}{8}}(\pi T_2)^{\frac{1}{8}} \cos(5\pi/8) \int_{0}^{\infty}d\tau \Bigg[\frac{\cosh(\tau+2\chi)}{\sinh^{\frac{9}{8}}(\tau+2\chi)\sinh^{\frac{1}{8}}(n\tau)} + \frac{\cosh(\tau)}{\sinh^{\frac{9}{8}}(\tau)\sinh^{\frac{1}{8}}(n\tau+2n\chi)} -\frac{1}{\tau^{\frac{9}{8}}\sinh^{\frac{1}{8}}(2n\chi)} \Bigg] \nonumber \\
&= \frac{\Gamma_1\Gamma_2}{4\sqrt{2}}(\pi T_1)^{\frac{1}{8}}(\pi T_2)^{\frac{1}{8}} \cos(3\pi/8) ~F_{\text{int}}(n,\chi)
\end{align}
where we defined the function $F_{\text{non-int}}(n,\chi)$ as in Eq.~(\ref{eq:func_F_int}). We thus arrive at the expression for the total thermal current due to anyon tunneling,
\begin{align}\label{eq:FP_Ising_full}
\langle I_T\rangle = \frac{(\pi T_1)^{\frac{1}{8}}(\pi T_2)^{\frac{1}{8}} \cos(3\pi/8) }{4\sqrt{2}} 
\Bigg[& (\Gamma_1^2+\Gamma_2^2) ~F_{\text{non-int}}(n) + \Gamma_1\Gamma_2 ~F_{\text{int}}(n,\chi) \Bigg]
\end{align}
The edge velocity was set to 1 above. Therefore the answer should be divided by $v^{1/4}$. 

\section{Tunneling in Mach-Zehnder geometry}\label{appendix:Mach-zehnder_phase}

The purpose of this appendix is to find the additional phase $\alpha$ in the Mach-Zehnder tunneling Hamiltonian, Eq.~(\ref{eq:mach-zehnder_tunneling-ham}). We will do this with three different methods, all of which produce the same result. The first approach involves the calculation of a partition function in conformal field theory (CFT) and will only be used for the Ising statistics. The second and third approaches will be applied to a general case. The second method builds on the algebraic theory of anyons. The third method uses detailed balance.

In our first approach we also use the principle of detailed balance, assuming that the temperatures of the edges are equal. For a stationary distribution, the transition probabilities $p_{\sigma a}^b$ satisfy the detailed balance equations, which state that the probabilities associated to the process $\sigma \times a \rightarrow b$ and the reverse process $\sigma \times b \rightarrow a$ are related. The two probabilities are given as,
\begin{subequations}
\begin{align}
\label{AE1}
    p_{\sigma a}^{b} &= \frac{2\pi}{\hbar} \sum_{nm} |\langle m| H_{T} |n \rangle |^2 \delta(E_{m}-E_{n})P_{n}^{a}(T,T) \\
    \label{AE2}
     p_{\sigma b}^{a} &= \frac{2\pi}{\hbar} \sum_{nm} |\langle {n} |H_{T}|m \rangle|^2 \delta(E_{n}-E_{m})P_{m}^{b}(T,T).
\end{align}
\end{subequations}
Notice that these two probabilities differ only by the partition functions in the Gibbs factors $P_{n}^a = \exp({-E_n/T})/Z_{a}(T)$, where the partition functions are calculated in the superselection sectors defined by the topological charge $a$. Now the ratio of the two probabilities in the above equations equals the ratio of the partition functions, which can be computed using conformal field theory. 

We first solve for the energy spectrum of fermions on each of the two edges.
This calls for the choice of the boundary conditions, periodic (Ramond), or anti-periodic (Neveu-Schwarz). The two are related by a twist field operator $\sigma$ of conformal dimension $h_{\sigma} = 1/16$, which is precisely the Ising field we worked with in Sec.~\ref{sec:Ising-singleQPC}. This then leads to the following spectra in the two sectors~\cite{difrancesco1997:conformal}:
\begin{align}
H_{\text{R.}} = \frac{2\pi}{L} \left[ \sum_{k>0} k c_{-k}c_{k} + \frac{1}{24} \right]  ~~~~~~~~ &k\in\mathbb{Z}~~(\text{Ramond}) \\
H_{\text{N.S.}} = \frac{2\pi}{L} \left[ \sum_{k>0} k c_{-k}c_{k} - \frac{1}{48} \right]  ~~~~~~~~ &k\in(\mathbb{Z}+1/2)~~(\text{Neveu-Schwarz})
\end{align}
The energy spectra in the two sectors can now be used to compute the partition functions of the system in the fermionic, $\psi$, and anyonic, $\sigma$, topological sectors. Notice that the above equations correspond to only one of the edges. To compute the partition function of the full system we also take into account the other edge independently of the first. We now calculate the partition function of one of the edges for the $\psi$ sector that corresponds to the Ramond boundary case, 
\begin{align}
Z_{\psi} = \text{Tr}\left( e^{-\beta H_{\text{R.}}}\right) &= \text{Tr} \left( \prod_{k\in \mathbb{Z}^+}\exp({-2\pi\beta k \hat{n}_{k}/L}) \exp({-2\pi\beta/24L})  \right) \nonumber\\
& = \prod_{k\in\mathbb{Z}^+} \exp({-2\pi\beta/24L}) \Big( 1+ \exp({-2\pi\beta k/L}) \Big).
\end{align}
For the $\sigma$ sector that corresponds to the Neveu-Schwarz boundary case, we get a similar expression except that the sum runs over $(\mathbb{Z}+1/2)$,
\begin{align}
Z_{\sigma} = \text{Tr}\left( e^{-\beta H_{\text{N.S.}}}\right) &= \text{Tr} \left( \prod_{k\in \mathbb{Z}^+\cup \{0\}}\exp({-2\pi\beta (k+1/2) \hat{n}_{k+\frac{1}{2}}/L}) \exp({2\pi\beta/48L})  \right) \nonumber\\
& = \prod_{k\in\mathbb{Z}^+\cup\{0\}} \exp({2\pi\beta/48L}) \Big( 1+ \exp({-2\pi\beta k/L})\exp({-\pi\beta/L}) \Big) 
\end{align}
The overall exponential factor in both partition functions  is irrelevant since in the thermodynamic limit $L\rightarrow \infty$, it gives $1$. We now compute the ratio of the remaining product terms,
\begin{align}
P &\equiv \frac{\prod_{k\in \mathbb{Z}^+}\left( 1+ \exp(-2\pi\beta k/L)  \right)}{\prod_{k\in \mathbb{Z}^+\cup\{0\}}\left( 1+ \exp{(-2\pi\beta k/L)}\exp(-\pi\beta/L) \right)} \\
\log{P} &= - \log(2) + \sum_{k=0}^{\infty} \left[ \log{(1+\exp(-2\pi\beta k/L))}-\log(1+\exp(-2\pi\beta k/L)\exp(-\pi\beta/L)) \right] 
\end{align} 
Since we are interested in the thermodynamic limit, we take the continuum limit. Substituting $2\pi\beta k/L = t$, we have $\sum_{k\in \mathbb{Z}\cup \{0\}} \rightarrow (L/2\pi\beta) \int_{0}^{\infty}dt$. Therefore we have,
\begin{align}
\log(P) &= -\log(2) - \frac{L}{2\pi\beta} \int_{0}^{\infty} dt\, \log \left( \frac{1+e^{-t}e^{-\pi\beta/L}}{1+e^{-t}} \right) \nonumber \\
&= -\log(2) + \frac{1}{2} \int_{0}^{\infty} dt\left(\frac{e^{-t}}{1+e^{-t}}\right)  + \mathcal{O}\left(\frac{1}{L}\right) \\
&= -\log(2) + \frac{1}{2} \log (2) + \mathcal{O}\left(\frac{1}{L}\right) \xrightarrow{L\rightarrow \infty} \log\left(\frac{1}{\sqrt{2}}\right)
\end{align}
As mentioned before, a free system is described by the sum of  free Hamiltonians on each edge of the interferometer and therefore, the total partition function will be the product of the two partition functions obtained above. Thus,
\begin{align}
\frac{Z^{\text{total}}_{\psi}}{Z^{\text{total}}_{\sigma}} = \frac{1}{2}.
\end{align}
This then gives the detailed balance condition $p_{\sigma\psi}^{\sigma} = 2p_{\sigma\sigma}^{\psi}$, and from the comparison with the tunneling probabilities from Sec.~\ref{subsec:Ising_MZ},
\begin{align}
    p_{\sigma\psi}^{\sigma} = p(T_1,T_2) |\Gamma_1-\Gamma_2 e^{i\alpha}|^2,~~~~~~~~~~ p_{\sigma\sigma}^{\psi} = \frac{p(T_1,T_2)}{2} |\Gamma_1-\Gamma_2 e^{i\alpha-i\pi/4}|^2,
\end{align}
we obtain the phase $\alpha = \pi/8+n\pi$. The choice of the integer $n$ has no effect on physics. As a consistency check, it can be seen that this choice of $\alpha$ also makes the tunneling Hamiltonian (\ref{eq:mach-zehnder_tunneling-ham}) Hermitian. Indeed this phase can also be directly computed by demanding Hermiticity condition as we now demonstrate.

\begin{figure}[htbp]
    \centering
    \includegraphics[scale=1]{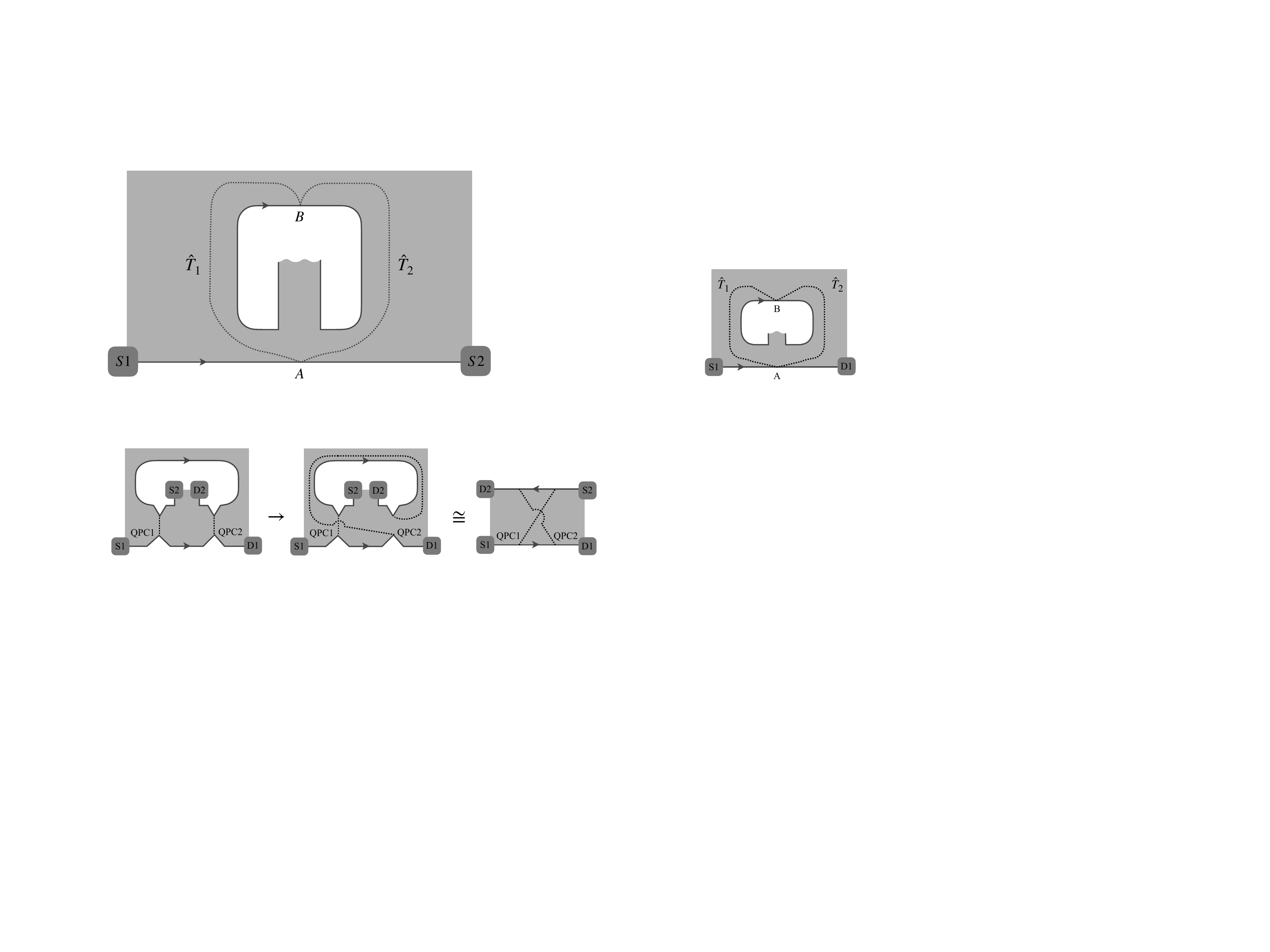}
    \caption{In the low-energy effective model of a Mach-Zehnder interferometer, the two tunneling operators $\hat{T}_{1,2}$ describe anyon transfer along the two dashed lines between points $A$ and $B$.}
    \label{fig:appendix-MZ_phase}
\end{figure}

 In this discussion we do not specialize to the case of the Ising statistics and only assume that the tunneling anyon $x$ is its own antiparticle.
We simply demand that the tunneling Hamiltonian $H_T=\Gamma_1\hat{T}_1+\Gamma_2\hat{T}_2e^{i\alpha}$ be Hermitian, and compare $H_T$ and $H_T^{\dagger}$. Note that this Hamiltonian corresponds to the case when the tunneling anyon $x$ is its own anti-particle. The  case $x\ne\bar x$  is treated in Secs.~\ref{sec-V-C-2} and \ref{subsec:non_abelian_not_anti}.

We observe that the statistical phase $\exp(i\phi_{ab}^c)$ describes the phase accumulated by an anyon $a$ on a full counterclockwise circle around an anyon $b$ under the assumption that the two anyons fuse to $c$. Restricting ourselves to the low-energy effective model, we see that the two tunneling operators $\hat{T}_{1,2}$, transfer anyons between the same points, taking them from point $A$ on the lower edge to point $B$ on the upper edge, see Fig.~\ref{fig:appendix-MZ_phase}. This forms a closed loop, and therefore allows us to relate the two tunneling operators, restricted to a particular topological sector, via a statistical phase $\exp(i\phi_{ab}^c)$. However, since we deal with non-Abelian anyons, we also need to restrict ourselves to a particular fusion channel. Therefore we introduce the projectors $\Pi_{a}=\Pi^{\dagger}_a$ that project to the sector with the trapped topological charge $a$. Using this we can now relate the two tunneling operators as,
\begin{align}%
\label{E12}
    \Pi_c \hat{T}_2 \Pi_b = \Pi_c \hat{T}_1 \Pi_b \exp(i\phi_{x b}^c).
\end{align}
The Hermiticity of the tunneling Hamiltonian requires $\hat{T}_2e^{i\alpha} = \hat{T}^{\dagger}_2e^{-i\alpha}$. We can now restrict this tunneling operator to a particular fusion channel,
\begin{align}
\label{E13}
    \Pi_c \hat{T}_2 \Pi_b e^{i\alpha} = (\Pi_b \hat{T}_2 \Pi_c)^{\dagger} e^{-i\alpha}.
\end{align}
We observe that the operator on the left hand side transfers an anyon $x$ from point $A$ on the lower edge to point $B$ on the upper edge when initially the trapped topological charge is $b$, and at the end, $x$ and $b$ fuse to $c$ inside the Mach-Zehnder interferometer. The operator on the right hand side however describes the reverse process. We now make use of the relation in Eq.~(\ref{E12}) and 
the relation
$\hat T_1=\hat T_1^\dagger$ 
to obtain,
\begin{align}
\label{E14}
    e^{2i\alpha} \Pi_c \hat{T}_1 \Pi_b = \theta_{x}^2 \Pi_c \hat{T}_1 \Pi_b ,
\end{align}
where we used the equation $\exp(i\phi_{x b}^c) = \theta_c/(\theta_{x}\theta_b)$. This fixes the phase $\exp(i\alpha) = \pm \theta_{x}$. 

We now turn to another method to compute the phase $\alpha$ again using the principle of detailed balance. The heat current goes to zero in the case when the temperatures of the two edges are equal, however, the tunneling probabilities are non-zero. For a stationary distribution, the transition probabilities $p_{x a}^b$  satisfy the detailed balance conditions between the processes $x\times a \rightarrow b$ and $x\times b \rightarrow a$.
The probabilities are given by Eqs.~(\ref{AE1}) and (\ref{AE2}) with $x$ in place of $\sigma$.
Notice that the probabilities of these processes differ only by their partition functions calculated in a particular superselection sector. We computed the ratio of the partition functions in different superselection sectors explicitly for the Ising anyon case and found it to be a constant that in turn gives the detailed balance condition. In fact we see that for a general case, the ratio of the partition functions, each computed in a particular superselection sector, is independent of the tunneling amplitudes $\Gamma_{1,2}$, and therefore the ratio of the probabilities $p_{x a}^b/p_{x b}^a$ is independent of $\Gamma_{1,2}$. This restriction allows us to compare the probabilities of the mutually reverse processes and fix the phase $\alpha$, provided the tunneling anyon, in this case anyon $x$, is other than a fermion or a boson. We may now compare the probabilities for the two processes $x \times a \rightarrow b$ and $x \times b \rightarrow a$, 
\begin{eqnarray}
\label{E16}
    p_{x a}^b = N^b_{xa}\frac{d_b}{d_x d_a}|\Gamma_1+\Gamma_2e^{i\alpha+i\phi_{xa}^b}|^2p(T,T), ~~~~~~~~ \text{and}~~~~~~~~ p_{x b}^{a} = N^a_{xb}\frac{d_a}{d_x d_b} |\Gamma_1+\Gamma_2e^{i\alpha+i\phi_{x b}^a}|^2p(T,T).
\end{eqnarray}
As argued above, their ratio should be independent of $\Gamma_{1,2}$, which amounts to saying that the ratio $\lambda \equiv \frac{|\Gamma_1+\Gamma_2\exp({i\varphi_1})|^2}{|\Gamma_1+\Gamma_2\exp({i\varphi_2})|^2}$ is independent of $\Gamma_{1,2}$, where we defined $\varphi_1 = \alpha + \phi_{x a}^b$ and $\varphi_{2} = \alpha+\phi_{x b}^a$. This gives a condition on the angles $\varphi_{1,2}$,
\begin{align}
    \cos(\varphi_1) - \lambda \cos(\varphi_2) = \frac{(\lambda-1)(\Gamma_1^2+\Gamma_2^2)}{2\Gamma_1\Gamma_2}.
\end{align}
We want this condition to be independent of $\Gamma_{1,2}$ for arbitrary $\Gamma_{1,2}$. This is only possible when $\lambda=1$.  This then gives, $\cos(\varphi_1)=\cos(\varphi_2)$, or $\varphi_1 = 2n\pi \pm \varphi_2$, where $n\in \mathbb{Z}$. The first case, $\varphi_1 = \varphi_2 \text{ mod } 2\pi$, puts no restriction on $\alpha$, when $\phi_{xa}^b=\phi_{xb}^a \text{ mod }2\pi$. This however implies that $\phi_{x \bm{1}}^x=\phi_{xx}^{\bm{1}}$.
This is only possible for $\theta_x=\pm 1$, i.e., for tunneling particles, which are bosons or fermions. 
Assuming that is not the case, we use the other condition, $\varphi_1 = - \varphi_2 \text{ mod }2\pi$, to obtain a restriction on $\alpha$,
\begin{align} 
    e^{i\alpha} = \pm e^{-i\frac{(\phi_{x a}^b+\phi_{x b}^a)}{2}}.
\end{align}
Upon using $\exp(i\phi_{ab}^c)=\frac{\theta_{c}}{\theta_{a}\theta_{b}}$, we get $e^{i\alpha}= \pm\theta_x$ consistent with the Hermiticity condition, as well as the phase found for the special case of Ising anyons.

\section{Topological order with unusual fusion rules}\label{appendix:anyon_example}
In Sec.~\ref{subsec:non_abelian_not_anti}, we consider anyon models having the following property: if $x\neq \bar{x}$, the fusion results of $ x \times x$ and $ a \times  \bar{a}$ share no common topological charge for all possible $ a $. It is not true that all anyon models satisfy this requirement. As a counterexample, an anyon model is constructed in this Appendix through the approach used in Ref.~\cite{bais1992:quantum_symmetries}.

We briefly summarize the approach in this paragraph.
Given a discrete group $ \bar{H} $, one can extend it to a quasi-triangular Hopf algebra $ D(\bar{H}) $ with the basis $ \{\begin{gathered}
\includegraphics{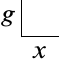}
\end{gathered} \}_{g,x\in \bar{H}} $. 
Different representations of the Hopf algebra $ D(\bar{H}) $ are interpreted as anyons with different topological charges 
(called superselection sectors in~\cite{bais1992:quantum_symmetries}), and the representations are labeled in two steps: (1) find the $ A $th conjugacy class $ {^A}C $ of $ H $, of which $ {^A}g_1 $ is an element; (2)  find the $ \alpha $th representation $ {^\alpha} \Gamma $ of the centralizer $ {^A}N $ of $ {^A}g_{1} $ in $ \bar{H} $.
We use $ |{^A}C, {^\alpha} \Gamma \rangle $ or $ \Pi_{\alpha}^{A} $ to denote the topological charges or representations. An explicit construction of a representation of $ D(\bar{H}) $ is as follows: 
(1) Let $ {^A}C=\{{^A}g_1,{^A}g_2,\dots,{^A}g_k\} $; (2) choose representatives $ \{{^A}x_1,{^A}x_2,\dots,{^A}x_k\} $ of the equivalence classes in $ \bar{H}/{^A}N $ under the requirement $ {^A}g_i = {^A}x_i {^A}g_1 {^A}x_{i}^{-1} $; (3) denote basis elements of the irreducible representation $ {^\alpha} \Gamma $ as $ {^\alpha}v_j $; (4) in the vector space $ V_{\alpha}^{A} $ spanned by $ |  {^A}g_i, {^\alpha} v_j \rangle $, where $ i=1,2,\dots,\dim{\alpha} $ and $ j=1,2,\dots,k $, a representation of $ D(\bar{H}) $ is
\begin{equation}
\Pi_{\alpha}^{A} (\begin{gathered}\includegraphics{basis_gx.pdf}\end{gathered})
|  {^A}g_i, {^\alpha} v_j \rangle
=
\delta_{g, x\, {^A}g_{i}\, x^{-1}}  | x\, ^{A} g_{i}\, x^{-1} , {^\alpha}\Gamma\left({^A}x_{l}^{-1} \, x\, {^A} x_{i}\right) {^\alpha} v_{j} \rangle,
\end{equation}
where the index $ l $ is chosen by letting $ {^A}g_{l} = x {^A}g_{i} x^{-1} $.
To find the fusion rules, we consider the tensor product of two representations of the algebra $ D(\bar{H}) $, $ \Pi_{\alpha}^{A} \otimes \Pi_{\beta}^{B} $.
Using comultiplication $ \Delta: D(\bar{H}) \to D(\bar{H}) \otimes D(\bar{H}) $, one interprets the product  as
another representation of $ D(\bar{H}) $. The decomposition of the tensor product into irreducible representations, 
\begin{equation}
\Pi_{\alpha}^{A} \otimes \Pi_{\beta}^{B} = N_{\alpha\beta\gamma}^{ABC} \Pi_{\gamma}^{C},
\end{equation}
is known as a fusion rule.

Here we only consider the sector whose conjugacy class is that of the identity element $ e $ of the group $ \bar{H} $, also known as the magnetic vacuum sector.
The conjugacy class is the set $ {^1}C = \{e\} $. 
The centralizer $ {^1}N $ is thus $ \bar{H} $, and $ \bar{H}/{^1}N $ contains only one equivalence class, namely $ \bar{H} $, whose representative is chosen to be $ e $. For a unitary irreducible representation $ {^\alpha} \Gamma $ of $ {^1}N = \bar{H} $, we can find the representation $  \Pi^{1}_{\alpha} $ of $ D(\bar{H}) $ on the vector space spanned by the basis $ | e, {^1}v_i\rangle $ as
\begin{equation}
\Pi_{\alpha}^{1} ( 
\begin{gathered}
\includegraphics{basis_gx.pdf} 
\end{gathered}  ) |e, {^\alpha} v_{i} \rangle  
= \delta_{g,e} |e, {^\alpha}\Gamma (x) {^\alpha} v_{i} \rangle  .
\end{equation}
The above representations of $ D(\bar{H}) $ are equivalent to those of $ \bar{H} $.
The tensor product of two representations of $ \bar{H} $ can be decomposed into irreducible representations of $ \bar{H} $. Hence, the fusion rule for $ \Pi^{1}_{\alpha} $,
$\Pi^{1}_{\alpha} \otimes \Pi^{1}_{\beta} = N_{\alpha \beta 1}^{1 1 \gamma} \Pi^{1}_{\gamma} $,
is equivalent to 
${^\alpha}\Gamma \otimes {^\beta}\Gamma = N_{\alpha \beta 1}^{1 1 \gamma} {^\gamma}\Gamma $.

\begin{table}[htbp]
\begin{tabular}{c|cccccccccc} 
\toprule
class & 1 & 2 & 4A & 4B & 7A & 7B & 7C & 7D & 7E & 7F   \\
size & 1 & 7 & 28 & 28 & 64 & 64 & 64 & 64 & 64 & 64  \\
\midrule
$\rho_{1}$ & 1 & 1 & 1 & 1 & 1 & 1 & 1 & 1 & 1 & 1  \\
$\rho_{2}$ & 1 & 1 & 1 & 1 & $\zeta_{7}^{4}$ & $\zeta_{7}^{6}$ & $\zeta_{7}^{2}$ & $\zeta_{7}^{5}$ & $\zeta_{7}$ & $\zeta_{7}^{3}$  \\
$\rho_{3}$ & 1 & 1 & 1 & 1 & $\zeta_{7}^{2}$ & $\zeta_{7}^{3}$ & $\zeta_{7}$ & $\zeta_{7}^{6}$ & $\zeta_{7}^{4}$ & $\zeta_{7}^{5}$  \\
$\rho_{4}$ & 1 & 1 & 1 & 1 & $\zeta_{7}^{5}$ & $\zeta_{7}^{4}$ & $\zeta_{7}^{6}$ & $\zeta_{7}$ & $\zeta_{7}^{3}$ & $\zeta_{7}^{2}$  \\
$\rho_{5}$ & 1 & 1 & 1 & 1 & $\zeta_{7}^{3}$ & $\zeta_{7}$ & $\zeta_{7}^{5}$ & $\zeta_{7}^{2}$ & $\zeta_{7}^{6}$ & $\zeta_{7}^{4}$  \\
$\rho_{6}$ & 1 & 1 & 1 & 1 & $\zeta_{7}$ & $\zeta_{7}^{5}$ & $\zeta_{7}^{4}$ & $\zeta_{7}^{3}$ & $\zeta_{7}^{2}$ & $\zeta_{7}^{6}$  \\
$\rho_{7}$ & 1 & 1 & 1 & 1 & $\zeta_{7}^{6}$ & $\zeta_{7}^{2}$ & $\zeta_{7}^{3}$ & $\zeta_{7}^{4}$ & $\zeta_{7}^{5}$ & $\zeta_{7}$  \\
$\rho_{8}$ & 7 & 7 & $-1$ & $-1$ & 0 & 0 & 0 & 0 & 0 & 0  \\
$\rho_{9}$ & 14 & $-2$ & $-2 i$ & $2 i$ & 0 & 0 & 0 & 0 & 0 & 0  \\
$\rho_{10}$ & 14 & $-2$ & $2 i$ & $-2 i$ & 0 & 0 & 0 & 0 & 0 & 0  \\
\bottomrule
\end{tabular}
\caption{Character table of the group $ C_2^3.F_8 $, adapted from Ref.~\cite{dokchitser}. This group has 10 conjugacy classes, each of them corresponds to one column in the table. There are 10 irreducible representations, whose characters are shown in the rows. $\zeta_{7}$ is used to denote $\exp(2\pi i/7)$ in this table.}
\label{tab:char_table}
\end{table}

Now we provide an example where $ \bar{H} = C_2^3.F_8  $ is a group of order 448. Table~\ref{tab:char_table} is its character table. 
Anyons corresponding to the representations in the table are denoted as $ |e, \rho_{i} \rangle $.
Note that this should not be confused with the basis $ | e, {^\alpha} v_{i} \rangle$ used above.

Observe that $ \rho_9 $ and $ \rho_{10} $ are conjugate representations and  their tensor product is
$\rho_{9} \times \rho_{10} = \rho_{1} + \rho_{2} + \dots + \rho_{7}
+ 3 \rho_{8} + 6 \rho_{9} + 6 \rho_{10}$.
The tensor product of $ \rho_{9} $ with itself is
$\rho_{9} \times \rho_{9} = 4 \rho_{8} + 6 \rho_{9} + 6 \rho_{10}$.
Identifying $ |e, \rho_{1} \rangle $ as the vacuum, we find that $ |e, \rho_{9} \rangle $ and $ |e, \rho_{10} \rangle $ are mutual antiparticles, and further $ |e, \rho_{9} \rangle \times |e, \rho_{9} \rangle \to |e, \rho_{8} \rangle $, $ |e, \rho_{10} \rangle \times |e, \rho_{9} \rangle \to |e, \rho_{8} \rangle $.
In the tunneling problem we considered, this means that both $ |e, \rho_{9} \rangle $ and its antiparticle contribute to the process where the topological charge on edge 2 changes from $ |e, \rho_{9} \rangle $ to $  |e, \rho_{8} \rangle  $. Hence, in general, one cannot separate the operators $ T $ and $ T^{\dagger} $ as done in Eq.~(\ref{eq:current_golden_not_anti}). 
However, as seen in this example, anyon models with such a property are complicated, so we omit the discussion of those models in the main text.

\section{Calculation of the average noise}\label{appendix:noise}
This appendix provides details of the noise calculation from Sec.~\ref{sec:noise}. Specifically, we calculate the expectation value of some expressions containing the heat current with respect to the random variables $ s_{2k-1} $ and $ t_k $. First, consider the expectation value of the heat current $ I_{T}(t) $, Eq.~(\ref{eq:heat_current_rand}),
\begin{equation}
\langle I_{T}(t) \rangle = I_0 + \Delta I \sum_{k=1}^{\infty} \langle s_k\rangle \theta(t-t_k).
\end{equation}
After averaging over the Bernoulli random variables $ s_{2k-1} $, the $ \Delta I $ term vanishes, so we find that $ \langle I_{T}(t) \rangle = I_0 $.
Then we consider the expression $ I_T(0)I_T(t) $ in the definition of the noise (Eq.~(\ref{eq:noise_def})),
\begin{equation}
I_T(0)I_T(t) = 
I_0^2 + I_0 \Delta I \Bigg( \sum_{k=1}^{\infty} s_k \theta(-t_k) + \sum_{k=1}^{\infty} s_k \theta(t-t_k)  \Bigg)
+ (\Delta I)^2  \Bigg( \sum_{k=1}^{\infty} s_k \theta(-t_k) \Bigg) \Bigg( \sum_{k=1}^{\infty} s_k \theta(t-t_k)  \Bigg).
\end{equation}
To find the average, we choose to average over the Bernoulli random variables $ s_{2k-1} $ first, then we take the average over the Poisson process $ t_k $.
First, when averaging over all $s_{2k-1}$'s, only terms containing $s_{2k-1}s_{2k}=-1$ or $(s_k)^2=1$ are non-zero.
\begin{equation}
\langle I_{T}(0)I_{T}(t) \rangle_{\mathrm{avg.}}  
= I_0^2 + (\Delta I)^2 \Bigg( \sum_{k=1}^{\infty}  -\theta(-t_{2k-1}) \theta(t-t_{2k})-\theta(-t_{2k})\theta(t-t_{2k-1})   + \sum_{l=1}^{\infty} \theta(-t_l)\theta(t-t_l) \Bigg).
\end{equation}
Following the definition in Eq.~(\ref{eq:number_events_def}), $ N(0) $ is used to denote the total number of tunneling events up to the time $ t=0 $, and our subsequent analysis will depend on the parity of $ N(0) $.
When $ N(0) $ is odd, we have $t_{2n-1}\leq 0 < t_{2n}$. 
The coefficient of $(\Delta I)^2$ is found to be %
\begin{equation}
-\sum_{k=1}^{n}\theta(-t_{2k-1}) \theta(t-t_{2k})-\sum_{k=1}^{n-1}\theta(-t_{2k})\theta(t-t_{2k-1})  + \sum_{l=1}^{2n-1} \theta(-t_l)\theta(t-t_l)  = \theta(t-t_{2n-1}) - \theta(t-t_{2n}).
\end{equation}
The following integral appearing in the definition of the zero-frequency noise gives,
\begin{equation}
\lim_{\omega\to 0  } \int d t\, e^{i\omega t} \big[ \langle I_{T}(0)I_{T}(t) \rangle_{\mathrm{avg.}} - \langle I_{T}(t) \rangle \langle I_{T}(0) \rangle  \big]
= (\Delta I)^2 (t_{2n}-t_{2n-1}).
\end{equation}
On the other hand, $N(0)$ being even is equivalent to the condition $t_{2n}\leq 0 < t_{2n+1}$. The coefficient of $(\Delta I)^2$ vanishes in such case:
\begin{equation}
  -\sum_{k=1}^{n}\theta(-t_{2k-1}) \theta(t-t_{2k})-\sum_{k=1}^{n}\theta(-t_{2k})\theta(t-t_{2k-1})  + \sum_{l=1}^{2n} \theta(-t_l)\theta(t-t_l)  = 0.
\end{equation}
Thus, the corresponding integral in the definition of the noise also vanishes.

Finally we can find the average integral under the Poisson process given by the $ \tau_{k} $'s. Since we know that 
only odd $N(0)$ contribute, the expectation value could be written as
\begin{equation}
(\Delta I)^2 \sum_{\substack{\text{odd } n,\\n>0}} P(N(0)=n)  \int_{t_0}^{0} d {t_n} \int_{0}^{\infty} d t_{n+1}\, f(t_{n+1}, t_n) \,   (t_{n+1}-t_{n}) , 
\end{equation}
where $ f(t_{n+1}, t_n) $ is the joint conditional probability distribution function for $ t_{n} $ and $ t_{n+1} $ under the condition $ N(0) = n $.
Poisson process has the property that it is memoryless:
given $N(0)=n$, $t_{n+1}$ is exponentially distributed with the parameter $\lambda$ and independent of the history up to time $0$. 
Furthermore the distributions of $t_{n+1}$ and $t_{n}$ are independent, and we know that%
 \begin{equation}
 E(t_{n+1}|N(0)=n) = \tau, \qquad E(t_{n}|N(0)=n) = \frac{t_0}{n+1}.
 \end{equation}
So the average coefficient of $(\Delta I)^2$ is
\begin{align}
\sum_{\substack{\text{odd } n,\\n>0}}   e^{\lambda t_0} \frac{(-\lambda t_0)^n}{n!} \Big(\tau - \frac{t_0}{n+1}\Big) 
&=  e^{\lambda t_0} \tau \sinh(-\lambda t_0)  + e^{\lambda t_0} \frac{1}{\lambda} \sum_{\text{odd } n} \frac{(-\lambda t_0)^{n+1}}{(n+1)!} \\
&=  e^{\lambda t_0} \tau \sinh(-\lambda t_0) + e^{\lambda t_0} \tau [\cosh(-\lambda t_0)-1] \\
&= \tau (1- e^{\lambda t_0} ).
\end{align}
It is now straightforward to obtain Eq.~(\ref{eq:noise_result}) in Sec.~\ref{sec:noise}.

\section{Dependence of the tunneling heat current on the size of the interferometer}\label{appendix:inter_size}
In this appendix, we discuss the dependence of the tunneling heat current on the interferometer size. 
We consider topological orders allowing a single edge mode. One of such orders is the Ising order in Kitaev liquids. For simplicity, we assume that the edge velocity is coordinate-independent and identical on both edges. 
It is easy to generalize our results to coordinate-dependent velocities. 
We also assume that the tunneling operators at the two constrictions have precisely the same structure except for an overall amplitude multiplying the tunneling operator. The latter assumption is true for many topological orders as long as it is legitimate to focus on only the most relevant tunneling operator. As claimed in Sec.~\ref{sec:model_FP} and \ref{sec:model_mach-zehnder},
the tunneling heat current in a
Fabry-P\'erot interferometer depends on the sum of the distances
between the tunneling contacts along the two edges, $ L_1 + L_2$.  The heat current in a Mach-Zehnder interferometer depends on the difference of the distances $ |L_1 - L_2| $. 

We consider the tunneling of anyons of type $ x $ with an arbitrary statistics: $  \hat{x}_{1,2} (y,t) $ is the operator that creates an anyon with topological charge $ x $ on edge 1 or 2 of the interferometer at position $ y $ and time $ t $, and $  \hat{\bar{x}}_{1,2} (y,t) $ creates its antiparticle.
The two tunneling operators across one constriction are Hermitian conjugate to each other; we use $ \hat{T}\sim \hat{x}_{2} (y,t) \hat{\bar{x}}_{1} (y,t) $ and $ \hat{T}^\dagger \sim \hat{x}_{1} (y,t) \hat{\bar{x}}_{2} (y,t) $ to denote them.
We adopt the following convention: when considering two separate edges, lower and upper, the coordinate axes on the two edges are always chosen to be in the same right-moving direction.

For a Fabry-P\'erot geometry, on the lower edge (edge 1), the two constrictions are labeled by their coordinates $ y_1 $ and $ y_2 $, and the distance between them is $ L_1 = y_2 - y_1  $. 
We choose the coordinates on the upper edge (edge 2) to be $ y_1 $ and $ y_2+\ell $, then $ L_2 = L_1 +\ell $.
The tunneling Hamiltonian is given by 
\begin{equation}
H_T = \Gamma_1 \hat{T}_1 + \Gamma_2 \hat{T}_2 + \mathrm{H.c.} 
= \Gamma_1 \hat{x}_{2} (y_1,t) \hat{\bar{x}}_{1} (y_1,t) + \Gamma_2 \hat{x}_{2} (y_2+\ell,t) \hat{\bar{x}}_{1} (y_2,t) + \mathrm{H.c.}
\end{equation}
By the same argument as in Appendix~\ref{appendix:sigma}, the operator for the tunneling heat current is
\begin{equation}
I_T = \Gamma_1 \hat{T'}_1 + \Gamma_2 \hat{T'}_2 + \mathrm{H.c.} 
= - \Gamma_1 \hat{x}_{2} (y_1,t) \partial_{t} \hat{\bar{x}}_{1} (y_1,t) - \Gamma_2 \hat{x}_{2} (y_2+\ell,t) \partial_{t}\hat{\bar{x}}_{1} (y_2,t) + \mathrm{H.c.},
\end{equation}
where $ \hat{T'}_{1,2} = - \hat{x}_{2} (y_1,t) \partial_{t} \hat{\bar{x}}_{1} (y_1,t) $ contains the time derivative on $ \hat{\bar{x}}_1 $.
The derivative should be computed in the theory without tunneling.
One can find the interference terms in the expectation value of the heat current as
\begin{equation}\label{eq:FP_current_size}
-i \int_{-\infty}^{t} dt'\, \left\{ 
\Gamma_1 \Gamma_2^* \langle[ \hat{T'}_1(t), \hat{T}_2^\dagger(t')]\rangle + \Gamma_1 \Gamma_2^* \langle[\hat{T'}_2^\dagger(t), \hat{T}_1(t')]\rangle - \mathrm{H.c.} 
\right\}.
\end{equation}
Due to the translational symmetry, the two-point correlation functions can be expressed as
\begin{align}
\langle \hat{x}_1(y_1, t_1) \hat{\bar{x}}_1(y_2, t_2) \rangle &= \langle \hat{\bar{x}}_1(y_1, t_1) \hat{x}_1(y_2, t_2) \rangle = G_1(t_1-t_2-y_1+y_2),\\
\langle \hat{x}_2(y_1, t_1) \hat{\bar{x}}_2(y_2, t_2) \rangle &= \langle \hat{\bar{x}}_2(y_1, t_1) \hat{x}_2(y_2, t_2) \rangle = G_2(t_1-t_2+y_1-y_2).
\end{align}
Since the correlation functions of the form $ \langle \hat{x}_2(y_1) \hat{\bar{x}}_1(y_2) \hat{x}_1(y_3) \hat{\bar{x}}_2(y_4) \rangle $ can be decomposed into conformal blocks $ \langle \hat{x}_2(y_1) \hat{\bar{x}}_2(y_4) \rangle \langle \hat{\bar{x}}_1(y_2) \hat{x}_1(y_3) \rangle $, we find the following relations,
\begin{align}
\langle\hat{T'}_1(t) \hat{T}_2^\dagger(t')\rangle - \langle \hat{T}_1(t') \hat{T'}_2^\dagger(t)\rangle \propto
G'_1(t-t'+y_{21})G_2(t-t'-y_{21}-\ell)+(t\leftrightarrow t'),
\end{align}
where $ G' $ denotes the derivative of $ G $, and 
\begin{align}
\langle \hat{T'}_2^\dagger(t) \hat{T}_1(t') \rangle - \langle \hat{T}_2^\dagger(t') \hat{T'}_1(t)\rangle \propto
G'_1(t-t'-y_{21}) G_2(t-t'+y_{21}+\ell)+(t\leftrightarrow t').
\end{align}
Hence, the integral in Eq.~(\ref{eq:FP_current_size}) can be written as
\begin{align}
\int_{-\infty}^{t} dt'\, \left\{
\langle[ \hat{T'}_1(t), \hat{T}_2^\dagger(t')]\rangle + \langle[\hat{T'}_2^\dagger(t), \hat{T}_1(t')]\rangle
\right\} \propto
\int_{-\infty}^{\infty} d\tau\,
[G'_1(\tau)G_2(\tau-2y_{21}-\ell)
+ G'_1(\tau)G_2(\tau+2y_{21}+\ell)].
\end{align}
Combining this equation with Eq.~(\ref{eq:FP_current_size}), we find that the tunneling heat current in a Fabry-P\'erot interferometer depends on $ L_1 + L_2 = 2 y_{21}+\ell $.

For the Mach-Zehnder geometry, we make the same choice for the coordinates of the two constrictions on the upper and lower edges.
Naively, the tunneling Hamiltonian can be written as,
\begin{equation}
H_T = \Gamma_1 \hat{T}_1 + \Gamma_2 \hat{T}_2 + \mathrm{H.c.} = 
\Gamma_1 \hat{x}_{2} (y_1,t) \hat{\bar{x}}_{1} (y_1,t) + \Gamma_2 \hat{x}_{2} (y_2+\ell,t) \hat{\bar{x}}_{1} (y_2,t)
 + \mathrm{H.c.}
\end{equation}
This naive Hamiltonian is actually incorrect though it will be useful to us below. The problem is that the above Hamiltonian violates locality: 
the two tunneling operators do not commute. The issue can be fixed in a systematic way with the help of Klein factors. Besides fixing the commutativity problem, the Klein factors keep track of the confined topological charge.

\begin{figure}[htbp]
    \centering
    \includegraphics{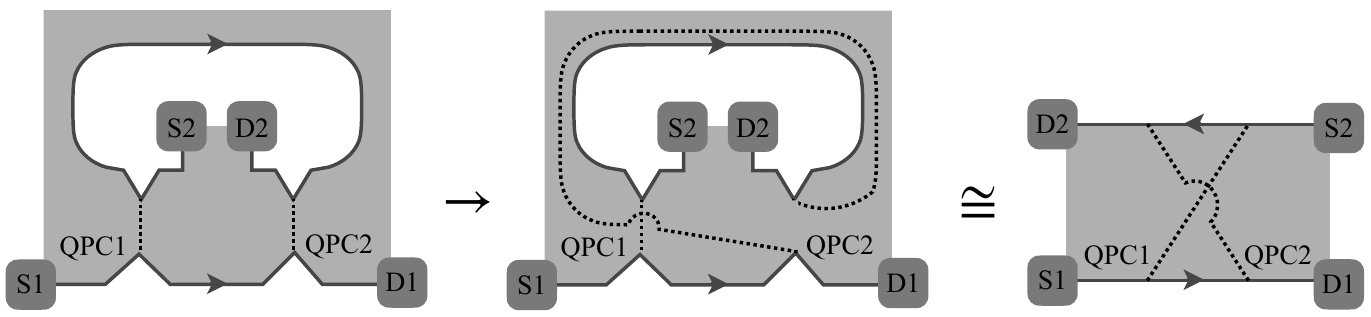}
    \caption{Topology of a Mach-Zehnder interferometer. Dotted lines show tunneling contacts. The paths through QPC2 in the left panel can be deformed as in Appendix~\ref{appendix:Mach-zehnder_phase}. This results in the configuration from the right panel.}
    \label{fig:MZ_topology}
\end{figure}

An alternative approach is based on Appendix~\ref{appendix:Mach-zehnder_phase}. The tunneling paths in the two constrictions run on the two sides of the hole in the interferometer. This leads to a number of technical challenges. In particular, it becomes impossible to connect the upper and lower edges as in Appendix~\ref{appendix:ising-correlation}. 
The problem can be solved by flipping one of the two tunneling paths to the other side of the hole (Fig.~\ref{fig:MZ_topology}) as in Appendix~\ref{appendix:Mach-zehnder_phase}. 
Klein factors are no longer needed after that since the topological charge between the tunneling paths no longer changes after each tunneling event. Also, the tunneling operators no longer have to commute
since the two tunneling paths cross (Fig.~\ref{fig:MZ_topology}). 
Of course, this comes at the price of the model being useful for computing the tunneling probability only for a given value of the trapped topological charge in a given fusion channel with the tunneling anyon. This is enough, however, for the purposes of this appendix and justifies the use of the above non-local Hamiltonian.

In contrast to the Fabry-P\'erot geometry, the two edges are co-propagating.
The correlation function on edge 2 should be of the form,
\begin{align}
\langle \hat{x}_2(y_1, t_1) \hat{\bar{x}}_2(y_2, t_2) \rangle 
= \langle \hat{\bar{x}}_2(y_1, t_1) \hat{x}_2(y_2, t_2) \rangle = G_2(t_1-t_2-y_1+y_2)
\end{align}
We focus on the interference contribution again, 
\begin{align}
\langle\hat{T'}_1(t) \hat{T}_2^\dagger(t')\rangle - \langle \hat{T}_1(t') \hat{T'}_2^\dagger(t)\rangle \propto
G'_1(t-t'+y_{21})G_2(t-t'+y_{21}+\ell)+(t\leftrightarrow t'),
\end{align}
and 
\begin{align}
\langle \hat{T'}_2^\dagger(t) \hat{T}_1(t') \rangle - \langle \hat{T}_2^\dagger(t') \hat{T'}_1(t)\rangle \propto
G'_1(t'-t-y_{21})G_2(t'-t-y_{21}-\ell)+(t\leftrightarrow t').
\end{align}
Comparing with the expression for the interference contribution to the tunneling heat current in the Fabry-P\'erot geometry, we can obtain the result for a Mach-Zehnder interferometer by replacing the parameter $ 2 y_{21} +\ell$ with $ \ell$.
We have consequently shown that the tunneling heat current in Mach-Zehnder interferometers  depends on the difference of the distances between the two point contacts on the two edges, i.e. $ |L_1 - L_2| $.

\twocolumngrid

\end{document}